\documentclass[11pt,a4paper]{article}
\usepackage{jheppub}
\usepackage{amsmath}
\usepackage{latexsym}
\title{\Large{A Unified Hidden-Sector Electroweak Model, Para-photons and the X-boson}}

\author{M. J. Neves $^{a}$}
\author{and J. A. Helay\"el-Neto $^b$}

\affiliation{$^a$Departamento de F\'{\i}sica, \\
Universidade Federal Rural do Rio de Janeiro\\
BR 465-07, 23890-971, Serop\'edica, Rio de Janeiro, Brazil \\\\
$^b$Centro Brasileiro de Pesquisas F\' isicas (CBPF),\\
Rua Dr. Xavier Sigaud 150, Urca, \\
CEP  22290-180, Rio de Janeiro, Brazil\\\\
\bigskip
\today\\}
\emailAdd{mariojr@ufrrj.br, helayel@cbpf.br}
\abstract{
Our contribution sets out to investigate the phenomenology of a gauge model based on an $SU_{L}(2) \times U_{R}(1)_{J} \times U(1)_{K}$-symmetry group.
The model can accommodate in its distinct phases - by virtue of different symmetry-breaking patterns - 
a candidate to a heavy $Z'$-boson at the TeV-scale and the recently discussed $17 \, \mbox{MeV}$ X-boson that may bring about a new physics at
the $\mbox{MeV}$-scale. Furthermore, the para-photon of the dark matter sector and the candidate to the so-called dark photon can also be
described in the scenario we are endeavoring to set up in this paper. In the $Z'$-scenario, the extended Higgs sector introduces a heavy scalar whose mass lies in the region
$1.2-3.7 \, \mbox{TeV}$, while that the lower energy scale scenario presents a light scalar with mass in the interval  $1.9-19 \, \mbox{GeV}$.
The fermionic sector includes an exotic candidate to dark matter that mixes with the right-neutrino component in the
Higgs sector, so that the whole field content yields the cancellation of the U(1)-anomaly.
The masses are fixed according to the particular way the symmetry breaking takes place. In view of the possible symmetry breakdown patterns, we study the phenomenological
implications in both scenarios of high- and low-energy scales. As examples, we consider the scattering of Standard Model matter into dark sector constituents
intermediated by the $Z'$-boson in the $\mbox{TeV}$-scale, and the corresponding scattering intermediated by a lightest boson in the lower scale physics.
Furthermore, we obtain the magnetic dipole momentum (MDM) of the exotic fermion and the transition MDM due to the mixing with the right-neutrino.
}

\usepackage{graphicx}
\usepackage{pstricks,pst-coil}
\usepackage{pst-all}
\usepackage{color}
\usepackage{latexsym}
\usepackage{amsmath}
\usepackage{amssymb}
\usepackage{eufrak}
\usepackage{euscript}
\usepackage{graphics}
\usepackage{picture}
\def\uma{\rm 1\!\!\hskip 1 pt l}
\begin{document}
%%%%%%%%%%%%%%%%%%%%%%%%%%%%%%%%%%%%%%%%%%%%%%%

\maketitle

\pagestyle{myheadings}
\markright{TeV- and MeV- scale physics : The $Z'$- and $X$-boson phenomenology}

%
%
%%%%%%%%%%%%%%%%%%%%%%%%%%%%%%%%%%%%%%%%%%%%%%%%%%%%%%%%%%%%%%%%%%%%%%%%%%%%%%%%%%%%%%%%%%%%%%%
%

%
\section{Introduction}
\renewcommand{\theequation}{1.\arabic{equation}}
\setcounter{equation}{0}

The search for new particles and interactions beyond the Standard Model (SM) has been challenging High-Energy Physics, both Theoretical and
Experimental, over the recent decades. The results from the LHC's ATLAS- and CMS-Collaborations may point to the existence of a (new) fifth
interaction \cite{Atlas,CMS}. Analysing data of $pp$-collisions at a center-of-mass energy scale of $\sqrt{s}= 13 \, \mbox{TeV}$ has revealed
peaks of masses that may correspond to the hypothetical heavy $W^{\, \prime}$- and $Z^{\, \prime}$-bosons \cite{CMS20171,CMS20172},
which may be an evidence for new Physics communicated from very high energies to the $\mbox{TeV}$-scale.
We should keep in mind that the introduction of an extra Higgs may be needed to explain the heavy mass of the new bosons at the $\mbox{TeV}$-scale.
These masses are associated with a new range of vacuum expectation values (VEVs) of supplementary Higgs scalars.

A well-known model in this direction is based on an $SU_{L}(2)\times SU_{R}(2) \times U(1)_{B-L}$-gauge symmetry \cite{DobrescuJHEP2015,DobrescuPRL2015,DobrescuPRD2015,DobrescuJHEP2016,DevPRL2015,Patra2016,HongGuPRD2017,Dev2017}.
The extra $SU_{R}(2)$-subgroup is introduced to account for the hypothetical $W^{\, \prime}$- and $Z^{\, \prime}$-bosons,
besides the already known $W^{\pm}$- and $Z^{0}$-weak mediators of the Glashow-Salam-Weinberg (GSW) model.
The right-handed sector also introduces fermion doublets with a charged lepton and
its associated neutrino of right chirality. Two scalar doublets constitute the Higgs sector responsible for the chain of symmetry breakings. The first
doublet is introduced to break the $SU_{R}(2) \times U(1)_{B-L}$-gauge symmetry at the scale set up by the VEV $u$, above the SM Electroweak scale $u > v = 246 \, \mbox{GeV}$. As a consequence, the hypercharge, $Y$, appears as a combination of generators of $SU_{R}(2) \times U(1)_{B-L}$.
Next, the Standard Model Higgs breaks the remaining gauge symmetry down to the eletromagnetic Abelian symmetry, $U(1)_{em}$.
%The scheme of these two spontaneous symmetry breaking is represented by
%
%\begin{eqnarray*}
%SU_{L}(2)\! \times \! SU_{R}(2) \! \times \! U(1)_{B-L}\! \stackrel{u}{\longmapsto}\!
%\nonumber \\
%SU_{L}(2) \! \times \! U_{Y}(1)
%\!\stackrel{v}{\longmapsto}\! U(1)_{em} \, ,
%\end{eqnarray*}
%
%where $u$ is a VEV bringing the $\mbox{TeV}$-scale of the LHC experiments.
Therefore, the recent CMS simulation indicates
%$\mbox{TeV}$-Scale establishes
that the estimate for masses of hypothetical $W'$ and $Z'$ is around the $2 \, \mbox{TeV}$ \cite{CMS20171,CMS20172}.
%
%\begin{eqnarray}
%m_{W'} \!=\! 2.0 \, \mbox{TeV}
%\; ,
%\nonumber \\
%\hspace{0.2cm} \mbox{and} \hspace{0.2cm}
%2.0 \, \mbox{TeV} <
%m_{Z'} = 2.5 \, \mbox{TeV} \; .
%\end{eqnarray}
%

The well-known model in the literature that describes only the $Z'$-heavy boson is based on the gauge symmetry $SU_{L}(2) \times
U_{Y}(1) \times U(1)_{B-L}$ \cite{LangackerRMP2009,Kanemura2011,MorettiarXiv2017}, which includes a $U(1)$-extra group in the GSW model.
The gauge sector has just one extra boson, while the Higgs is extended to include a doublet and a singlet, or bi-doublets of scalars fields.
The fermion sector is enlarged to guarantee anomaly cancellation; for example, the introduction of right-neutrinos components and exotic
fermions that could be candidates to dark matter content.

In 2016, in a particular Nuclear Physics experiment, anomalies in the decay of the excited state of $8\, \mbox{Be}^{\ast}$ to its ground state
has suggested the existence of a new neutral boson, called $X$, through the decay mode $8 \, \mbox{Be}^{\ast} \rightarrow 8 \, \mbox{Be} + X$
\cite{KrasPRL2016}. The $X$-boson immediately decays into an electron-positron pair $X \rightarrow e^{+} \, + \, e^{-}$.
It is a vector-type spin-1 neutral particle, with mass around $m_{X}=17 \, \mbox{MeV}$ that mixes with the SM photon through a kinetic term.
Its origin could, in principle, be traced back to an extra gauge symmetry, $U(1)_{X}$, in addition to the symmetries of the SM. The discovery of the X-boson may be pointing to the existence of a fifth fundamental interaction in Nature. This is an interesting scenario in which the hypothetical fifth interaction would be accessible at the $\mbox{MeV}$-scale.

The effective Lagrangian proposed to describe this extra $X$-boson is given below \cite{JFengPRD2017}
\begin{equation}\label{Leff}
{\cal L}_{eff}= - \frac{1}{4} \, F_{\mu\nu}^{\, 2}
- \frac{1}{4} \, X_{\mu\nu}^{\, 2}
+ \frac{\chi}{2} \, X_{\mu\nu} F^{\mu\nu}
%\nonumber \\
%&&
%\hspace{-0.5cm}
+ \frac{1}{2} \, m_{X}^{\, 2} \, X_{\mu}^{\, 2} - J_{\mu} X^{\mu} \, ,
\end{equation}
where $J^{\mu}$ is the current weakly coupled to $X^{\mu}$
\begin{eqnarray}\label{Jproto}
J^{\mu}= \!\!\!\!\! \sum_{\Psi \, = \, e \, , \, u \, , \, d \, , \, ...} \!\!\!\! e \, \chi_{\Psi} \, \bar{\Psi} \, \gamma^{\mu} \, \Psi \; ,
\end{eqnarray}
with $\Psi$ standing for any fermion of the SM. This vector current exhibits a protophobic character
described by the $\chi_{p}$-parameter whose magnitude for the proton and neutron satisfies the condition $\chi_{p} \ll \chi_{n}$,
whenever the $X$-boson interacts with the nucleon. We list some estimates
of $\chi_{\Psi}$ for the fermions of the SM, following the phenomenology of the $X$-boson \cite{FengPRL2016} :
\begin{eqnarray}\label{chiconstraints}
2 \times 10^{-4} < &|\chi_{e}|& < 1.4 \times 10^{-3} \; ,
\nonumber \\
|\chi_{n}| & < & 2.5 \times 10^{-2} \; ,
\nonumber \\
 \sqrt{|\chi_{\nu} \, \chi_{e}|} & \lesssim & 7 \times 10^{-5} \; .
\end{eqnarray}
while that for the $u$- and $d$-quarks, the extreme protophobic limit, $(\chi_{p}=0)$,
parameterizes $\chi_{u}$ and $\chi_{d}$ as it follows below :
\begin{eqnarray}
\chi_{u} &=& -\frac{\chi_{n}}{3} \simeq \pm \, 3.7 \times 10^{-3}
\nonumber \\
\chi_{d} &=& +\frac{2 \, \chi_{n}}{3} \simeq \mp \, 7.4 \times 10^{-3} \; .
\end{eqnarray}
The $X^{\mu}$-boson can also have a chiral interaction with the SM leptons via an
axial current, see \cite{GuHe2016}. For a complete review on the anomaly in Beryllium decays,
see \cite{JFengPRD2017}.
%
%\begin{eqnarray}
%e \, \chi_{\Psi} \, \bar{\Psi} \, \gamma^{\mu} \, \gamma_{5} \, \Psi \; .
%\end{eqnarray}
%
Still in the Lagrangian (\ref{Leff}), the parameter $\chi$ mixes the $X^{\mu}$-boson with the usual electromagnetic (EM) photon,
$A^{\mu}$, where $X^{\mu\nu}=\partial^{\mu}X^{\nu}-\partial^{\nu}X^{\mu}$ is the field-strength tensor for $X^{\mu}$, and
$F^{\mu\nu}=\partial^{\mu}A^{\nu}-\partial^{\nu}A^{\mu}$, the corresponding tensor field for the photon.
It is clear that the massive term spoils the $U(1)_{X}$-symmetry, and the Lagrangian exhibits only EM gauge symmetry, $U_{em}(1)$.
Thus, the presence of the mass in (\ref{Leff}) motivates us to search for a physical mechanism that underlies its generation. There has emerged an important
connection between the $U(1)_{X}$-symmetry and dark matter in a model with spontaneous symmetry breaking (SSB); for that, see the recent paper of Ref.
\cite{KitaharaPRD2017}.

In another context, the search for single photons in $53 \, \mbox{fb}^{-1}$, in the $e^{+} \, e^{-}$ CM-collision with
the production of a spin-1 particle, $A'$, referred to as dark photon, is associated to the process $e^{+} \, e^{-} \, \rightarrow \, \gamma \, A'$ \cite{BaBarPRL2017}.
Immediately, the dark photon decays into invisible fermion matter $A \, \rightarrow \, \bar{\zeta} \, \zeta$, that can be a candidate to dark matter constituent.
The $A'$ dark-photon has therefore a similar phenomenology to the $X$-Boson, with a kinetic mixing of order $10^{-3}$,
with the EM photon. The dominant decay mode for the lowest-mass of the $\zeta$-state fixes the condition $m_{A'} \, > \, 2 \, m_{\zeta}$,
considering that the $m_{A'}$-dark-photon mass is bounded from above by $m_{A'} \leq 8 \, \mbox{GeV}$.

In the literature, there is also a great deal of interest in the activity related to the phenomenology of hidden sector para-photons \cite{Ahlers2007,Jaeckel2008,Arias2010}. The para-photon is a neutral vector boson with a sub-eV mass and with the property of electromagnetic interactions with coupling constants referred to as millicharges. The para-photons are characterized by a mixed mass term with the photon and they appear with an extra $U(1)$ gauge factor in the full symmetry group. The particles like the boson-X, dark-photon or para-photon that interact with a
current like (\ref{Jproto}) are known in the literature as Weakly Interacting Massive Particles (WIMP). This idea connects the matter content of the SM
with possible new dark matter particles via scattering processes.

In the present paper, we re-assess the $SU_{L}(2) \times U_{R}(1)_{J} \times U(1)_{K}$ model whose
spontaneous symmetry breaking (SSB) mechanism can be applied in two scenarios: with a higher (TeV) or a lower (MeV)
energy scale; for details, consult \cite{MJNeves2017Annalen}. We study its phenomenology such that the new bosons, $Z'$ or
$X$, can connect the world of SM particles with dark matter particles in particular scattering processes.
Furthermore, the introduction of a mixing involving a Dirac right-neutrino component with exotic fermions,
{\it i. e.}, heavy and neutral fermions, can unveil new magnetic properties of these new fermions. We pursue this investigation and
we obtain the transition Magnetic Dipole Momentum (MDM) and the MDM for the exotic fermion that depend on its mass,
as it happens in the neutrino case. Sectors of fermions and scalar bosons are introduced with quantum numbers consistent with
the gauge invariance and the chiral anomalies associated to the Abelian sectors cancel out.

The organization of this paper follows the outline below: in Section II, we review the $Z'$-model based on an $SU_{L}(2) \times U_{R}(1)_{J} \times U(1)_{K}$-symmetry presented in details in \cite{MJNeves2017Annalen}. In Section III, we study in details the diagonalization of the dark fermion and right-neutrino sector. Section IV is presented in three subsections, where we study the $Z'$-decay modes into fermions, the decay modes of the extra Higgs and the scattering processes quark-quark and dark fermion mediated
by the $Z'$ particle. In the Section V, we obtain the transition MDM and the MDM of the dark fermion.
Section VI contains a review of SSB in association with the $X$-boson and the dark photon scenario of \cite{MJNeves2017Annalen}. In Section VII, we study the decay modes of the dark photon (first subsection) and we work out the scattering process for the cascade effect $e^{+} \, e^{-} \, \rightarrow \, X \, \rightarrow \, \bar{\zeta} \, \zeta$ (second subsection). The third subsection is dedicated to form factor calculations in the correction to the QED vertex
due to the axial interaction of the X-boson with the leptons of the SM. Finally, our Concluding Comments are cast in Section VIII.
%
%
%%%%%%%%%%%%%%%%%%%%%%%%%%%%%%%%%%%%%%%%%%%%%55
%

\section{A review of the $Z'$-model}
\renewcommand{\theequation}{2.\arabic{equation}}
\setcounter{equation}{0}

In this Section, we present a short review of the $Z^{\prime}$-model; the details may be found in the work of Ref. \cite{MJNeves2017Annalen}.
Here, we introduce the model in the renormalizable $R_{\xi}$-gauge. The sector of fermions and gauge fields of the model
$SU_{L}(2) \times U_{R}(1)_{J} \times U(1)_{K}$-model is described by the Lagrangian below :
\begin{equation}\label{Lleptons}
{\cal L}_{f}=\bar{\Psi}_{L}\, i \, \, \slash{\!\!\!\!D} \, \Psi_{L}
+\bar{\Psi}_{R} \, i \, \, \slash{\!\!\!\!D} \, \Psi_{R}
+\bar{\nu}_{iR} \, i \, \, \slash{\!\!\!\!D} \, \nu_{iR}
+\bar{\zeta}_{i} \, i \, \, \slash{\!\!\!\!D} \, \zeta_{i}  \; ,
\end{equation}
and
\begin{equation}\label{Lgauge}
{\cal L}_{gauge}=-\frac{1}{2} \,\mbox{tr}\left(F_{\mu\nu}^{\; 2}\right)
-\frac{1}{4} \, B_{\mu\nu}^{\; 2}
-\frac{1}{4} \, C_{\mu\nu}^{\; 2} \; .
\end{equation}
The slashed notation corresponds to the contraction of the covariant derivatives with the usual Dirac matrices.
The covariant derivatives acting on the fermions of the model are cast according to:
\begin{eqnarray}\label{DmuPsiLPsiR}
D_{\mu}\Psi_{L} &=& \left(\partial_{\mu}+i \, g \, A_{\mu}^{\, a} \, \frac{\sigma^{a}}{2} +i \, J_{L} \, g^{\prime} \, B_{\mu}
+ i \, K_{L} \, g^{\prime \prime} \, C_{\mu} \! \right) \! \Psi_{L} \; ,
\nonumber \\
D_{\mu}\Psi_{R} &=& \left( \phantom{\frac{1}{2}} \!\!\!\! \partial_{\mu} + i \, J_{R} \, g^{\prime} \, B_{\mu}+i \, K_{R} \, g^{\prime \prime} \, C_{\mu} \right) \Psi_{R} \; ,
\nonumber \\
D_{\mu}\nu_{iR} &=& \left( \phantom{\frac{1}{2}} \!\!\!\! \partial_{\mu} + i \, J_{\nu_{R}} \, g^{\prime} \, B_{\mu}+i \, K_{\nu_{R}} \, g^{\prime \prime} \, C_{\mu} \right) \nu_{iR} \; ,
\nonumber \\
D_{\mu}\zeta_{i} &=& \left( \phantom{\frac{1}{2}} \!\!\!\! \partial_{\mu}+ i \, J_{\zeta} \, g^{\prime} \, B_{\mu} + i \, K_{\zeta} \, g^{\prime \prime} \, C_{\mu} \right) \zeta_{i}  \; ,
\end{eqnarray}
where $A^{\mu \, a}\!=\!\left\{ \, A^{\mu \, 1} \, , \, A^{\mu \, 2} \, , \, A^{\mu \, 3} \, \right\}$ are the gauge fields of $SU_{L}(2)$, $B^{\mu}$ is the Abelian gauge field of $U_{R}(1)_{J}$, and $C^{\mu}$ the similar one to $U(1)_{K}$. Here, we have chosen the symbol $J$ to represent the generator of $U_{R}(1)_{J}$, $K$ is the generator of $U(1)_{K}$, the generators of $SU_{L}(2)$ are the Pauli matrices $\frac{\sigma^{a}}{2} \, (a=1,2,3)$, and
$g$, $g^{\, \prime}$ and $g^{\, \prime\prime}$ are dimensionless gauge couplings. The fermionic field content is given in the sequel.
The notation $\Psi_{L}$ indicates the usual doublet of neutrinos/leptons $L_{i}=(\nu_{i} \, \, \, \ell_{i})_{L}^{t}$,
or quarks $Q_{iL}$ $(i=1,2,3)$ left-handed of the SM, it turn out in the fundamental representation of $SU_{L}(2)$.
The label $\ell_{iL}=\left( \, e_{L} \, , \, \mu_{L} \, , \, \tau_{L} \, \right)$ indicates the leptons family displayed in the doublet, and
the $\nu_{iL}$-label for neutrinos is defined by $\nu_{iL}=(\nu_{e_{L}},\nu_{\mu_{L}},\nu_{\tau_{L}})$.
The $\Psi_{R}$-fermion is any right-handed of the SM, {\it i. e.}, it can be the lepton
$\ell_{iR}=\left( \, e_{R} \, , \, \mu_{R} \, , \, \tau_{R} \, \right)$ or right-handed quarks
$Q_{iR}=\left\{ \, u_{R} \, , \, c_{R} \, , \, t_{R} \, \right\}$ and $q_{iR}=\left\{ \, d_{R} \, , \, s_{R} \, , \, b_{R} \, \right\}$.
The new content of fermions beyond the SM is represented by the Dirac Right-Neutrino $\nu_{iR}=(\nu_{e_{R}},\nu_{\mu_{R}},\nu_{\tau_{R}})$
and the exotic neutral $\zeta_{i}$-fermion $(i=1,2,3)$ that we have introduced it associated with the $U(1)_{K}$-group:
\begin{eqnarray}
%L_{i}
%\!=\!
%\left(
%\begin{array}{c}
%\nu_{i} \\
%\ell_{i} \\
%\end{array}
%\right)_{L}
%\hspace{0.2cm}
L_{i}&=&
\left(
\begin{array}{c}
\nu_{e} \\
\ell_{e} \\
\end{array}
\right)_{L} ,
%\hspace{-0.1cm}
%L_{2}
%\!=\!
\left(
\begin{array}{c}
\nu_{\mu} \\
\ell_{\mu} \\
\end{array}
\right)_{L} ,
%\hspace{-0.1cm}
%L_{3}
%\!=\!
\left(
\begin{array}{c}
\nu_{\tau} \\
\ell_{\tau} \\
\end{array}
\right)_{L}
\hspace{-0.1cm}
: \left(\underline{{\bf 2}}, 0, -\frac{1}{2} \right) \, ,
\nonumber \\
Q_{iL}&=&
\left(
\begin{array}{c}
u \\
d \\
\end{array}
\right)_{L} ,
%\hspace{-0.1cm}
%Q_{2L}\!=\!
\left(
\begin{array}{c}
c \\
s \\
\end{array}
\right)_{L} ,
%\hspace{-0.1cm}
%Q_{3L}\!=\!
\left(
\begin{array}{c}
t \\
b \\
\end{array}
\right)_{L}
\hspace{-0.1cm}
: \left(\underline{{\bf 2}}, 0, +\frac{1}{6} \right) \, ,
%\hspace{0.45cm}
\nonumber
\end{eqnarray}
\vspace{-0.5cm}
\begin{eqnarray}
\ell_{iR} &=& \left\{ \, e_{R} \, , \, \mu_{R} \, , \, \tau_{R} \, \right\} : \left(\underline{{\bf 1}}, -\frac{1}{2}, -\frac{1}{2} \right) \, ,
\nonumber \\
Q_{iR} &=& \left\{ \, u_{R} \, , \, c_{R} \, , \, t_{R} \, \right\} : \left(\underline{{\bf 1}}, +\frac{1}{2}, +\frac{1}{6} \right) \, ,
\nonumber \\
q_{iR} &=& \left\{ \, d_{R} \, , \, s_{R} \, , \, b_{R} \, \right\} : \left(\underline{{\bf 1}}, -\frac{1}{2}, +\frac{1}{6} \right) \, ,
\nonumber \\
\nu_{iR} &=& \left\{ \, \nu_{eR} \, , \, \nu_{\mu R} \, , \, \nu_{\tau R} \, \right\} : \left(\underline{{\bf 1}}, +\frac{1}{2}, -\frac{1}{2} \right) \, ,
\nonumber \\
\zeta_{iL} &=& \left\{ \, \zeta_{1L} \, , \, \zeta_{2L} \, , \, \zeta_{3L} \, \right\} : \left(\underline{{\bf 1}}, +\frac{1}{2}, -\frac{1}{2} \right) \, ,
\nonumber \\
\zeta_{iR} &=& \left\{ \, \zeta_{1R} \, , \, \zeta_{2R} \, , \, \zeta_{3R}  \, \right\} : \left(\underline{{\bf 1}}, -\frac{1}{2}, +\frac{1}{2} \right) \, .
\end{eqnarray}
All these fields are singlets that undergo Abelian transformations under the  $U_{R}(1)_{J}$- and $U(1)_{K}$-groups.
%as follows :
%As usual, $L$ describes a doublet with a charged lepton and its neutrino partner (both left-handed). It transforms under the fundamental representation of $SU_{L}(2)$ as
%
%The $\zeta$-fermion can be interpreted as a hidden (dark matter) fermion.
%where
%
%Under the Abelian factors, $U_{R}(1)_{J}$ and $U(1)_{K}$, these fermions transform as follows:
%
%\begin{eqnarray}
%U_{R}(1)_{J}
%\!-\!\mbox{symmetry}
%\left\{
%\begin{array}{llll}
%U_{R}(1)_{J}
%\!-\!\mbox{symmetry}
%:
%\left\{
%\begin{array}{ll}
%\Psi_{R} \longmapsto \Psi_{R}^{\, \prime}(x)= e^{ \, i \, g^{\prime} \, J_{R} \, f(x) } \, \Psi_{R}(x)
\\
%\nu_{R} \longmapsto \nu_{R}^{\, \prime}(x)= e^{ \, i \, g^{\prime} \, J_{\nu_R} \, f(x) } \, \nu_{R}(x)
\\
The corresponding field-strength tensors in the sector of gauge fields are defined by
\begin{eqnarray}\label{Fmunu}
F_{\mu\nu} &=& \partial_{\mu}A_{\nu}-\partial_{\nu}A_{\mu}+i \, g \, \left[ \, A_{\mu} \, , \, A_{\nu} \, \right]  \; ,
\nonumber \\
B_{\mu\nu} &=& \partial_{\mu}B_{\nu}-\partial_{\nu}B_{\mu} \; ,
C_{\mu\nu}= \partial_{\mu}C_{\nu}-\partial_{\nu}C_{\mu} \; ,
\end{eqnarray}
%
%
%\begin{eqnarray}
%\left[D_{L\mu},D_{L\nu} \right]_{\star}=ig_{1}F_{\mu\nu}+i J \, g_{1}^{\prime} \, H_{\mu\nu} \, \openone_{2} \; ,
%\end{eqnarray}
%

%

%

%
%where $\chi$ is a real parameter that mixes the Abelian gauge fields of the $U_{R}(1)_{J} \times U(1)_{K}$-subgroup.
%Its currently estimated value is $10^{-3} < \chi < 10^{-6}$ for models that discuss hidden photons as dark matter \cite{Arias2010}.
%With the field transformations shown above, the action is clearly $SU_{L}(2) \times U_{R}(1)_{J} \times U(1)_{K}$-invariant.
%The mixing term of $X^{\mu}-B^{\mu}$
%can be eliminated by the unitary rotation
%
%
%\begin{eqnarray}
%X^{\mu}=\tilde{X}^{\mu}-\chi \; B^{\mu} \; .
%\end{eqnarray}
%
%Since is real parameter $\chi \ll 1$, so the Lagrangian (\ref{Lgauge}) can be written as
%
%\begin{eqnarray}\label{LGaugeXYtil}
%{\cal L}_{Gauge}=-\frac{1}{2} \,\mbox{tr}\left(F_{\mu\nu}^{\; 2}\right)
%-\frac{1}{4} \, \, \tilde{X}_{\mu\nu}\tilde{X}^{\mu\nu}
%-\frac{1}{4} \, \, B_{\mu\nu}B^{\mu\nu} \; .
%\end{eqnarray}
%
%The covariant derivatives in terms of $\tilde{X}^{\mu}$ and $\tilde{B}^{\mu}$ also redefine
%the coupling constant attached to $\tilde{B}^{\mu}$ as
%
%\begin{eqnarray}
%K \, \tilde{g}^{\prime \prime}:=\frac{K g^{\prime}-J \, \chi \, g^{\prime \prime}}{\sqrt{1-\chi^2}} \; .
%\end{eqnarray}
%

The Higgs sector is essential to introduce the masses,
the physical fields and the charges for the particle content
of the model. The framework of Higgs sector is setting
up by two independent scalar fields; the first is singlet
scalar, $\Xi$, that breaks the Abelian subgroup to generate
mass to the new gauge boson, that in this scenario we
call it $Z′$-boson. The second Higgs field, $\Phi$, is a $SU_{L}(2)$-
doublet to break the residual electroweak symmetry and,
consequently, it yields the known masses for $W^{\pm}$ and
$Z^{0}$. Finally, we end up with the exact electromagnetic
symmetry
\begin{equation}
SU_{L}(2) \times U_{R}(1)_{J} \times U(1)_{K} \stackrel{\langle \Xi \rangle_{0}}{\longmapsto}
%\nonumber \\
SU_{L}(2) \times U_{Y}(1) \stackrel{\langle \Phi \rangle_{0}}{\longmapsto} U_{em}(1) \; ,
\end{equation}
where the $U_{Y}(1)$-group comes out as the mixing of the $U_{R}(1)_{J}$- and $U(1)_{K}$-subgroups.
To accomplish this SSB pattern, we start off from the Higgs Lagrangian below:
\begin{eqnarray}\label{LHiggs}
{\cal L}_{Higgs} &=&
\left(D_{\mu}\Xi\right)^{\dagger} D^{\mu} \Xi
-\mu_{\Xi}^{\, 2} \, |\Xi|^{2} -\lambda_{\Xi} \, |\Xi|^{4}
\nonumber \\
&&
\hspace{-1cm}
+\left(D_{\mu}\Phi\right)^{\dagger} D^{\mu} \Phi
-\mu_{\Phi}^{\, 2} \, |\Phi|^{2}-\lambda_{\Phi} \, |\Phi|^{4}
%\nonumber \\
%&&
%\hspace{-0.5cm}
-\lambda \, |\Xi|^{\, 2} \, |\Phi|^{\, 2} \, ,
\nonumber \\
&&
\hspace{-0.6cm}
- \, G_{ij}^{(\ell)} \, \bar{L}_{i} \, \Phi \, \ell_{jR}
- \, G_{ij}^{(\ell)\,\ast} \, \bar{\ell}_{iR} \, \Phi^{\dagger} \, L_{j}
\nonumber \\
&&
\hspace{-0.6cm}
- \, G_{ij}^{(d)} \, \bar{Q}_{iL} \, \Phi \, q_{jR}
- G_{ij}^{(d)\ast} \bar{q}_{iR} \, \Phi^{\dagger} Q_{jL}
\nonumber \\
&&
\hspace{-0.6cm}
- \, G_{ij}^{(u)} \, \bar{Q}_{iL} \, \widetilde{\Phi} \, Q_{jR}
- G_{ij}^{(u)\ast} \bar{Q}_{iR} \, \widetilde{\Phi}^{\dagger} Q_{jL}
\nonumber \\
&&
\hspace{-0.5cm}
- \, X_{ij} \, \bar{L}_{i} \, \widetilde{\Phi} \, \nu_{j R}
- \, X_{ij}^{\, \ast} \, \bar{\nu}_{i R} \, \widetilde{\Phi}^{\dagger} \, L_{j}
\nonumber \\
&&
\hspace{-0.5cm}
- \, Y_{ij} \, \bar{L}_{i} \, \widetilde{\Phi} \, \zeta_{jR}
- \, Y_{ij}^{\ast} \, \bar{\zeta}_{iR} \, \widetilde{\Phi}^{\dagger} \, L_{j}
\nonumber \\
&&
\hspace{-0.5cm}
- \, Z_{ij} \, \bar{\zeta}_{iL} \, \Xi \, \nu_{j R}
- \, Z_{ij}^{\ast} \, \bar{\nu}_{i R} \, \Xi^{\dagger} \, \zeta_{j L}
\nonumber \\
&&
\hspace{-0.6cm}
- \, W_{ij} \, \bar{\zeta}_{iL} \, \Xi \, \zeta_{jR}
- \, W_{ij}^{\, \ast} \, \bar{\zeta}_{iR} \, \Xi^{\dagger} \, \zeta_{jL}
\, .
\end{eqnarray}
In (\ref{LHiggs}), $\left\{ \, \mu_{\Xi} \, , \, \mu_{\Phi} \, , \, \lambda_{\Xi} \, , \, \lambda_{\Phi} \, , \, \lambda\, \right\}$ are real parameters,
$\left\{ \, G_{ij}^{(\ell)} \, , \, G_{ij}^{(u)} \, , \, G_{ij}^{(d)} \, , \, X_{ij} \, , \, Y_{ij} \, , \, Z_{ij} \, , \, W_{ij} \, \right\}$
are Yukawa (complex) coupling parameters needed for the fermions to acquire non-trivial masses. In general, these Yukawa couplings
set non-diagonal matrices $3 \times 3$,
%that satisfy the condition $A_{ji}^{\ast}=A_{ij}^{\ast}$.
and as usual, the $\widetilde{\Phi}$-field is defined as $\widetilde{\Phi}=i \, \sigma_{2} \, \Phi^{\ast}$ to ensure the gauge invariance.

The covariant derivatives of (\ref{LHiggs}) act on the $\Xi$- and $\Phi$-Higgs as follows:
\begin{eqnarray}\label{DmuPhi1}
D_{\mu} \, \Xi(x) &=& \left( \phantom{\frac{1}{2}} \!\!\!\! \partial_{\mu}+i \, J_{\Xi} \, g^{\prime} B_{\mu}+i \, K_{\Xi} \, g^{\prime\prime}
C_{\mu} \right)\Xi(x)
\nonumber \\
D_{\mu} \Phi(x) &=& \left(\partial_{\mu}
+ i g \, A_{\mu}^{\, a} \, \frac{\sigma^{a}}{2}+ i g'  \, J_{\Phi} \, B_{\mu} \right) \Phi(x) \; ,
\end{eqnarray}
where the $J_{\Phi}$ is the generator of $\Phi$-Higgs corresponding to the $U_{R}(1)_{J}$-subgroup.
The $\Xi$-field is a scalar singlet of $SU_{L}(2)$, with transformations under $U_{R}(1)_{J} \times U(1)_{K}$.
%given below:
%
%\begin{eqnarray}\label{transfXi}
%\Xi \, \longmapsto \, \Xi^{\prime}(x) &=& e^{\, i \, J_{\Xi} \, g^{\prime} \, f(x)} \, \Xi(x) \; ,
%\nonumber \\
%\Xi \, \longmapsto \, \Xi^{\prime}(x) &=& e^{\, i \, K_{\Xi} \, g^{\prime \prime} \, h(x)} \, \Xi(x) \; .
%\end{eqnarray}
%
%Using the previous gauge transformations, it can be readily checked that $D_{\mu}\Xi$ has the same transformation as (\ref{transfXi}).
%
The second $\Phi$-scalar field is a doublet $\Phi=\left( \, \phi^{ \, (+)} \; \;
\phi^{\, (0)} \, \right)^{t}$ that turns out in the fundamental representation of $SU_{L}(2)$, and it also transforms under $U_{R}(1)_{J}$ subgroup.   %couples to the gauge fields through the covariant derivatives which read as:
%
%\begin{eqnarray}\label{DmuPhi2}
%D_{\mu} \Phi \!\!&=&\!\! \left(\partial_{\mu}
%+ i g \, A_{\mu}^{\, a} \, \frac{\sigma^{a}}{2}+ i g'  \, J_{\Phi} \, B_{\mu} \right) \Phi \; ,
%\end{eqnarray}
%
%The $\Phi$-Higgs is a complex scalar doublet that has the gauge transformation under $SU_{L}(2)$ given by
%
%\begin{eqnarray}\label{transfPhi2}
%\Phi\!=\!
%\left(
%\begin{array}{c}
%\phi^{ \, (+)} \\
%\phi^{\, (0)} \\
%\end{array}
%\right)
%\!\longmapsto\! \Phi^{\prime}(x)=e^{i \, g \, \frac{\sigma^{a}}{2} \, \omega^{a}(x)} \Phi(x) \, ,
%\hspace{0.4cm}
%\end{eqnarray}
%
%and under $U_{R}(1)_{J}$ subgroup, it has the transformation
%
%\begin{eqnarray}
%\Phi \; \; \longmapsto \; \; \Phi^{\prime}(x)=e^{\, i \, J_{\Phi} \, g' \, f(x)} \, \Phi(x) \; .
%\end{eqnarray}
%
%The Yukawa interactions of also guarantee the invariance under $U_{R}(1)_{J}$,
%whenever the generators satisfy the relations $-J_{L}+J_{\Phi}+J_{R}=0$ and $-J_{L}+J_{\Phi}+J_{\zeta_{R}}=0$.
%
We choose the parametrization of the $\Xi$- and $\Phi$-complex fields as
\begin{eqnarray}\label{PhiGaugeparametrization}
\Xi(x) = \frac{u+\tilde{F}(x)}{\sqrt{2}} \, \, e^{\, i \, \frac{\eta(x)}{u}}
\hspace{0.2cm} , \hspace{0.2cm}
\Phi(x)=\exp\left[\, \frac{i}{v}
\left(
\begin{array}{cc}
\chi^{3} & \sqrt{2} \, \chi^{-} \\
\sqrt{2} \, \chi^{+} & -\chi^{3} \\
\end{array}
\right)
\right]
\left(
\begin{array}{c}
0 \\
\frac{v+\tilde{H}(x)}{\sqrt{2}} \\
\end{array}
\right) \, ,
\hspace{0.5cm}
\end{eqnarray}
where ˜$\tilde{F}$, $˜\tilde{H}$ and $\{ \, \eta \, , \, \chi^{\pm} \, , \, \chi_{3} \, \}$ (four Goldstone
bosons) are real functions. The minima of the Higgs potential are given by the non-trivial VEVs
$\{ \, u \, , \, v \, \}$ listed below:
\begin{eqnarray}
\langle \Xi \rangle_{0}=\frac{u}{\sqrt{2}} &\simeq& \sqrt{-\frac{\mu_{\Xi}^{2}}{2\lambda_{\Xi}}} \left(1- \frac{\lambda}{4 \, \lambda_{\Phi}}
\frac{\mu_{\Phi}^{2}}{\mu_{\Xi}^{2}}  \right) \, ,
\nonumber \\
\langle \Phi \rangle_{0}=\frac{v}{\sqrt{2}} &\simeq& \sqrt{-\frac{\mu_{\Phi}^{2}}{2\lambda_{\Phi}}}  \left( 1- \frac{\lambda}{4 \, \lambda_{\Xi}}
\frac{\mu_{\Xi}^{2}}{\mu_{\Phi}^{2}} \right) \, ,
\end{eqnarray}
where the following conditions are satisfied : $\mu_{\Xi}^{2}<0$ , $\mu_{\Phi}^{2}<0$ and $\lambda \ll 1$.
It is important to emphasize that, in this
$Z^{\prime}$-approach, the necessary condition $u \gg v$ between the
VEVs must be satisfied, such that $u$-scale generates mass
for the heavy $Z^{\prime}$-boson, while the $v = 246 \, \mbox{GeV}$ is the electroweak scale of the SM.

After the SSBs, the gauge sector is given by
\begin{eqnarray}\label{massesBWZ}
{\cal L}_{mass} &=&
%-\frac{1}{2} \, W_{\mu\nu}^{+}W^{\mu\nu \, -} \! +
m_{W}^{\, 2} \, W_{\mu}^{+}W^{\mu-}+
%\nonumber \\
%&&
%\hspace{-1cm}
%-\frac{1}{4}\left(\partial_{\mu}A_{\nu}^{\,3}-\partial_{\nu}A_{\mu}^{\,3} \right)^{2}
%-\frac{1}{4} \, B_{\mu\nu}^{\, 2}
%-\frac{1}{4} \, C_{\mu\nu}^{\, 2}
%\nonumber \\
%&&
%\hspace{-1cm}
%+
\frac{u^2}{2} \left( \phantom{\frac{1}{2}} \!\!\!\!\! J_{\Xi} \, g^{\prime} B_{\mu}+ K_{\Xi} \, g^{\prime \prime} C_{\mu} \right)^2
%\nonumber \\
%&&
%\hspace{-1cm}
+\frac{v^{2}}{2} \left( \phantom{\frac{1}{2}} \hspace{-0.3cm} g^{\prime} \, J_{\Phi} \, B_{\mu}-\frac{1}{2} \, g \, A_{\mu}^{3} \right)^{2}
%\nonumber \\
%&&
%\hspace{-1cm}
%+\frac{1}{2}\left(\partial_{\mu}\eta\right)^{2}
%+ \partial_{\mu}\chi^{+} \partial^{\mu}\chi^{-}+\frac{1}{2} \left(\partial_{\mu}\chi^{3} \right)^2
\nonumber \\
&&
%\hspace{-1cm}
+u \, \partial_{\mu}\eta \left( \phantom{\frac{1}{2}} \!\!\!\! J_{\Xi} \, g^{\prime} \, B^{\mu} + K_{\Xi} \, g^{\prime \prime} C^{\mu} \right)
%\nonumber \\
%&&
%\hspace{-1cm}
+m_{W} \left( \phantom{\frac{1}{2}} \!\!\!\! \partial_{\mu}\chi^{+} W^{\mu \, -}+ \partial_{\mu}\chi^{-} W^{\mu \, +} \right)
\nonumber \\
&&
%\hspace{-1cm}
-v \, \partial_{\mu}\chi^{3}\! \left( -\frac{1}{2} \, g \, A^{\mu\,3} + g' J_{\Phi} \, B^{\mu} \right)
\; .
\end{eqnarray}
The $W^{\pm}$-particles are identified as the combination $\sqrt{2} \, W_{\mu}^{\pm}:=A_{\mu}^{1} \, \mp \, i \, A_{\mu}^{2}$,
%in which the field strength is $W_{\mu\nu}^{\pm}:=\partial_{\mu}W_{\nu}^{\pm}-\partial_{\nu}W_{\mu}^{\pm}$.
and it $W^{\pm}$- mass is like in GSW-model $m_{W}= gv/2$. Thus, the charged Goldstone bosons are $\sqrt{2} \, \chi^{\pm}:=\chi^{1} \, \mp \, i \, \chi^{2}$. The $u^2$- and $v^2$-mass terms suggest to introduce the orthogonal $SO(2)$-transformations :
\begin{eqnarray}\label{transfA0CGY}
B_{\mu} &=& \cos\alpha \, \tilde{Z}'_{\mu}+ \sin\alpha \, Y_{\mu}
\nonumber \\
C_{\mu} &=& - \, \sin\alpha \, \tilde{Z}'_{\mu}+ \cos\alpha \, Y_{\mu} \; ,
\nonumber \\
A_{\mu}^{3} &=& \cos\theta_{W} \, \tilde{Z}_{\mu} + \sin\theta_{W} \, A_{\mu}
\nonumber \\
Y_{\mu} &=& -\sin\theta_{W} \, \tilde{Z}_{\mu} + \cos\theta_{W} \, A_{\mu} \; ,
\end{eqnarray}
where $\alpha$ is a mixing angle, $\theta_{W}$ is known as the Weinberg’s
angle $\sin^2 \theta_{W} = 0.22$ \footnote{We are considering the Weinberg's angle value of
$\sin^{2}\theta_{W} = 0.22$, taking into account the radiative corrections of the Electroweak Theory.},
such that it satisfies the parametrization in terms of the fundamental charge $e^2=4\pi/137\simeq 0.09$
\begin{eqnarray}\label{gYgg}
g_{Y} &=& g^{\prime} \, \sin\alpha= g^{\prime\prime} \, \cos\alpha \; ,
\nonumber \\
e &=& g \, \sin\theta_{W}= g_{Y} \, \cos\theta_{W} \; .
\end{eqnarray}
The $Y^{\mu}$ is the massless hypercharge gauge field, and ˜$\tilde{Z}^{'\mu}$
sets a massive boson associated with the VEV $u$-scale.
The ˜$\tilde{Z}$- and $˜\tilde{Z}'$ are not the fields that represent the physical
$Z^0$- and $Z′$-bosons yet due to the mixing they present
in the Lagrangian. Another diagonalization must be introduced to obtain the physical
masses for the condition $u \gg v$. Therefore, the masses
of the $W$, $Z$ and $Z'$ bosons in terms of the fundamental
parameters of the model are given by
\begin{eqnarray}\label{massesWZZ'}
m_{W} &\simeq& 80.2 \, \mbox{GeV}
\nonumber \\
M_{Z^{0}} & \simeq & 91.1 \left( 1 - \frac{v^{2}}{32 \, u^2} \, \cos^{4}\alpha \right) \, \mbox{GeV} \; ,
\nonumber \\
M_{Z'} & \simeq & \frac{0.34 \, u}{\sin\alpha \cos\alpha} \! \left( 1 + \frac{v^{2}}{32 \, u^{2}} \, \cos^{4}\alpha \right) \; ,
\end{eqnarray}
in which the experimental data have been accounted for \cite{PDG2016}.
The parameters $u$-scale and $\alpha$-angle were not determined in the previous expressions for the $Z$- and $Z'$-masses.
The ratio between the masses from (\ref{massesWZZ'}) is given by
\begin{eqnarray}
\frac{M_{Z'}}{M_{Z^{0}}} \simeq \frac{u}{\sin(2\alpha)} \, \frac{4\sin\theta_{W}}{v} \simeq \frac{u}{\sin(2\alpha)} \, 0.007 \, \mbox{GeV}^{-1} .
\end{eqnarray}
The recent papers of the CMS Collaboration point to the hypothetical $Z'$  upper limits
that excludes up to $95\%$ confidence level masses below the $2.0 \, \mbox{TeV}$ \cite{CMS20172}.
Therefore, we fix the $Z$- and $Z'$-masses as $M_{Z'}=2.0 \, \mbox{TeV}$ and $M_{Z^{0}}=91 \, \mbox{GeV}$
to estimate the ratio of the $u$-scale by the $\alpha$-angle, {\it i. e.}, $u \simeq 2.8 \times \sin(2\alpha) \, \mbox{TeV}$.
Thereby, the maximum value for the VEV-scale is $u=2.8 \, \mbox{TeV}$, when $\alpha=45^{o}$.
To eliminate the mixed terms $A^{\mu \, 3}-\chi^{3}$, $W^{\pm}-\chi^{\pm}$ and $Z'-\eta$ in (\ref{massesBWZ}),
we introduce the $R_{\xi}$-gauge fixing Lagrangian
\begin{eqnarray}\label{LgfAChi}
{\cal L}_{gf} &=& -\frac{1}{2\alpha} \left( \partial_{\mu}A^{\mu}\right)^{2}
-\frac{1}{2\gamma} \left( \partial_{\mu}Z^{\mu} - \, \gamma \, M_{Z^{0}} \, \chi^{3}  \right)^{2}
\nonumber \\
&&
%\hspace{-0.6cm}
-\frac{1}{2\beta} \left(\! \partial_{\mu}Z^{\prime\mu} \! + \frac{\beta}{2} \, M_{Z'} \, \eta-\beta M_{Z'} \sin\theta_{W} \chi^{3}\! \right)^{2}
\nonumber \\
&&
%\hspace{-0.6cm}
-\frac{1}{\delta} \left(\partial_{\mu}W^{\mu +}\!\!-\delta \, m_{W} \chi^{+}\right)\left(\partial_{\nu}W^{\nu -}\!\!-\delta \, m_{W} \chi^{-}\right) \, ,
%\hspace{0.5cm}
%\left(\partial_{\nu}W^{\nu \, -}
%-\delta \, m_{W} \, \chi^{-}  \right) \; ,
\end{eqnarray}
where $\{\alpha,\beta,\gamma,\delta\}$ are real parameters. Then, we obtain all the terms with the gauge fields in the renormalizable
$R_{\xi}$-gauge :
\begin{eqnarray}\label{FreeLACZW}
&&{\cal L}_{gauge}^{\, 0} = -\frac{1}{4} \, F_{\mu\nu}^{\, 2}
-\frac{1}{2\alpha} \left(\partial_{\mu}A^{\mu} \right)^2
\nonumber \\
&&
-\frac{1}{2} \, W_{\mu\nu}^{+}W^{\mu\nu-} \!\! - \frac{1}{\delta} \, \partial_{\mu}W^{\mu+}\partial_{\nu}W^{\nu-}\!\!+m_{W}^{\, 2} W_{\mu}^{+}W^{\mu-}
\nonumber \\
&&
-\frac{1}{4} \, Z_{\mu\nu}^{\, 2}
-\frac{1}{2\gamma} \left(\partial_{\mu}Z^{\mu} \right)^2
+\frac{1}{2} \, M_{Z^{0}}^{\, 2} \, Z_{\mu}^{\, 2}
\nonumber \\
&&
-\frac{1}{4} \, Z_{\mu\nu}^{\prime \, 2}-\frac{1}{2\beta}\left(\partial_{\mu}Z^{\prime\mu}\right)^{2}
+\frac{1}{2}\, M_{Z'}^{\, 2} \, Z_{\mu}^{\prime \, 2} \; .
\hspace{0.5cm}
\end{eqnarray}
The Higgs sector, after the SSBs, is reduced to the scalar fields ˜$\tilde{F}$ and $\tilde{H}$, whose
diagonalization yields the following masses :
\begin{eqnarray}\label{MFH}
M_{H} &\simeq& \sqrt{ 2 \, \lambda_{\Phi} \, v^{2}}
\, \left( 1-\frac{\lambda^2}{8 \, \lambda_{\Xi}\lambda_{\Phi}} \right)=125 \, \mbox{GeV}
\nonumber \\
M_{F} &\simeq& \sqrt{ 2 \, \lambda_{\Xi} \, u^{2}}
\, \left( 1+\frac{\lambda^2}{8 \, \lambda_{\Xi}^2} \, \frac{v^{2}}{u^{2}}  \right)
\; ,
\end{eqnarray}
%
%The interactions between scalars $F$ and $H$ are given by
%
%\begin{eqnarray}
%&&
%{\cal L}_{Higgs}^{\, \, int} = -\frac{\lambda_{\Xi}}{4} \, F^{4}-\frac{\lambda_{\Phi}}{4} \, H^{4}
%-\lambda_{\Xi} \, u \, F^{3}-\lambda_{\Phi} \, v \, H^{3}
%\nonumber \\
%&&
%-\frac{\lambda_{\Xi \, \Phi}}{2} \, u \, F \, H^{2}-\frac{\lambda_{\Xi \, \Phi}}{2} \, v \, H \, F^{2}
%-\frac{\lambda_{\Xi \, \Phi}}{4} \, F^{2} \, H^{2} \; .
%\end{eqnarray}
%
%
where $M_{F} \gg M_{H}$ as consequence of the VEV-scales condition. The estimation for $M_{F}$ is
\begin{eqnarray}\label{MassH1u8GEV}
1.24 \, \mbox{TeV} < M_{F} < 3.7 \, \mbox{TeV} \; ,
\end{eqnarray}
in which the maximum value $3.7 \, \mbox{TeV}$ correspond to $\alpha$-angle of $45^{o}$.

The sector of interaction of the $Z$-, $Z'$-bosons and photon
with fermions of the model is
\begin{equation}\label{LintAZC}
{\cal L}^{\, int}= - \, e\, Q_{em} \, \bar{\Psi} \, \, \slash{\!\!\!\!A} \, \Psi
-e\, Q_{Z} \, \bar{\Psi} \, \, \slash{\!\!\!\!Z} \, \Psi
%\nonumber \\
%&&
-e\, Q_{Z'} \, \bar{\Psi} \,\, \slash{\!\!\!\!Z'} \, \Psi \; .
\end{equation}
The electric charge content of model is
\begin{eqnarray}
Q_{em}=I^{3}+J+K \; ,
\end{eqnarray}
and the generators $Q_{Z}$ and $Q_{Z'}$ are defined by the relation
\begin{eqnarray}\label{Nishijima}
Q_{Z} &:=& \frac{1}{\sin\theta_{W}\cos\theta_{W}} \left( \phantom{\frac{1}{2}} \!\!\!\! I^{3}- Q_{em} \, \sin^{2}\theta_{W} \right) \; ,
\nonumber \\
Q_{Z'} &:=& \frac{2}{\sin(2\alpha) \cos\theta_{W}} \left( \phantom{\frac{1}{2}} \!\!\!\!\!\! - J \, + Y \, \sin^{2}\alpha \right) \; .
\end{eqnarray}
The interaction of neutrinos-leptons with the $W^{\pm}$ are, like in the GSW model, reobtained here :
\begin{eqnarray}\label{intWnuell}
{\cal L}^{int}= - \frac{g}{\sqrt{2}} \, \, \bar{\nu}_{iL} \, \, \slash{\!\!\!W}^{+} \, \ell_{iL}
-\frac{g}{\sqrt{2}} \, \, \bar{\ell}_{iL} \, \, \slash{\!\!\!W}^{-} \, \nu_{iL} \; .
\end{eqnarray}
The $Q_{Z}$-generator emerges as in the usual Electroweak Model, but we have here another charge, $Q_{Z'}$, of the interaction between the fermions
and the $Z'$-boson. The values of $Q_{em}$, $I^{3}$, $Y$ and the primitive charges $J$ and $K$ are summarized in the table below.
These values are due to a possible solution for the Abelian (chiral) anomaly to cancel out. Since the model is based on two Abelian subgroups,
there are six triangle graphs of the $J$- and $K$-symmetries that contribute for the anomaly : $J$ , $K$ , $J^2 \, K$ , $J \, K^2$ , $J^3$ , $K^3$.
Therefore, the sum of all the charges $J$ and $K$ in the table (\ref{Table1}) yield the cancellation of the six triangle graphs. The necessary condition for an 
anomaly-free model is that the $\zeta_{i}$-fermions have no electric charge, {\it i. e.}, $Y=0$, with $J=-K=+1/2$ for Left-component and $J=-K=-1/2$ for Right-component.
Furthermore, the neutrino right-component must also be added to ensure the model to be free from anomalies.
Thereby, the Yukawa interactions introduced in the Higgs sector are gauge invariant and it is satisfied by the $J$- and $K$-charges in the table. As consequence, $\zeta_{i}$-fermions do not interact with $Z$-boson and photon of the SM, but it do interact with the $Z'$-boson. All the interactions of the Electroweak Standard Model are reproduced in the model. The right-neutrino components and $\zeta_{i}$-fermions do not interact with the EW $Z$-boson, as can be verified by the charges.
\begin{table}
\centering
\begin{tabular}{|l|l|l|l|l|l|}
\hline
%after \\: \hline or \cline{col1-col2} \cline{col3-col4} ...
\mbox{Fields} \& \mbox{particles} & $Q_{em}$ & $I^{3}$ & $Y$ & $J$ & $K$ \\
\hline
\mbox{lepton-left} & $-1$ & $-1/2$ & $-1/2$ & $ 0 $ & $-1/2 $  \\
\hline
\mbox{neutrino-left} & $0$ & $+1/2$ & $-1/2$ & $ 0 $ & $ -1/2 $ \\
\hline
\mbox{lepton-right} & $-1$ & $0$ & $-1$ & $-1/2$ & $ -1/2 $ \\
\hline
\mbox{neutrino-right} & $0$ & $0$ & $0$ & $+1/2$ & $-1/2$ \\
\hline
$\zeta_{i}$-\mbox{fermions left} & $0$ & $0$ & $0$ & $-1/2$ & $+1/2$ \\
\hline
$\zeta_{i}$-\mbox{fermions right} & $0$ & $0$ & $0$ & $+1/2$ & $-1/2$ \\
\hline
\mbox{u-quark-left} & $+2/3$ & $+1/2$ & $+1/6$ & $0$ & $+1/6$  \\
\hline
\mbox{d-quark-left} & $-1/3$ & $-1/2$ & $+1/6$ & $0$ & $+1/6$ \\
\hline
\mbox{u-quark-right} & $+2/3$ & $0$ & $+2/3$ & $+1/2$ & $+1/6$ \\
\hline
\mbox{d-quark-right} & $-1/3$ & $0$ & $-1/3$ & $-1/2$ & $+1/6$ \\
\hline
$W^{\pm}$-\mbox{bosons} & $\pm \, 1$ & $\pm \, 1$ & $0$ & $0$ & $0$ \\
\hline
\mbox{neutral bosons} & $0$ & $0$ & $0$ & $0$ & $0$  \\
\hline
$\Xi$-\mbox{Higgs} & $0$ & $0$ & $0$ & $-1$ & $+1$ \\
\hline
$\Phi$-\mbox{Higgs} & $0$ & $-1/2$ & $+1/2$ & $+1/2$ & $0$ \\
\hline
\end{tabular}
\caption{The particle content for the $Z'$-model candidate at the TeV-scale physics.
The $J$- and $K$-charges are such that anomalies cancel out.}\label{Table1}
\end{table}
%
%\vspace{0.3cm}
%
The new interactions of the leptons, neutrinos and the $\zeta$-fermion with the $Z'$-boson are displayed in what follows :
\begin{equation}\label{LintZ'}
{\cal L}^{int}_{Z'}=- \, g_{Z'} \,
\bar{f} \, \, \slash{\!\!\!\!Z}' \left(g^{f}_{V}-g^{f}_{A}\gamma_{5}\right) f \; ,
\end{equation}
where $g_{Z'}:=e\csc(2\alpha)\sec\theta_{W}$, and we rewrite the $Z'$-interaction with the Left- and Right-components of fermions $\Psi$ from (\ref{LintAZC}). Here, the $f$-fermions means the the fermion fields with no quiral components. Furthermore, we define the coefficients
$g^{f}_{V}$ and $g^{f}_{A}$ as
\begin{eqnarray}
g^{f}_{V} &=& -J^{f_{L}}-J^{f_{R}}+\left(Y^{f_{L}}+Y^{f_{R}}\right)\sin^2\alpha
\nonumber \\
g^{f}_{A} &=& -J^{f_{L}}+J^{f_{R}}+\left(Y^{f_{L}}-Y^{f_{R}}\right)\sin^2\alpha \; .
\end{eqnarray}
The $f$-sum in (\ref{LintZ'}) runs to all fermions (no quiral components) of the model.
Thereby, we list all values of $g^{f}_{V}$ and $g^{f}_{A}$ following the charges in the table (\ref{Table1}) :
\begin{eqnarray}
g^{\ell}_{V} &=& -1+\frac{3}{2} \cos^2\alpha
\hspace{0.3cm} , \hspace{0.3cm}
%\nonumber \\
g^{\ell}_{A}= -\frac{1}{2} \cos^2\alpha
\nonumber \\
g^{\nu}_{V} &=& -1+ \frac{1}{2} \cos^2\alpha
\hspace{0.3cm} , \hspace{0.3cm}
%\nonumber \\
g^{\nu}_{A}= \frac{1}{2} \cos^2\alpha
\nonumber \\
g^{u}_{V} &=& -\frac{1}{2}+\frac{5}{6} \sin^2\alpha
\hspace{0.3cm} , \hspace{0.3cm}
%\nonumber \\
g^{u}_{A}= \frac{1}{2} \cos^2\alpha
\nonumber \\
g^{d}_{V} &=& \frac{1}{2}-\frac{1}{6} \sin^2\alpha
\hspace{0.3cm} , \hspace{0.3cm}
%\nonumber \\
g^{d}_{A}= -\frac{1}{2} \cos^2\alpha
\nonumber \\
g^{\zeta}_{V} &=& 0
\hspace{0.3cm} , \hspace{0.3cm}
%\nonumber \\
g^{\zeta}_{A}= +1  \; .
\end{eqnarray}
The $Z'$-vertex useful to perform loop calculations is reads as
\begin{figure}[!h]%\label{figvm}
\begin{center}
\newpsobject{showgrid}{psgrid}{subgriddiv=1,griddots=10,gridlabels=6pt}
\begin{pspicture}(5,1)(11,2.5)
%\showgrid
\psset{arrowsize=0.2 2}
\psset{unit=0.8}
%
%%%%%%%%%%%%%%%%%%%% Vertice Boson X - Leptons %%%%%%%%%%%%%%%%%%%%%%%%%%%%%%%%
%
\pscoil[coilarm=0,coilaspect=0,coilwidth=0.2,coilheight=1.0,linecolor=black](6.5,1.05)(6.5,3)
\psline[linecolor=black,linewidth=0.5mm]{-}(5,1)(8,1)
\psline[linecolor=black,linewidth=0.5mm]{->}(5,1)(6,1)
\psline[linecolor=black,linewidth=0.5mm]{->}(7,1)(7.55,1)
\put(6.8,2.8){\large$Z^{\prime}$}
\put(4.95,1.2){\large$\bar{f}$}
\put(7.8,1.2){\large$f$}
\put(8.5,1.1){\large$\Gamma_{Z'}^{\, \mu}=- \, i \, g_{Z'} \, \gamma^{\mu}\left( \, g_{V}^{f}-g_{A}^{f} \, \gamma_{5} \, \right) .$}
\end{pspicture}
%
%\vspace{0.2cm}
%
%\caption{\scshape{The one loop correction of the vertex diagram.}}\label{figvm}
%
\end{center}
\end{figure}

\section{The mixing between right-neutrinos and $\zeta_{i}$-fermions}
\renewcommand{\theequation}{3.\arabic{equation}}
\setcounter{equation}{0}
After the SSB takes place, the neutrinos and the $\zeta$-fermion acquire
mass terms as displayed below:
\begin{eqnarray}\label{LpsichiM}
&&
{\cal L}_{\nu-\zeta}^{0}=\bar{\nu}_{i} \, i \, \slash{\!\!\!\partial} \; \nu_{i}
+\bar{\zeta}_{i}\, i \, \slash{\!\!\!\partial} \; \zeta_{i}
\nonumber \\
&&
- \frac{v}{\sqrt{2}} \left( \, X_{ij} \, \bar{\nu}_{iL} \, \nu_{jR}
+ X_{ij}^{\, \ast} \, \bar{\nu}_{iR} \, \nu_{jL} \right)
\nonumber \\
&&
- \frac{v}{\sqrt{2}} \left( \, Y_{ij} \, \bar{\nu}_{iL} \, \zeta_{jR}
+Y_{ij}^{\, \ast} \, \bar{\zeta}_{iR} \, \nu_{jL} \right)
\nonumber \\
&&
-\frac{u}{\sqrt{2}} \left( \, Z_{ij} \, \bar{\zeta}_{iL} \, \nu_{jR} \, + \, Z_{ij}^{\, \ast} \, \bar{\nu}_{iR} \, \zeta_{jL} \, \right)
\nonumber \\
&&
- \frac{u}{\sqrt{2}} \left( \, W_{ij} \, \bar{\zeta}_{iL} \, \zeta_{jR}
+ W_{ij}^{\, \ast} \, \bar{\zeta}_{iR} \, \zeta_{jL} \right) \; .
\end{eqnarray}
It can be cast in the matrix form
\begin{eqnarray}\label{LnuzetaMatrix}
{\cal L}_{\nu-\zeta}^{\, 0}=
\bar{\chi}_{L} \, i \, \slash{\!\!\!\partial} \; \chi_{L}
+\bar{\chi}_{R} \, i \, \slash{\!\!\!\partial} \; \chi_{R}
-\bar{\chi}_{L} \, M \, \chi_{R} +{\mbox h. \, c.} \; ,
%-\frac{v}{\sqrt{2}} \, \bar{\nu}_{L} \, X \, \nu_{R}
%-\frac{v}{\sqrt{2}} \, \bar{\nu}_{L} \, Z \, \zeta_{R}
%\nonumber \\
%&&
%-\frac{u}{\sqrt{2}} \, \bar{\zeta}_{L} \, T \, \nu_{R}
%-\frac{u}{\sqrt{2}} \, \bar{\zeta}_{L} \, W \, \zeta_{R} \; ,
\end{eqnarray}
where $\chi_{L(R)}^{t}:=\left( \, \nu_{L(R)} \; \; \zeta_{L(R)} \, \right)$ and the mass matrix
is given by
\begin{eqnarray}\label{MassMatrixM}
M=\frac{1}{\sqrt{2}}
\left(
\begin{array}{cc}
v \, X & v \, Y
\\
\\
u \, Z & u \, W
\\
\end{array}
\right) \; .
\end{eqnarray}
Here, $X$, $Y$, $Z$ and $W$ are $3 \times 3$ matrices of elements $\{ \, X_{ij} \, , \, Y_{ij} \, , \, Z_{ij} \, , \, W_{ij} \, \}$, and thus,
the mass matrix is $6 \times 6$. We introduce the unitary transformation
\begin{eqnarray}\label{chiLRtransf}
\chi_{L(R)} \, \longmapsto \, \chi'_{L(R)} \!&=&\! S \, \chi_{L(R)} \; ,
%\nonumber \\
%\nu_{R} \, \rightarrow \, \nu'_{R} \!&=&\! K_{\nu} \, \nu_{L}
%\nonumber \\
%\zeta_{L} \, \rightarrow \, \zeta'_{L} \!&=&\! U_{\zeta} \, \zeta_{L}
%\nonumber \\
%\zeta_{R} \, \rightarrow \, \zeta'_{R} \!&=&\! K_{\zeta} \, \zeta_{R} \; ,
\end{eqnarray}
where $S$ is the unitary matrix, $S^{\dagger} S={\uma}$. The sector neutrino-$\zeta$-fermion (\ref{LnuzetaMatrix})
is diagonal such that the mass matrix, after the transformation (\ref{chiLRtransf}), satisfies the relation
\begin{eqnarray}\label{MD}
M_{D}=S \, M \, S^{\dagger}=
\left(
\begin{array}{cc}
M^{(\nu)} & 0
\\
0 & M^{(\zeta)} \\
\end{array}
\right)
\; .
\end{eqnarray}
To obtain the form of $S$, we begin with the most general form of a $U(2)$-matrix
\begin{eqnarray}
S=\left(
\begin{array}{cc}
e^{i \, \alpha} \, \cos\theta & e^{i \, \beta} \, \sin\theta \\
-e^{i \, (\gamma-\beta)} \, \sin\theta & e^{i \, (\gamma-\alpha)} \, \cos\theta \\
\end{array}
\right) \; ,
\end{eqnarray}
with four independents parameters : $\theta$ is mixing angle between the families of Right-Neutrino and $\zeta$-fermion, and three phases $\{ \, \alpha \, , \, \beta \, , \, \gamma \, \}$. The $M_{D}$-matrix is diagonal if the angles $\left\{ \, \theta \, , \, \alpha \, , \, \beta \, \right\}$ satisfy the relation
\begin{equation}\label{tantheta}
\left( v \, X - u \, W \right)\tan(2\theta)= 2 \, v \, Y \, e^{i \, (\alpha-\beta)}=2 \, u \, Z \, e^{-i \, \left(\alpha-\beta\right)} \; ,
%\nonumber \\
%v \, Z \!\!&=&\!\! u \, T \, e^{-i \, 2 \, \left(\alpha-\beta\right)} \; ,
\end{equation}
and the diagonal of (\ref{MD}) is formed by the two $3 \times 3$ matrices. In terms of $\theta$-angle, these matrices are given by
\begin{eqnarray}\label{MnuMzeta}
M^{(\nu)} &=& \frac{v \, X}{\sqrt{2}} + \frac{\tan\theta}{2\sqrt{2}}\left( v \, X -u \, W \right)\tan\left(2\theta\right) \; ,
\nonumber \\
M^{(\zeta)} &=& \frac{u \, W}{\sqrt{2}} - \frac{\tan\theta}{2\sqrt{2}}\left( v \, X -u \, W \right)\tan\left(2\theta\right)  \, .
\end{eqnarray}
The diagonalization does not depend on the $\gamma$-phase, so we can eliminate it.
We can also eliminate the $\theta$-mixing angle in (\ref{MnuMzeta}) in terms of $u$, $v$, $Z$ and $Y$;
after some algebra, we obtain the $3 \times 3$ mass matrices
\begin{eqnarray}\label{MnuMzeta}
M^{(\nu)} &=& \frac{ v \, X + u \, W - \sqrt{\left(v \, X - u \, W \right)^{2}+4 v \, u \, Z \, Y }}{2\sqrt{2}}
\nonumber \\
M^{(\zeta)} &=& \frac{ v \, X + u \, W + \sqrt{\left(v \, X - u \, W \right)^{2}+4 v \, u \, Z \, Y }}{2\sqrt{2}} \; .
\end{eqnarray}
It is immediate that, whenever $Y=Z=0$, we obtain the mass matrices $M^{(\nu)}=vX/\sqrt{2}$ and $M^{(\zeta)}=uW/\sqrt{2}$, respectively.
Other way to obtain the same result is rewrite (\ref{LnuzetaMatrix}) in terms of
In termos of the Left- and Right-projectors, the $\chi$-spinor is split as $\chi_{i}^{\, t}=\left( \, \nu_{i} \, \, \, \, \zeta_{i} \, \right)^{t}$;
then, the mass matrix (\ref{MassMatrixM}) is rewritten as
\begin{equation}\label{Lleptonsmatrix}
M_{\nu-\zeta} =\frac{1}{\sqrt{2}}
\left(
\begin{array}{cc}
v \, X & Y \, v \, R + Z \, u \, L
\\
\\
Y \, v \, L + Z \, u \, R & u \, W
\end{array}
\right) \; ,
\end{equation}
where the eigenvalues have the same results as in (\ref{MnuMzeta}).

%$v \, X/\sqrt{2}$ and $u \, W/\sqrt{2}$ are the mass matrices of the neutrinos and $\zeta$-fermion if
%the mass matrix is diagonal. Here, $R$ and $L$ are the right- and left- projectors, that satisfy the relations
%$RL=LR=0$, $L^{2}=L$, $R^{2}=R$ and $L+R={\uma}$.
%Thus, the
%%neutrino/$\zeta$-fermion matrix can be diagonalized by a unitary transformation so that the masses are given by the
%$M_{\nu-\zeta}^{(\pm)}$-eigenvalues of (\ref{Lleptonsmatrix}) yield the sub-matrices of (\ref{MD}) :
%
%\begin{eqnarray}
%m_{\ell} \simeq \frac{|y_{\ell}| \, v}{\sqrt{2}}
%\hspace{0.5cm} , \hspace{0.5cm}
%m_{\zeta} \simeq \frac{|g_{\zeta}| \, u}{\sqrt{2}} \; ,
%\end{eqnarray}
%\begin{eqnarray}\label{mlzeta}
%M_{\nu-\zeta}^{(\pm)}= \frac{ v \, X + u \, W \!\pm\! \sqrt{\left(v \, X - u \, W \right)^{2}+4 v \, u \, X \, W \, |g_{f}|} }{2\sqrt{2}} \; ,
%\nonumber \\
%m_{\zeta} \!\!&=&\!\! -\frac{|y_{\ell}| \, v}{2\sqrt{2}}\left[ 1-\sqrt{1+4 \, \frac{|x_{\ell}||z_{\ell}|}{|y_{\ell}|^{2}} \,  \frac{u}{v}} \; \right] \, .
%\nonumber \\
%\hspace{0.3cm}
%\end{eqnarray}
%
%where
%
%\begin{eqnarray}
%$m_{\ell}\!=\!|y_{\ell}| \, v/\sqrt{2}$
%and
%$m_{\zeta}\!=\!|w| \, u/\sqrt{2}$
%\end{eqnarray}
%
%are the masses of the $\ell$- and $\zeta$-fermions when we make $|z_{\ell}|=|x_{\ell}| \rightarrow 0$,
%and $|g_{f}|$ is defined by the Yukawa coupling constants
%
%\begin{eqnarray}
%|g_{f}|= \sum_{i,j=1,2,3} \frac{|Z_{ij}||Y_{ij}|}{|W_{ij}||X_{ij}|} \; .
%\end{eqnarray}
%
Here, we are in the $Z'$-scenario where the $u$- VEV scale satisfies
the condition $u \gg v=246 \, \mbox{GeV}$; so, we hope that the set of fermions
$\zeta_{i}=\left\{ \, \zeta_{1} \, , \, \zeta_{2} \, , \, \zeta_{3} \, \right\}$
should describe three particles heavier than any neutrino of the SM,
{\it i. e.}, it is reasonable to consider that $u \, W \gg v \, X$. Furthermore,
the elements $\{ \, Y_{ij} \, , \, Z_{ij} \, \}$ can be considered weaker coupling constants
due to the mixing of right-neutrinos with the sector of $\zeta_{i}$-fermions. Under these conditions,
the mass matrices (\ref{MnuMzeta}) can be written as corrections of
%of (\ref{mlzeta}) are given by corrections of $|g_{f}|$ :
%
\begin{eqnarray}\label{autoveloresmfermions}
M^{(\nu)} &\simeq& \frac{v \, X}{\sqrt{2}} \left( \, {\uma} - \frac{ZY}{XW} \, \right)
%\hspace{0.1cm} , \hspace{0.1cm}
\nonumber \\
M^{(\zeta)} &\simeq& \frac{u \, W}{\sqrt{2}} \left( \, {\uma} +  \frac{v}{u} \, \frac{ZY}{W^2} \, \right)
\simeq \frac{u \, W}{\sqrt{2}} \; .
\end{eqnarray}
%
%where we have used that $|g_{f}|v/u \simeq 0$. Under these conditions, the $\theta$-mixing angle satisfies the matricial relation
%$\tan\theta \, {\uma} \simeq -2v \, W \, Z^{-1} \, e^{i\left(\alpha-\beta\right)}/u$.
%

%
The neutrino-$\zeta$-fermion sector is now free from mixed terms $\tilde{\nu}-\tilde{\zeta}$ in the basis $\tilde{\chi}_{L(R)}$ :
\begin{equation}\label{Lnuzetalinha}
{\cal L}_{\nu-\zeta}^{\, 0}=\bar{\tilde{\nu}}_{i} \, i \, \slash{\!\!\!\partial} \; \tilde{\nu}_{i}
+\bar{\tilde{\zeta}}_{i} \, i \, \slash{\!\!\!\partial} \; \tilde{\zeta}_{i}
-\bar{\tilde{\nu}}_{iL} \, M_{ij}^{\, (\nu)} \, \tilde{\nu}_{jR}-\bar{\tilde{\zeta}}_{iL} \, M_{ij}^{\, (\zeta)} \, \tilde{\zeta}_{jR}
+{\mbox h. \, c.} \; .
\end{equation}
Therefore, the mass matrices in (\ref{Lnuzetalinha}) can be independently diagonalized. The basis $\chi_{L(R)}$ can be
written in terms of $\tilde{\chi}_{L(R)}$ by the inverse transformation of (\ref{chiLRtransf}), {\it i. e.}, the transformation
$\left( \, \nu_{L(R)} \, , \, \zeta_{L(R)} \, \right) \mapsto \left(\, \tilde{\nu}_{L(R)} \, , \, \tilde{\zeta}_{L(R)} \, \right)$
\begin{eqnarray}
\nu_{iL(R)} &=& e^{-i \, \alpha} \, \cos\theta \, \tilde{\nu}_{iL(R)} - e^{-i \, \beta} \, \sin\theta \, \tilde{\zeta}_{iL(R)}
\nonumber \\
\zeta_{iL(R)} &=& e^{-i \, \beta} \, \sin\theta \, \tilde{\nu}_{iL(R)} + e^{i \, \alpha} \, \cos\theta \, \tilde{\zeta}_{iL(R)} \; .
\end{eqnarray}
The interactions $\nu-\ell-W^{\pm}$ in (\ref{intWnuell}) are written in the new basis $\left(\, \tilde{\nu}_{L(R)} \, , \, \tilde{\zeta}_{L(R)} \, \right)$, such that the neutrino-lepton-$W^{\pm}$ interaction is given by
\begin{equation}\label{LintnnuWell}
{\cal L}_{\tilde{\nu}-\ell-W}^{int}=-\frac{g \cos\theta}{\sqrt{2}} \, \, \bar{\tilde{\nu}}_{iL} \, \, \slash{\!\!\!\!W}^{+} \, \ell_{iL}
-\frac{g \cos\theta}{\sqrt{2}} \, \, \bar{\ell}_{iL} \, \, \slash{\!\!\!\!W}^{-} \, \tilde{\nu}_{iL} \; ,
\end{equation}
in which the $\alpha$-phase has been absorbed into the left-neutrino field, and $\cos\theta\simeq 1$ for $\theta \ll 1$.
Therefore, this changing of basis does not affect the already-known neutrino-lepton-$W^{\pm}$ interaction of the GSW model.
The mass basis of the Dirac neutrinos is introduced via a unitary transformation of the $\nu_{L(R)}$-fields,
then the Pontecorvo-Maki-Nakagawa-Sakata (PMNS) emerges in the interaction (\ref{LintnnuWell}) with
only one Dirac phase for Dirac Right-Neutrinos. Furthermore, there are three independent angles in the PMNS matrix to mix the neutrino fields, as usual.
If we introduce the unitary transformation $\tilde{\nu}_{L(R)} \, \longmapsto \, \nu'_{L(R)}=S_{L(R)} \, \tilde{\nu}_{L(R)}$,
the neutrino mass matrix in (\ref{Lnuzetalinha}) is diagonal in the mass basis $\nu'_{L(R)}$, in which we have $\bar{\tilde{\nu}}_{L} \, M^{(\nu)} \, \tilde{\nu}_{R}=\bar{\nu}'_{L} \, M_{D}^{(\nu)} \, \nu'_{R}$, where $M_{D}^{(\nu)}$ is given by
\begin{equation}\label{MDneutrinos}
M_{D}^{(\nu)}=S_{L}\frac{v \, X}{\sqrt{2}} \left( \, {\uma} - \frac{ZY}{XW} \, \right)S_{R}^{\dagger}
=\left(
\begin{array}{ccc}
M_{\nu'_{e}} & 0 & 0 \\
0 & M_{\nu'_{\mu}} & 0 \\
0 & 0 & M_{\nu'_{\tau}} \\
\end{array} \right) \; .
\end{equation}
Since the neutrino masses are measured through their oscillations, the transition probabilities depend on the subtraction of the
squared masses. In the case of the electron- and muon-neutrinos, this subtraction is $\Delta M_{\nu'_{e}-\nu'_{\mu}}^{2}:=|M_{\nu'_{e}}^2-M_{\nu'_{\mu}}^2|\simeq \left( \, 7.53 \, \pm \, 0.18 \, \right) \times 10^{-5} \, \mbox{eV}^{2}$ \cite{ArakiPRL2005}. Thus, the subtraction of squared coupling constants are extremely weak,
$\Delta X_{\nu'_{e}-\nu'_{\mu}}^{2}:=||X_{\nu'_{e}}|^2-|X_{\nu'_{\mu}}|^2|\simeq 2.5 \times 10^{-27}$.
The mixing between $\tau$- and muon- neutrino yield the squared subtraction
$\Delta M_{\nu'_{\tau}-\nu'_{\mu}}^{2}:=|M_{\nu'_{\tau}}^2-M_{\nu'_{\mu}}^2| \simeq \left( \, 2.44 \, \pm \, 0.06 \, \right) \times 10^{-3} \, \mbox{eV}^{2}$, then we estimate $\Delta X_{\nu'_{\tau}-\nu'_{\mu}}^{2}:=||X_{\nu'_{\tau}}|^2-|X_{\nu'_{\mu}}|^2|\simeq 8 \times 10^{-26}$.
These estimation helps us to obtain the corresponding values of mixed Yukawa coupling constants $Y$ and $Z$,
but we need define a range for $\zeta_{i}$-masses.
The $\zeta_{i}$-fermions interact weakly with the leptons according to
\begin{equation}\label{Lintzetatilell}
{\cal L}_{\tilde{\zeta}-\ell-W}^{int}=\frac{g \sin\theta}{\sqrt{2}} \, \, \bar{\tilde{\zeta}}_{iL} \, \, \slash{\!\!\!\!W}^{+} \, \ell_{iL}
+\frac{g \sin\theta}{\sqrt{2}} \, \, \bar{\ell}_{iL} \, \, \slash{\!\!\!\!W}^{-} \, \tilde{\zeta}_{iL} \; ,
\end{equation}
where the $\beta$-phase has been absorbed into the $\tilde{\zeta}_{iL}$-fields. The leptonic sector is diagonalized like
in the SM. The mass term is $-v \, \bar{\ell}_{L} \, G^{(\ell)} \, \ell_{R}/\sqrt{2}$, that can be diagonalized by means of the
unitary transformations
%
%\begin{eqnarray}\label{chiLRtransf}
$\ell_{L(R)} \, \longmapsto \, \ell'_{L(R)}=U_{L(R)} \, \ell_{L(R)}$,
%\nonumber \\
%\nu_{R} \, \rightarrow \, \nu'_{R} \!&=&\! K_{\nu} \, \nu_{L}
%\nonumber \\
%\zeta_{L} \, \rightarrow \, \zeta'_{L} \!&=&\! U_{\zeta} \, \zeta_{L}
%\nonumber \\
%\zeta_{R} \, \rightarrow \, \zeta'_{R} \!&=&\! K_{\zeta} \, \zeta_{R} \; ,
%\end{eqnarray}
%
in which $U_{L}^{\dagger} \, U_{L}=U_{R}^{\dagger} \, U_{R}={\uma}$, the mass matrix diagonal is
$M_{D}^{(\ell)}=U_{L} \, v \, G^{(\ell)}/\sqrt{2} \, U_{R}^{\dagger}=\mbox{diag}\left( \, M_{e} \, , \, M_{\mu} \, , \, M_{\tau} \, \right)$.
Analogously, the diagonalization of $\zeta_{i}$- mass matrix is performed by another unitary transformation,
that we denote by $\tilde{\zeta}_{L(R)} \, \longmapsto \, \zeta'_{L(R)}=V_{L(R)} \, \tilde{\zeta}_{L(R)}$, where $V_{L}^{\dagger} \, V_{L}=V_{R}^{\dagger} \, V_{R}={\uma}$, and we obtain the diagonal mass matrix :
\begin{eqnarray}
M_{D}^{\, (\zeta)}=V_{L} \, \frac{u \, W}{\sqrt{2}} \, V_{R}^{\dagger}
= \left(
\begin{array}{ccc}
M_{\zeta'_{1}} & 0 & 0 \\
0 & M_{\zeta'_{2}} & 0 \\
0 & 0 & M_{\zeta'_{3}} \\
\end{array} \right) \; .
\end{eqnarray}
So, we can write the interactions (\ref{Lintzetatilell})
in the mass basis $\left\{ \, \zeta'_{L} \, , \, \ell'_{L} \, \right\}$ as below
\begin{equation}\label{Intzeta'ell'W}
{\cal L}_{\zeta'-\ell'-W}^{int}=\frac{g \sin\theta}{\sqrt{2}} \, \, \bar{\zeta}'_{iL} \, \, \slash{\!\!\!\!W}^{+} \, V_{ij} \, \ell'_{jL}
+\frac{g \sin\theta}{\sqrt{2}} \, \, \bar{\ell}'_{iL} \, \, \slash{\!\!\!\!W}^{-} \, V_{ij}^{\dagger} \, \zeta'_{jL} \; ,
\end{equation}
where the most general $V$-unitary matrix $V:=V_{L} \, U_{L}^{\dagger}$ is parameterized by
\begin{equation}\label{PMNSMatrix}
V=\left(
\begin{array}{ccc}
c_{12}c_{13} & s_{12}c_{13} & s_{13} \, e^{-i\delta} \\
-s_{12}c_{23}-c_{12}s_{23}c_{13}e^{i\delta} & c_{12}c_{23}-s_{12}s_{23}s_{13}e^{i\delta} & s_{23}c_{13} \\
s_{12}s_{23}-c_{12}c_{23}s_{13}e^{i\delta} & -c_{12}s_{23}-s_{12}c_{23}s_{13}e^{i\delta} & c_{23}c_{13} \\
\end{array}
\right) .
\end{equation}
It has the same structure as the Cabibbo-Kobayashi-Maskawa (CKM) matrix : it displays three mixing angles
$\left\{ \, \theta_{12} \, , \, \theta_{13} \, , \, \theta_{23} \, \right\}$ and only one Dirac $\delta$-phase,
since we do not introduce $\zeta_{i}$-Majorana fermions. In (\ref{PMNSMatrix}), we simply the
sines and cosines of the angles as : $\cos\theta_{ij}=c_{ij}$ and $\sin\theta_{ij}=s_{ij}$.
The $\zeta'_{i}$-masses depend on the $u$-VEV scale, so it must exhibit a heavier fermion content
in comparison with the SM fermions. The recent simulations of CMS-Collaboration point out to
dark matter fermion content with mass of order $0.55 \, \mbox{TeV}$ \cite{CMS20172}. Therefore,
we take here $M_{\zeta'_{1}}=0.5 \, \mbox{TeV}$, and if we use $u=2.8 \, \mbox{TeV}$,
the estimation for $W_{\zeta'_{1}}$-Yukawa constant is $W_{\zeta'_{1}}\simeq 0.28$.
The two others heavy fermions $\zeta'_{2}$ and $\zeta'_{3}$ can be particles in the mass range of
$>0.5 \, \mbox{TeV}$, so we choose $M_{\zeta'_{2}}=0.8 \, \mbox{TeV}$ and $M_{\zeta'_{3}}=1 \, \mbox{TeV}$,
so the correspondent coupling constants has the values $W_{\zeta'_{2}}\simeq 0.4$ and $W_{\zeta'_{3}}\simeq 0.5$, respectively.
Using the uncertainty on $\Delta M_{\nu'_{e}-\nu'_{\mu}}^{2}$, the correction to the neutrinos masses give the upper bound
\begin{eqnarray}
\frac{\Delta Y_{\nu'_{e}-\nu'_{\mu}}\Delta Z_{\nu'_{e}-\nu'_{\mu}}}{\Delta X_{\nu'_{e}-\nu'_{\mu}}W_{\zeta'_{1}}} \lesssim 0.15 \; ,
\end{eqnarray}
and $\Delta Y \simeq \Delta Z$, we obtain $\Delta Y_{\nu'_{e}-\nu'_{\mu}}\simeq\Delta Z_{\nu'_{e}-\nu'_{\mu}}\simeq 8.3 \, \times \, 10^{-8}$.
Under these conditions and the previous bounds, the $\theta$-mixing angle in (\ref{tantheta}) turns out to be extremely small :
$\tan\theta\simeq-9\, \times \, 10^{-8}$.

The interaction of the $\zeta_{i}$-fermions with the $Z'$-boson is not affected by the change of basis dictated by the masses.
Other important fact is that the interaction (\ref{Intzeta'ell'W}) connects the fermion sector of the SM
with a set of fermions candidate to dark sector via $W^{\pm}$-bosons. Since $\theta \ll 1$, the $\theta$-angle
rules the magnitude of (\ref{Intzeta'ell'W}), {\it i. e.}, $\sin\theta \simeq \theta \simeq-9\, \times \, 10^{-8}$.
This vertex is represented by the diagram below.
\begin{figure}[!h]%\label{figvm}
\begin{center}
\newpsobject{showgrid}{psgrid}{subgriddiv=1,griddots=10,gridlabels=6pt}
\begin{pspicture}(5,1)(11,2.5)
%\showgrid
\psset{arrowsize=0.2 2}
\psset{unit=0.8}
%
%%%%%%%%%%%%%%%%%%%% Vertice Zeta - Leptons %%%%%%%%%%%%%%%%%%%%%%%%%%%%%%%%
%
\pscoil[coilarm=0,coilaspect=0,coilwidth=0.2,coilheight=1.0,linecolor=black](6.5,1.05)(6.5,3)
\psline[linecolor=black,linewidth=0.5mm]{-}(5,1)(8,1)
\psline[linecolor=black,linewidth=0.5mm]{->}(5,1)(6,1)
\psline[linecolor=black,linewidth=0.5mm]{->}(7,1)(7.55,1)
\put(6.8,2.8){\large$W^{\pm}$}
\put(4.95,1.3){\large$\bar{\zeta}'_{i}$}
\put(7.8,1.3){\large$\ell'_{j}$}
\put(8.5,1.1){\large $\Gamma_{ij}^{\, \mu}=- \, \frac{i \, g \,\theta \, V_{ij}}{2\sqrt{2}} \gamma^{\mu}\left(1- \gamma_{5}\right) .$}
\end{pspicture}
%
%\vspace{0.2cm}
%
%\caption{\scshape{The one loop correction of the vertex diagram.}}\label{figvm}
%
\end{center}
\end{figure}
\noindent
Therefore, this vertex yields an important contribution to $\zeta_{i}$-fermions magnetic dipole momentum at the one-loop approximation.
The weakly coupling constant that emerges here is $g \, \theta \simeq-6 \times 10^{-9}$. The interaction (\ref{Intzeta'ell'W})
also violates the CP-symmetry due to the $\delta$-phase in the mixing matrix (\ref{PMNSMatrix}).

%
%The $m_{\nu_{\ell}-\zeta}^{(-)}$-eigenvalue gives us the neutrinos masses with the correction of the $|g_{f}|$-coupling constant.
%The neutrinos masses to the squared are constraints by the subtraction between itself. For example,
%

%
\section{The $Z'$-phenomenology}
\renewcommand{\theequation}{4.\arabic{equation}}
\setcounter{equation}{0}

\subsection{The $Z'$-decay into fermions : $Z' \, \rightarrow \, \bar{f} \, f$}

The recent $Z'$-phenomenology points to the cascade effects at the tree-level using the CMS
data for the pp-collision at $\sqrt{s}=13 \, \mbox{TeV}$. We will obtain an expression for the
$Z'$-decay width into the any $f$-fermion of the model.
Then, using the previous rules and quantum field-theoretic results,
the decay width of $Z'$ into any $f$-fermion is given by
\begin{eqnarray}
\Gamma(Z' \rightarrow \bar{f} \, f )= \frac{g_{Z'}^{2} M_{Z'} }{24\pi}
\left( |g^{f}_{V}|^2+|g^{f}_{A}|^2 \right)
%\times
%\nonumber \\
%&&
%\times \,
\sqrt{1-\frac{4M_{f}^{\, 2}}{M_{Z'}^{\, 2}}} \left( 1
- \frac{3M_{f}^{\, 2}}{4M_{Z'}^{\, 2}} \right) \, ,
\end{eqnarray}
where $M_{Z'} > 2 M_{f}$, for $f=\left\{ \, \ell_{i} \, , \, \nu_{i} \, , \, \zeta_{i} \, \right\}$ or quarks.
The decay width into the lepton pair is
\begin{eqnarray}\label{Z'Decayll}
\Gamma(Z' \rightarrow \bar{\ell}_{i} \, \ell_{i} ) \!&=&\! \frac{g_{Z'}^2 M_{Z'} }{24\pi}
%\times
%\nonumber \\
%&&
%\times
\left(\frac{1}{2}-2\sin^2\alpha+\frac{5}{2} \sin^4\alpha \right) .
\hspace{0.3cm}
\end{eqnarray}
In the neutrino case, the Left- and Right-components provide the following contributions
\begin{equation}
\Gamma(Z' \rightarrow \bar{\nu}_{i} \, \nu_{i} )=\frac{g_{Z'}^2 M_{Z'} }{48\pi}
\left( 1+ \sin^4\alpha \right) \, .
\end{equation}
Notice that we have used that $M_{Z'}\gg \left\{ \, 2M_{\ell} \, , \, 2M_{\nu} \, \right\}$ for leptons and neutrinos.
The processes of the $Z'$-decay can be useful to search the dark matter through the mono-V jets channels associated
with the electroweak bosons $W$ or $Z$. The observation of these final states could be interpreted as a dark matter
particle content, that here we identify as the $\zeta'_{i}$-fermions.
The diagram for this effect is illustrated in the figure (\ref{Z'Decay}) :
%
%
%%%%%%%%%%%%%%%%% The Z'-cascade into W or Z %%%%%%%%%%%%%%%%%%%%%%%%%
%
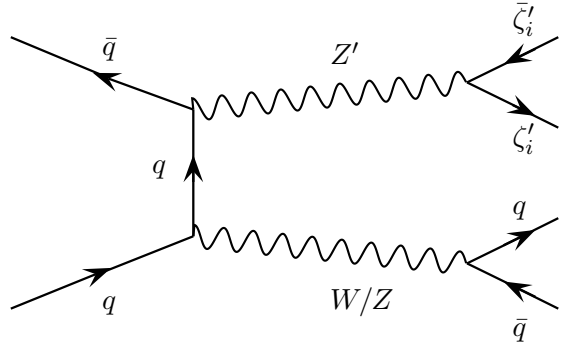
\begin{figure}[!h]%\label{figvm}
\begin{center}
\newpsobject{showgrid}{psgrid}{subgriddiv=1,griddots=10,gridlabels=6pt}
\begin{pspicture}(0,-1)(2.5,3.2)
%\showgrid
\psset{arrowsize=0.2 2}
\psset{unit=1.2}
%
%%%%%%%%%%%%%%%%%%%%%%%%%%% Bosons Z' Z W %%%%%%%%%%%%%%%%%%%%%%%%%%%%%%%%%%
%
\pscoil[coilaspect=0,coilarm=0,coilwidth=0.25,coilheight=1.3,linecolor=black](0,0.3)(2.99,0)
\pscoil[coilaspect=0,coilarm=0,coilwidth=0.25,coilheight=1.3,linecolor=black](0,1.7)(2.99,2)
\put(1.5,2.2){$Z'$}
\put(1.5,-0.5){$W/Z$}
%
%\psline[linecolor=black,linewidth=0.3mm]{->}(5,2)(6.5,3.5)
%
\psline[linecolor=black,linewidth=0.3mm]{->}(0,0.3)(0,1.2)
\psline[linecolor=black,linewidth=0.3mm]{-}(0,0.9)(0,1.7)
\put(-0.45,1){$q$}
%
%%%%%%%%%%%%%%%%%%%%%%%%%%%%%%%%%%%%%%
%
\psline[linecolor=black,linewidth=0.3mm]{->}(-2,-0.5)(-0.9,-0.05)
\psline[linecolor=black,linewidth=0.3mm]{-}(-1,-0.1)(0,0.3)
\put(-1,2.3){$\bar{q}$}
\psline[linecolor=black,linewidth=0.3mm]{->}(0,1.7)(-1.1,2.105)
\psline[linecolor=black,linewidth=0.3mm]{-}(-1,2.08)(-2,2.5)
\put(-1,-0.5){$q$}
%
%%%%%%%%%%%%%%%%%%%%%%%%%%%%%%%%%%%%
%
\psline[linecolor=black,linewidth=0.3mm](3,2)(3.75,2.37)
\psline[linecolor=black,linewidth=0.3mm]{<-}(3.4,2.2)(4,2.5)
\put(3.5,2.6){$\bar{\zeta}'_{i}$}
\psline[linecolor=black,linewidth=0.3mm]{->}(3,2)(3.75,1.63)
\psline[linecolor=black,linewidth=0.3mm](3.5,1.75)(4,1.5)
\put(3.5,1.3){$\zeta'_{i}$}
%
%%%%%%%%%%%%%%%%%%%%%%%%%%%%%%%%%%%%%%%%%%%%%%%%%%%%
%
\psline[linecolor=black,linewidth=0.3mm]{->}(3,0)(3.75,0.37)
\psline[linecolor=black,linewidth=0.3mm](3.4,0.2)(4,0.5)
\put(3.5,0.55){$q$}
\psline[linecolor=black,linewidth=0.3mm](3,0)(3.75,-0.37)
\psline[linecolor=black,linewidth=0.3mm]{<-}(3.4,-0.2)(4,-0.5)
\put(3.5,-0.8){$\bar{q}$}
%
%\psline[linecolor=black,linewidth=0.3mm]{<-}(0,1)(-1.5,-0.5)
%\psline[linecolor=black,linewidth=0.3mm]{->}(1,2)(-0.4,3.4)
%
%
%Momenta
%
%\psline[linecolor=black,linewidth=0.3mm]{->}(2.2,2.7)(3.8,2.7)
%\put(3,2.6){\Large$X^{\mu}$}
%
%\psline[linecolor=black,linewidth=0.3mm]{->}(7.8,1.2)(7.8,2.8)
%\put(3,1){\Large$k$}
%
%\put(5.5,3.6){\Large$e^{-}$}
%\put(5.4,-0.1){\Large$e^{+}$}
%\put(6.7,3.1){\Large$q$}
%\put(6.7,0.7){\Large$q^{\prime}$}
%
%\put(-0.2,3.6){\Large$e^{-}$}
%\put(-0.4,-0.1){\Large$e^{+}$}
%\put(-1.1,3.1){\Large$q$}
%\put(-1.1,0.7){\Large$q^{\prime}$}
%
%\put(0.6,-0.3){$p$}
%
%\put(1.3,0){$\Gamma_{\mu}^{(3)\;abc}(p)=g_{1}f^{abc}p_{\mu}$}
%
%
\end{pspicture}
%
%\vspace{0.2cm}
%
\caption{\scshape{The $Z'$-decay into any pair $\bar{\zeta}'_{i}-\zeta'_{i}$ of the $\zeta'_{i}$-fermions set.
The cascade effect as a possible dark matter detection via $W$- or $Z$-monojets.}} \label{Z'Decay}
\end{center}
\end{figure}

\noindent
Therefore, the result for the total decay width of $Z'$ into the $\zeta_{i}$-family is given by the sum
\begin{eqnarray}
\Gamma(Z' \rightarrow \bar{\zeta}' \, \zeta')=\sum_{i=1}^{3}\Gamma\left( \, Z' \rightarrow \bar{\zeta}'_{i} \, \zeta'_{i} \, \right) \; ,
\end{eqnarray}
in which for a particular $\zeta'_{i}$, it is shown to be given by
%$Z' \rightarrow \ell^{+} \, \ell^{-}$,
%$Z' \rightarrow \bar{\nu}_{\ell} \, \nu_{\ell}$ and
%
\begin{eqnarray}
\Gamma(Z' \rightarrow \bar{\zeta}'_{i} \, \zeta'_{i})=\frac{g_{Z'}^2 M_{Z'}}{24\pi}
%\times
%\nonumber \\
%\times \,
\sqrt{1-\frac{4M_{\zeta'_{i}}^{\, 2}}{M_{Z'}^{\, 2}}} \left( 1
- \frac{3M_{\zeta'_{i}}^{\, 2}}{4M_{Z'}^{\, 2}} \right) \, .
\end{eqnarray}
Here, the condition $M_{Z'} > 2 \, M_{\zeta'_{i}}$ must be satisfied for any $\zeta'_{i}$-fermion.
Using the previous values $M_{Z'}= 2 \, \mbox{TeV}$ and $M_{\zeta'_{1}}=0.55 \, \mbox{TeV}$,
the $Z'$-width decay rate is
\begin{eqnarray}
\Gamma(Z' \rightarrow \bar{\zeta}'_{1} \, \zeta'_{1}) \simeq \frac{0.018}{\sin^2(2\alpha)} \, \mbox{TeV} \; .
\end{eqnarray}
%
%This decay width is plotted in the figure (\ref{Z'zetaDecay}).
%
%\begin{figure}[h]
%\centering
%\includegraphics[scale=0.44]{ZDecaySinalpha.eps}
%\caption{The decay width of the process $Z' \rightarrow \bar{\zeta}_{1} \, \zeta_{1}$ plotted as function of the $\alpha$-mixing angle.}
%\label{Z'zetaDecay}
%\end{figure}
%
In the case of $\alpha=45^{o}$, the decay width is
$\Gamma(Z' \rightarrow \bar{\zeta}'_{1} \, \zeta'_{1}) \simeq 0.018 \, \mbox{TeV}$, and the $Z'$-decay time in this process is estimated by
\footnote{We have used the conversion formula $1 \, \mbox{TeV}=1.52 \times 10^{27} \, \mbox{s}^{-1}$ in the natural units $\hbar=c=1$. }
\begin{eqnarray}
\tau(Z' \rightarrow \bar{\zeta}'_{1} \, \zeta'_{1})=\frac{1}{\Gamma(Z' \rightarrow \bar{\zeta}'_{1} \, \zeta'_{1})}\simeq 3.7 \times 10^{-26} \, \mbox{s} \; .
\end{eqnarray}

The possible $Z'$-decays into quarks, {\it i. e.}, $Z' \rightarrow \bar{q} \, q$, has also a phenomenological analysis at the CMS Collaboration,
see \cite{CMS20172}. The $Z'$-decay cases into the first generation, {\it i. e.},  $Z' \, \rightarrow \, \bar{u} \, u$ and $Z' \, \rightarrow \, \bar{d} \, d$,
have the decays width below :
\begin{eqnarray}
\Gamma(Z' \rightarrow \bar{u} \, u) &=& \frac{g_{Z'}^{2} M_{Z'}}{24\pi}
%\, \times
%\nonumber \\
%\times \,
\left( \frac{1}{2}-\frac{4}{3} \sin^2\alpha+\frac{17}{18} \sin^4\alpha \right)
\; ,
\nonumber \\
\Gamma(Z' \rightarrow \bar{d} \, d) &=& \frac{g_{Z'}^2 M_{Z'}}{24\pi}
%\, \times
%\nonumber \\
%\times \,
\left( \frac{1}{2}-\frac{2}{3} \sin^2\alpha+\frac{5}{18} \sin^4\alpha \right) \, ,
\hspace{0.8cm}
\end{eqnarray}
where we have used that $M_{Z'} \gg m_{u}$ and $M_{Z'} \gg m_{d}$. Using the $\alpha$-angle of $\alpha=45^{o}$, we obtain the decay widths at the $\mbox{GeV}$-scale :
\begin{eqnarray}
\Gamma(Z' \rightarrow \bar{u} \, u) \simeq 2 \, \mbox{GeV}
\hspace{0.2cm} , \hspace{0.2cm}
%\nonumber \\
\Gamma(Z' \rightarrow \bar{d} \, d) \simeq 7 \, \mbox{GeV} \; .
\end{eqnarray}
\subsection{The $F$-Higgs decays}
%$F \, \rightarrow \, Z' \, Z' $}
%

The $Z'$-decay into scalars has a phenomenological interest in the
study of $Z'$-resonance to the final four-lepton state \cite{CMS20171}.
On the other hand, the $F$-Higgs decays into the leptons pairs $Z' \, \rightarrow \, 4 \, \ell_{i}$.
The process is illustrated at the tree-level as shown below:
%
%%%%%%%%%%%%%%%%% The Z'-cascade into FF-4 leptons %%%%%%%%%%%%%%%%%%%%%%%%%
%
\begin{figure}[!h]%\label{figvm}
\begin{center}
\newpsobject{showgrid}{psgrid}{subgriddiv=1,griddots=10,gridlabels=6pt}
\begin{pspicture}(-3,-0.7)(6,3.4)
%\showgrid
\psset{arrowsize=0.2 2}
\psset{unit=1.2}
%
%%%%%%%%%%%%%%%%%%%%%%%%%%% Boson Z' %%%%%%%%%%%%%%%%%%%%%%%%%%%%%%%%%%
%
%\pscoil[coilaspect=0,coilarm=0,coilwidth=0.25,coilheight=1.3,linecolor=black](0,0.3)(2.99,0)

%
\pscoil[coilaspect=0,coilarm=0,coilwidth=0.25,coilheight=1.3,linecolor=black](0,0.9)(1.5,1)
\put(0.8,1.3){$Z'$}
%
%\put(1.5,-0.5){$W/Z$}
%
%
%\put(-0.45,1){$q$}
%
%%%%%%%%%%%%%%%%%%  Quarks %%%%%%%%%%%%%%%%%%%%
%
\psline[linecolor=black,linewidth=0.3mm](-0.65,0.3)(0,0.9)
\psline[linecolor=black,linewidth=0.3mm]{->}(-1.5,-0.5)(-0.5,0.45)
\put(-0.8,1.9){$\bar{q}$}
\psline[linecolor=black,linewidth=0.3mm]{->}(0,0.9)(-0.85,1.7)
\psline[linecolor=black,linewidth=0.3mm](-0.45,1.33)(-1.5,2.3)
\put(-0.8,-0.1){$q$}
%
%%%%%%%%%%%%%%%%%%%%%%%%%%%%%%%%%%%%%%%%%%%%
%
%%%%%%%%%%%%%%%%% Scalars %%%%%%%%%%%%%%
%
\psline[linestyle=dashed,linewidth=0.5mm](1.5,1)(3,2)
\psline[linestyle=dashed,linewidth=0.5mm](1.5,1)(3,0)
\put(2.1,1.7){$F$}
\put(2.1,0){$F$}
%
%%%%%%%%%%%%%%%%%%%%%%%%%%%%%%%%%%%%%%%
%
\psline[linecolor=black,linewidth=0.3mm]{->}(3,2)(3.75,2.37)
\psline[linecolor=black,linewidth=0.3mm](3.4,2.2)(4,2.5)
\put(3.5,2.6){$\ell_{i}^{-}$}
\psline[linecolor=black,linewidth=0.3mm](3,2)(3.75,1.63)
\psline[linecolor=black,linewidth=0.3mm]{<-}(3.5,1.75)(4,1.5)
\put(3.5,1.2){$\ell_{i}^{+}$}
%
%%%%%%%%%%%%%%%%%%%%%%%%%%%%%%%%%%%%%%%%%%%%%%%%%%%%
%
\psline[linecolor=black,linewidth=0.3mm]{->}(3,0)(3.75,0.37)
\psline[linecolor=black,linewidth=0.3mm](3.4,0.2)(4,0.5)
\put(3.5,0.55){$\ell_{i}^{-}$}
\psline[linecolor=black,linewidth=0.3mm](3,0)(3.75,-0.37)
\psline[linecolor=black,linewidth=0.3mm]{<-}(3.4,-0.2)(4,-0.5)
\put(3.5,-0.8){$\ell_{i}^{+}$}
%
%\psline[linecolor=black,linewidth=0.3mm]{<-}(0,1)(-1.5,-0.5)
%\psline[linecolor=black,linewidth=0.3mm]{->}(1,2)(-0.4,3.4)
%
%
%Momenta
%
%\psline[linecolor=black,linewidth=0.3mm]{->}(2.2,2.7)(3.8,2.7)
%\put(3,2.6){\Large$X^{\mu}$}
%
%\psline[linecolor=black,linewidth=0.3mm]{->}(7.8,1.2)(7.8,2.8)
%\put(3,1){\Large$k$}
%
%\put(5.5,3.6){\Large$e^{-}$}
%\put(5.4,-0.1){\Large$e^{+}$}
%\put(6.7,3.1){\Large$q$}
%\put(6.7,0.7){\Large$q^{\prime}$}
%
%\put(-0.2,3.6){\Large$e^{-}$}
%\put(-0.4,-0.1){\Large$e^{+}$}
%\put(-1.1,3.1){\Large$q$}
%\put(-1.1,0.7){\Large$q^{\prime}$}
%
%\put(0.6,-0.3){$p$}
%
%\put(1.3,0){$\Gamma_{\mu}^{(3)\;abc}(p)=g_{1}f^{abc}p_{\mu}$}
%
%
\end{pspicture}
%
%\vspace{0.2cm}
%
\caption{\scshape{The leading order Feynman diagram for the cascade decay of the $Z'$-resonance into a four-lepton final state.}} \label{Z'4leptons}
\end{center}
\end{figure}
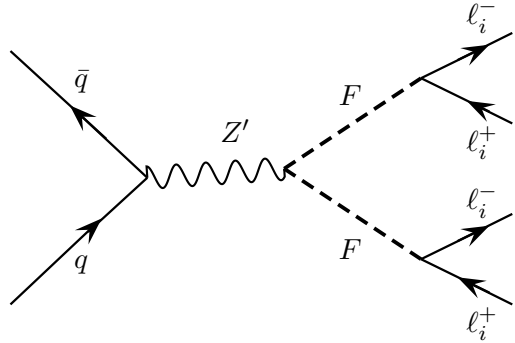

\noindent

Moreover, the decay process $Z' \, \rightarrow \, F \, F$ can not be described by the model due to definition of covariant derivative in the Higgs sector.
For example, after the SSB, the interaction of $F$-Scalar field with the $Z'$-boson is given by
\begin{eqnarray}
{\cal L}_{F-Z'}^{\, int}=\frac{M_{Z'}^{2}}{u} \, F \, Z_{\mu}^{\prime} Z^{\prime \mu}
+\frac{1}{2} \, \frac{M_{Z'}^{2}}{u^2} \, F^{2} \, Z_{\mu}^{\prime} Z^{\prime \mu} \; ,
\end{eqnarray}
where we have the possible vertex $F \, Z' \, Z'$ and $F \, F \, Z' \, Z'$. Using the usual rules of QFT, a process possible described by the $F-Z'$ sector is the decay
$F \, \rightarrow \, Z' \, Z'$, that has the following decay width :
\begin{eqnarray}
\Gamma(F \, \rightarrow \, Z' \, Z') \!&=&\! \frac{g_{Z'}^{2} M_{Z'}}{16\pi} \frac{M_{Z'}}{M_{F}}
%\times
%\nonumber \\
%&&
%\hspace{-1.8cm}
%\times
\left(1-\frac{4 M_{Z'}^{2}}{M_{F}^2} \right)^{1/2} \!
\left(3-\frac{M_{F}^{2}}{M_{Z'}^2}+\frac{M_{F}^{4}}{4M_{Z'}^4} \right) \, ,
\hspace{0.4cm}
\end{eqnarray}
where it is restricted by the condition $M_{F} > 2 \, M_{Z'}$.

However, other cascade effects can be described by the $Z'-F$-interaction. For example, the decay process in which the $Z'$ decays indirectly
into two final states of leptons and others two final states of $\zeta_{i}$-fermions,{\it i. e.}, $Z' \, \rightarrow \, 2 \, \ell_{i} + 2 \, \zeta'_{i}$.
This process is illustrated in the figure (\ref{Z'2leptons2zeta}).
%
%%%%%%%%%%%%%%%%% The Z'-cascade into FF-4 leptons %%%%%%%%%%%%%%%%%%%%%%%%%
%
\begin{figure}[!h]%\label{figvm}
\begin{center}
\newpsobject{showgrid}{psgrid}{subgriddiv=1,griddots=10,gridlabels=6pt}
\begin{pspicture}(-3,-1)(6,3.3)
%\showgrid
\psset{arrowsize=0.2 2}
\psset{unit=1.2}
%
%%%%%%%%%%%%%%%%%%%%%%%%%%% Boson Z' %%%%%%%%%%%%%%%%%%%%%%%%%%%%%%%%%%
%
%\pscoil[coilaspect=0,coilarm=0,coilwidth=0.25,coilheight=1.3,linecolor=black](0,0.3)(2.99,0)
%
\pscoil[coilaspect=0,coilarm=0,coilwidth=0.25,coilheight=1.3,linecolor=black](0,0.9)(1.5,1)
\put(0.8,1.3){$Z'$}
%
%\put(1.5,-0.5){$W/Z$}
%
%
%\put(-0.45,1){$q$}
%
%%%%%%%%%%%%%%%%%%  Quarks %%%%%%%%%%%%%%%%%%%%
%
\psline[linecolor=black,linewidth=0.3mm](-0.65,0.3)(0,0.9)
\psline[linecolor=black,linewidth=0.3mm]{->}(-1.5,-0.5)(-0.5,0.45)
\put(-0.8,1.9){$\bar{q}$}
\psline[linecolor=black,linewidth=0.3mm]{->}(0,0.9)(-0.85,1.7)
\psline[linecolor=black,linewidth=0.3mm](-0.45,1.33)(-1.5,2.3)
\put(-0.8,-0.1){$q$}
%
%%%%%%%%%%%%%%%%%%%%%%%%%%%%%%%%%%%%%%%%%%%%
%
%%%%%%%%%%%%%%%%% Scalars %%%%%%%%%%%%%%
%
\psline[linestyle=dashed,linewidth=0.5mm](1.5,1)(3,0)
\pscoil[coilaspect=0,coilarm=0,coilwidth=0.25,coilheight=1.3,linecolor=black](1.5,1)(3,2)
%
%\psline[linestyle=dashed,linewidth=0.5mm](1.5,1)(3,0)
%
\put(2.1,1.8){$Z'$}
\put(2.1,0){$F$}
%
%%%%%%%%%%%%%%%%%%%%%%%%%%%%%%%%%%%%%%%
%
\psline[linecolor=black,linewidth=0.3mm]{->}(3,2)(3.75,2.37)
\psline[linecolor=black,linewidth=0.3mm](3.4,2.2)(4,2.5)
\put(3.5,2.6){$\ell_{i}^{-}$}
\psline[linecolor=black,linewidth=0.3mm](3,2)(3.75,1.63)
\psline[linecolor=black,linewidth=0.3mm]{<-}(3.5,1.75)(4,1.5)
\put(3.5,1.2){$\ell_{i}^{+}$}
%
%%%%%%%%%%%%%%%%%%%%%%%%%%%%%%%%%%%%%%%%%%%%%%%%%%%%
%
\psline[linecolor=black,linewidth=0.3mm]{->}(3,0)(3.75,0.37)
\psline[linecolor=black,linewidth=0.3mm](3.4,0.2)(4,0.5)
\put(3.5,0.55){$\zeta_{i}^{'-}$}
\psline[linecolor=black,linewidth=0.3mm](3,0)(3.75,-0.37)
\psline[linecolor=black,linewidth=0.3mm]{<-}(3.4,-0.2)(4,-0.5)
\put(3.5,-0.8){$\zeta_{i}^{'+}$}
%
%\psline[linecolor=black,linewidth=0.3mm]{<-}(0,1)(-1.5,-0.5)
%\psline[linecolor=black,linewidth=0.3mm]{->}(1,2)(-0.4,3.4)
%
%
%Momenta
%
%\psline[linecolor=black,linewidth=0.3mm]{->}(2.2,2.7)(3.8,2.7)
%\put(3,2.6){\Large$X^{\mu}$}
%
%\psline[linecolor=black,linewidth=0.3mm]{->}(7.8,1.2)(7.8,2.8)
%\put(3,1){\Large$k$}
%
%\put(5.5,3.6){\Large$e^{-}$}
%\put(5.4,-0.1){\Large$e^{+}$}
%\put(6.7,3.1){\Large$q$}
%\put(6.7,0.7){\Large$q^{\prime}$}
%
%\put(-0.2,3.6){\Large$e^{-}$}
%\put(-0.4,-0.1){\Large$e^{+}$}
%\put(-1.1,3.1){\Large$q$}
%\put(-1.1,0.7){\Large$q^{\prime}$}
%
%\put(0.6,-0.3){$p$}
%
%\put(1.3,0){$\Gamma_{\mu}^{(3)\;abc}(p)=g_{1}f^{abc}p_{\mu}$}
%
%
\end{pspicture}
%
%\vspace{0.2cm}
%
\caption{\scshape{The leading order Feynman diagram for the cascate decay of $Z'$ resonance to a four-lepton final state.}} \label{Z'2leptons2zeta}
\end{center}
\end{figure}
\noindent
In this case, we have part of the final state described by the decay width (\ref{Z'Decayll}).
The other process in the final state is the $F$-decay into the $\zeta'_{i}$-fields, which we denote as
$F \, \rightarrow \, \bar{\zeta}' \, \zeta'$.  The $F$-scalar field interacts with the $\zeta'_{i}$-fields
by means of the expression
\begin{eqnarray}
{\cal L}^{int}_{F-\bar{\zeta}' \, \zeta'}=- \sum_{i,j=1}^{3} \frac{|W_{\zeta'_{i}}|}{\sqrt{2}} \, F \, \bar{\zeta}'_{i} \, \zeta'_{j} \; .
\end{eqnarray}
Thus, the total decay with is given by
\begin{eqnarray}
\Gamma(F \, \rightarrow \, \bar{\zeta}' \, \zeta')=\sum_{i=1}^{3} \, \Gamma\left(F \, \rightarrow \, \bar{\zeta}'_{i} \, \zeta'_{i}\right) \; ,
\end{eqnarray}
where we obtain the decay width for
\begin{eqnarray}
&&
\Gamma(F \, \rightarrow \, \bar{\zeta}'_{i} \, \zeta'_{i} )= M_{F} \, \frac{|W_{\zeta'_{i}}|^2}{8\pi}
\left[ \left(1- \frac{2M_{\zeta'_{i}}^2}{M_{F}^{2}}\right)
%\, \times
%\right.
%\nonumber \\
%&&
%\left.
%\times \,
\sqrt{1-\frac{4M_{\zeta'_{i}}^2}{M_{F}^{2}}}
-\frac{2 \, M_{\zeta'_{i}}^{2}}{M_{F}^{2}}\left(1- \frac{M_{\zeta'_{i}}^2}{M_{F}^{2}} \right)
\right] \; ,
\hspace{0.5cm}
\end{eqnarray}
where $M_{F} > 2 \, M_{\zeta'_{i}} \, (i=1,2,3)$.

%
%%%%%%%%%%%%%%%%%%%%%%%%%%%%%%%%
%
\subsection{The quark-quark scattering into the dark sector}
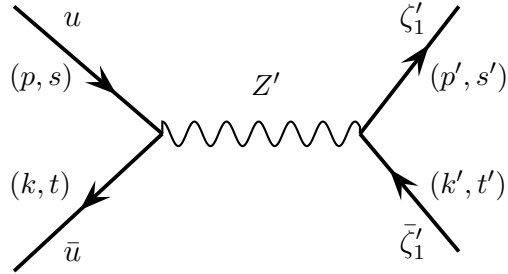
\begin{figure}[!h]%\label{figvm}
\begin{center}
\newpsobject{showgrid}{psgrid}{subgriddiv=1,griddots=10,gridlabels=6pt}
\begin{pspicture}(0,0.5)(2.5,4)
%\showgrid
\psset{arrowsize=0.2 2}
\psset{unit=1.3}
%
%%%%%%%%%%%%%%%%%%%%%%%%%%% Bosons Z' Z W %%%%%%%%%%%%%%%%%%%%%%%%%%%%%%%%%%
%
%\pscoil[coilaspect=0,coilarm=0,coilwidth=0.25,coilheight=1.3,linecolor=black](0,0.3)(2.99,0)
%
\pscoil[coilaspect=0,coilarm=0,coilwidth=0.25,coilheight=1.3,linecolor=black](0,1.7)(2,1.7)
\put(0.9,2.1){$Z'$}
%
%\put(1.5,-0.5){$W/Z$}
%
%\psline[linecolor=black,linewidth=0.3mm]{->}(5,2)(6.5,3.5)
%
%\psline[linecolor=black,linewidth=0.3mm]{->}(0,0.3)(0,1.2)
%\psline[linecolor=black,linewidth=0.3mm]{-}(0,0.9)(0,1.7)
%
%\put(-0.45,1){$q$}
%
%%%%%%%%%%%%%%%%%%%%%%%%%%%%%%%%%%%%%%
%
\psline[linecolor=black,linewidth=0.5mm,ArrowInside=->,ArrowInsidePos=0.5](0,1.7)(-1.5,0.3)
%\psline[linecolor=black,linewidth=0.3mm]{-}(-1,-0.1)(0,0.3)
%
\put(-1,2.8){\large$u$}
\psline[linecolor=black,linewidth=0.5mm,ArrowInside=->,ArrowInsidePos=0.65](-1.5,3)(0,1.7)
%\psline[linecolor=black,linewidth=0.3mm]{-}(-1,2.08)(-2,2.5)
%
\put(-1,0.4){\large$\bar{u}$}
%
%%%%%%%%%%%%%%%%%%%%%%%%%%%%%%%%%%%%
%
\psline[linecolor=black,linewidth=0.5mm,ArrowInside=->,ArrowInsidePos=0.65](2,1.7)(3,3)
%\psline[linecolor=black,linewidth=0.3mm]{<-}(3.4,2.2)(4,2.5)
%
\put(2.4,2.8){\large$\zeta'_{1}$}
\psline[linecolor=black,linewidth=0.5mm,ArrowInside=->,ArrowInsidePos=0.65](3,0.5)(2,1.7)
%\psline[linecolor=black,linewidth=0.3mm](3.5,1.75)(4,1.5)
%
\put(2.4,0.5){\large$\bar{\zeta}'_{1}$}
%
%%%%%%%%%%%%%%%%%%%%%%%%%%%%%%%%%%%%%%%%%%%%%%%%%%%%
%
%\Psline[linecolor=black,linewidth=0.3mm]{->}(3,0)(3.75,0.37)
%\psline[linecolor=black,linewidth=0.3mm](3.4,0.2)(4,0.5)
%
%\put(3.5,0.55){$q$}
%
%\psline[linecolor=black,linewidth=0.3mm](3,0)(3.75,-0.37)
%\psline[linecolor=black,linewidth=0.3mm]{<-}(3.4,-0.2)(4,-0.5)
%
%\put(3.5,-0.8){$\bar{q}$}
%
%\psline[linecolor=black,linewidth=0.3mm]{<-}(0,1)(-1.5,-0.5)
%\psline[linecolor=black,linewidth=0.3mm]{->}(1,2)(-0.4,3.4)
%
%
%Momenta
%
%\psline[linecolor=black,linewidth=0.3mm]{->}(2.2,2.7)(3.8,2.7)
%\put(3,2.6){\Large$X^{\mu}$}
%
%\psline[linecolor=black,linewidth=0.3mm]{->}(7.8,1.2)(7.8,2.8)
%\put(3,1){\Large$k$}
%
\put(-1.55,1.1){$(k,t)$}
\put(-1.55,2.2){$(p,s)$}
\put(2.7,1.1){$(k',t')$}
\put(2.7,2.2){$(p',s')$}
%
%\put(-0.2,3.6){\Large$e^{-}$}
%\put(-0.4,-0.1){\Large$e^{+}$}
%\put(-1.1,3.1){\Large$q$}
%\put(-1.1,0.7){\Large$q^{\prime}$}
%
%\put(0.6,-0.3){$p$}
%
%\put(1.3,0){$\Gamma_{\mu}^{(3)\;abc}(p)=g_{1}f^{abc}p_{\mu}$}
%
%
\end{pspicture}
%
%\vspace{0.2cm}
%
\caption{\scshape{The $Z'$-scattering $\bar{u} \, u \rightarrow Z' \rightarrow \bar{\zeta}'_{1} \, \zeta'_{1}$ that can connect the Standard model with the dark matter content.}} \label{ScatteringZ'DM}
\end{center}
\end{figure}

The cascade effect from (\ref{Z'Decay}) has the portal for dark matter scenario following the possible scattering
$\bar{u} \, u \rightarrow Z' \rightarrow \bar{\zeta}'_{1} \, \zeta'_{1}$, in which $\zeta'_{1}$ has the mass $M_{\zeta'_{1}}=0.5 \, \mbox{TeV}$.
It is illustrated at the tree-level in (\ref{ScatteringZ'DM}).
%
%%%%%%%%%%%%%%%%% The Z'-scatterring into DM %%%%%%%%%%%%%%%%%%%%%%%%%
%

\begin{figure}[!h]%\label{figvm}
\begin{center}
\newpsobject{showgrid}{psgrid}{subgriddiv=1,griddots=10,gridlabels=6pt}
\begin{pspicture}(-3,-1.1)(3,3.8)
%\showgrid
\psset{arrowsize=0.2 2}
\psset{unit=1.3}
%
%%%%%%%%%%%%%%%%%%%%%%%%%%% Bosons Z' Z W %%%%%%%%%%%%%%%%%%%%%%%%%%%%%%%%%%
%
%\pscoil[coilaspect=0,coilarm=0,coilwidth=0.25,coilheight=1.3,linecolor=black](0,0.3)(2.99,0)
%
%\pscoil[coilaspect=0,coilarm=0,coilwidth=0.25,coilheight=1.3,linecolor=black](0,1.7)(2,1.7)
%
\put(-0.5,1){\circle*{0.1}}
%
%\put(1.5,-0.5){$W/Z$}
%
%\psline[linecolor=black,linewidth=0.3mm]{->}(5,2)(6.5,3.5)
%
\psarc[linecolor=black,linewidth=0.2mm]{->}(-0.4,1){0.7}{0}{65}
%\psline[linecolor=black,linewidth=0.3mm]{-}(0,0.9)(0,1.7)
%
\put(0.4,1.4){$\beta$}
%
%%%%%%%%%%%%%%%%%%%%%%%%%%%%%%%%%%%%%%
%
\psline[linecolor=black,linewidth=0.3mm,ArrowInside=->,ArrowInsidePos=0.5](-3,1)(-0.8,1)
%\psline[linecolor=black,linewidth=0.3mm]{-}(-1,-0.1)(0,0.3)
%
\put(1,1.3){$u$}
\psline[linecolor=black,linewidth=0.3mm,ArrowInside=->,ArrowInsidePos=0.5](2,1)(-0.2,1)
%\psline[linecolor=black,linewidth=0.3mm]{-}(-1,2.08)(-2,2.5)
%
\put(-2.1,1.3){$\bar{u}$}
%
%%%%%%%%%%%%%%%%%%%%%%%%%%%%%%%%%%%%
%
\psline[linecolor=black,linewidth=0.3mm,ArrowInside=->,ArrowInsidePos=0.65](-0.35,1.3)(1,3)
%\psline[linecolor=black,linewidth=0.3mm]{<-}(3.4,2.2)(4,2.5)
%
\put(0.8,2.4){$\zeta'_{1}$}
\psline[linecolor=black,linewidth=0.3mm,ArrowInside=->,ArrowInsidePos=0.65](-0.6,0.7)(-2,-1)
%\psline[linecolor=black,linewidth=0.3mm](3.5,1.75)(4,1.5)
%
\put(-2,-0.3){$\bar{\zeta}'_{1}$}
%
%%%%%%%%%%%%%%%%%%%%%%%%%%%%%%%%%%%%%%%%%%%%%%%%%%%%
%
%\Psline[linecolor=black,linewidth=0.3mm]{->}(3,0)(3.75,0.37)
%\psline[linecolor=black,linewidth=0.3mm](3.4,0.2)(4,0.5)
%
%\put(3.5,0.55){$q$}
%
%\psline[linecolor=black,linewidth=0.3mm](3,0)(3.75,-0.37)
%\psline[linecolor=black,linewidth=0.3mm]{<-}(3.4,-0.2)(4,-0.5)
%
%\put(3.5,-0.8){$\bar{q}$}
%
%\psline[linecolor=black,linewidth=0.3mm]{<-}(0,1)(-1.5,-0.5)
%\psline[linecolor=black,linewidth=0.3mm]{->}(1,2)(-0.4,3.4)
%
%
%Momenta
%
%\psline[linecolor=black,linewidth=0.3mm]{->}(2.2,2.7)(3.8,2.7)
%\put(3,2.6){\Large$X^{\mu}$}
%
%\psline[linecolor=black,linewidth=0.3mm]{->}(7.8,1.2)(7.8,2.8)
%\put(3,1){\Large$k$}
%
\put(-2.7,0.5){$(E_{u},{\bf p})$}
\put(0.5,0.5){$(E_{u},-{\bf p})$}
\put(-0.9,2.2){$(E_{\zeta'_{1}},{\bf p}')$}
\put(-1.1,-0.3){$(E_{\zeta'_{1}},-{\bf p}')$}
%
%\put(-0.2,3.6){\Large$e^{-}$}
%\put(-0.4,-0.1){\Large$e^{+}$}
%\put(-1.1,3.1){\Large$q$}
%\put(-1.1,0.7){\Large$q^{\prime}$}
%
%\put(0.6,-0.3){$p$}
%
%\put(1.3,0){$\Gamma_{\mu}^{(3)\;abc}(p)=g_{1}f^{abc}p_{\mu}$}
%
%
\end{pspicture}
%
%\vspace{0.2cm}
%
\caption{\scshape{The collision in the center-of-mass frame for the process $\bar{u} \, u \, \rightarrow \, \bar{\zeta}'_{1} \, \zeta'_{1}$ .}} \label{FigCM}
\end{center}
\end{figure}
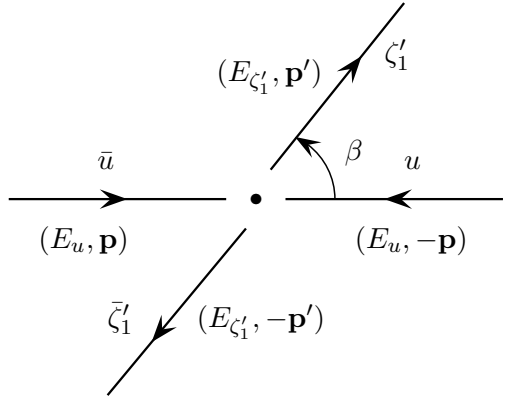

The scattering amplitude of the diagram (\ref{ScatteringZ'DM}) reads as follows:
\begin{eqnarray}\label{AmplitudeMZ}
%&&
i{\cal M}_{Z'}&=&\bar{v}(k,t) \, i \, g_{Z'} \, \gamma^{\mu} \left( g_{V}^{u}-g_{A}^{u} \, \gamma_{5} \right) u(p,s)
%\nonumber \\
%&&
%\times \,
\frac{-i \, \eta_{\mu\nu}}{(k+p)^2-M_{Z'}^{2}} \, \times
\nonumber \\
&&
\times \,
\bar{u}(p',s') \, i \, g_{Z'} \, \gamma^{\nu} \left( g_{V}^{\zeta}-g_{A}^{\zeta} \, \gamma_{5} \!\right) v(k',t') \; .
\end{eqnarray}
We consider the collision in the center-of-mass frame illustrated in the figure (\ref{FigCM}).
%
%%%%%%%%%%%%%%%%% The Z'-scatterring into DM %%%%%%%%%%%%%%%%%%%%%%%%%
%

\begin{figure}[h]
\centering
\includegraphics[scale=0.55]{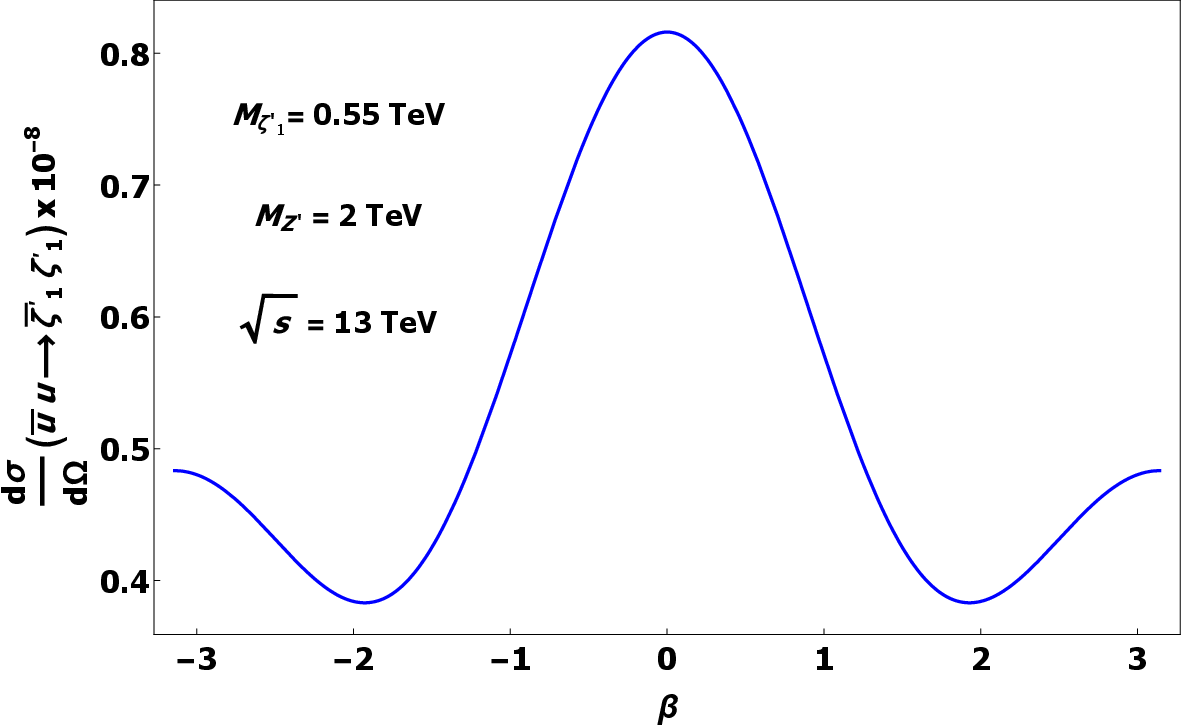}
\caption{The differential cross section of $\bar{u} \, u \rightarrow \bar{\zeta}'_{1} \, \zeta'_{1} $ as function of the $\beta$-angle.
We use here the values for masses $M_{\zeta'_{1}}=0.5 \, \mbox{TeV}$
and $M_{Z'}=2 \, \mbox{TeV}$, the CM-energy as $\sqrt{s}=13 \, \mbox{TeV}$, with a mixing angle of $\alpha=45^{o}$.}
\label{Secdiffbeta}
\end{figure}

\noindent
The squared amplitude (\ref{AmplitudeMZ}) is given by
\begin{eqnarray}\label{AmplitudeMZ2}
\frac{1}{4}\sum_{spins}|{\cal M}_{Z'}|^{2}&=&\frac{g_{Z'}^{4}}{4} \left(1-\frac{4M_{Z'}^{2}}{s} \right)^{-2}
%\, \times
%\nonumber \\
%\times \,
\left[\cos^{4}\alpha \, \left(1+\cos^{2}\beta\right)\left(1-\frac{4M_{\zeta'_{1}}^{2}}{s}\right)+
\right.
\nonumber \\
&&
\left.
+\left(1-\frac{5}{3}\sin^2\alpha\right)^2\left(1+\sqrt{1-\frac{4M_{\zeta'_{1}}^2}{s}} \cos\beta \right) \right] \; ,
\end{eqnarray}
where $s:=(p+k)^2=(p'+k')^2$, ${\bf p}+{\bf k}={\bf p}'+{\bf k}'={\bf 0}$
and $\beta$ is the angle between the $3$-momentum ${\bf p}'$ and ${\bf k}$, or between
${\bf p}$ and ${\bf k}'$, in the CM-collision scheme (\ref{FigCM}). Thus, we get the relation $s=4E_{u}^{2}=4E_{\zeta}^{2}$
by the energy conservation, and the previous amplitude can be written in terms of $s$-variable. We fix the condition $s>4 \, M_{\zeta'_{1}}^{2}$
to insure that (\ref{AmplitudeMZ2}) is real and positive, and we also use that $s \gg 4m_{u}^{2}$. Thus, using the
standard rules of QFT, the differential cross section for the CM-collision is
\begin{eqnarray}\label{diffsigma}
&&
\frac{d\sigma}{d\Omega}(\bar{u} u \rightarrow \bar{\zeta}'_{1} \zeta'_{1}) =\frac{\alpha_{e}^2}{16s}
\frac{1}{\cos^4\theta_{W}\sin^4(2\alpha)} \left(1-\frac{M_{Z'}^{2}}{s}\right)^{\!\!-2} \, \times
\nonumber \\
&&
%\hspace{-0.5cm}
\times \left[\cos^{4}\alpha \, \left(1+\cos^{2}\beta\right)\left(1-\frac{4M_{\zeta'_{1}}^{2}}{s}\right)
+\left(1-\frac{5}{3}\sin^2\alpha\right)^2\left(1+\sqrt{1-\frac{4M_{\zeta'_{1}}^2}{s}} \cos\beta \right) \right] .
\hspace{1cm}
\end{eqnarray}
If we fix $\alpha=45^{o}$, $\sqrt{s}=13 \, \mbox{TeV}$, $M_{Z}=2 \, \mbox{TeV}$
and $M_{\zeta'_{1}}=0.5 \, \mbox{TeV}$, the result (\ref{diffsigma}) appears as a function of $\beta$. This differential cross section in terms
of the $\beta$-angle is plotted in the figure (\ref{Secdiffbeta}).
%
%
%%%%%%%%%%%Figura da differential cross section %%%%%%
%
%

\begin{figure}[h]
\centering
\includegraphics[scale=0.55]{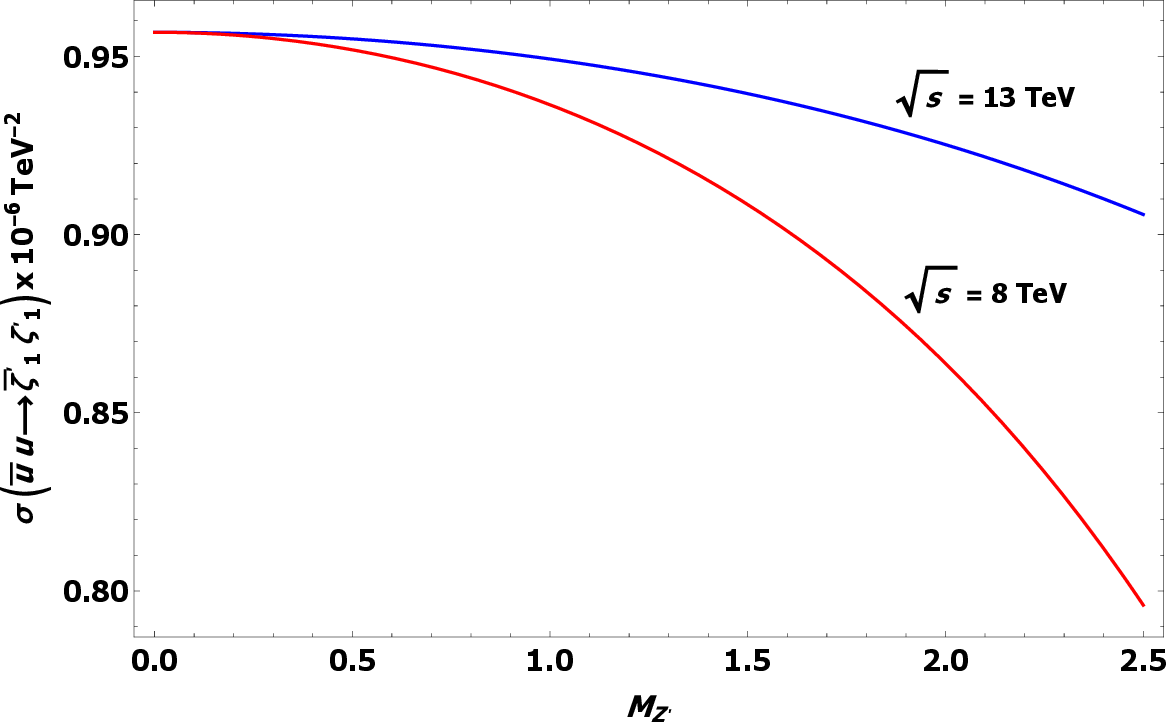}
\caption{The total cross section of $\bar{u} \, u \, \rightarrow \, \bar{\zeta}'_{1} \, \zeta'_{1} $ as function of the $Z'$-mass.
Here, we adopt the CM-energy values by $\sqrt{s}=13 \, \mbox{TeV}$ and $\sqrt{s}=8 \, \mbox{TeV}$. }
\label{SecTotalMZ'}
\end{figure}

\noindent
Returning to the expression (\ref{diffsigma}), the total cross section as a function of the $\alpha$-mixing angle is given by
\begin{eqnarray}
\sigma(\bar{u} \, u \, \rightarrow \, \bar{\zeta}'_{1} \, \zeta'_{1})&=&\frac{\pi \, \alpha_{e}^2}{4s}
\frac{1}{\cos^4\theta_{W}\sin^4(2\alpha)}\left(1-\frac{M_{Z'}^{2}}{s}\right)^{\!\!-2} \times
\nonumber \\
&&
\hspace{-0.5cm}
\times
\left[\frac{4}{3}\cos^{4}\alpha \left(1-\frac{4M_{\zeta'_{i}}^{2}}{s}\right)
+\left(1-\frac{5}{3}\sin^2\alpha\right)^2 \right] \; ,
\end{eqnarray}
in which the $\sigma$-function is illustrated in the figure (\ref{SecTotalMZ'}).
When $\alpha=45^{o}$, the total cross section assumes the value
\begin{eqnarray}
\sigma(\bar{u} \, u \, \rightarrow \, \bar{\zeta}'_{1} \, \zeta'_{1}) \simeq 3.76 \times 10^{-7} \, \, \mbox{TeV}^{-2}
%\simeq 2.8 \times 10^{4} \, \mbox{fb}
\; .
\end{eqnarray}
\section{The MDM for the $\zeta_{i}$-fermions}
\renewcommand{\theequation}{5.\arabic{equation}}
\setcounter{equation}{0}

In this Section, we investigate the magnetic properties of $\zeta_{i}$-fermions through
it Magnetic Dipole Momentum (MDM)
%and Electric Dipole Momentum (EDM).
The mixing of $\zeta_{i}$ with the right-neutrino components motivates us to understand if the $\zeta_{i}$-MDM
depends on its mass, as it happens in the case of neutrinos. For a review on the neutrinos' MDMs, go to the references \cite{BellPRL2005,BellPLB2006,Balantekin2006,Bhattacharya2004,StudenikinJP2016}. We start off with the field equations
in the mixed basis to obtain the so-called Transition Dipole Momenta (TDM). The Dirac equations for the
$\nu_{i}$-neutrinos and $\zeta_{i}$-fermions from (\ref{LpsichiM}) in the momentum space are given by
\begin{eqnarray}\label{eqnuzeta}
\left( \, \slash{\!\!\!p} - m_{\nu_{i}} \, \right) u_{\nu_{i}}(p)-M_{ij}^{RL} \, u_{\zeta_{j}}(p) &=& 0
\nonumber \\
\left( \, \slash{\!\!\!p} - m_{\zeta_{i}} \, \right) u_{\zeta_{i}}(p)-M_{ij}^{LR} \, u_{\nu_{j}}(p) &=& 0
\; ,
%\hspace{0.7cm}
\end{eqnarray}
where $m_{\nu_{i}}$ and $m_{\zeta_{i}}$ are the masses if we make the mixed coupling constants $Y_{ij}=Z_{ij} \, \rightarrow \, 0$.
The functions $u_{\nu_{i}}(p)$ and $u_{\zeta_{i}}(p)$ are the wave plane amplitudes of $\nu_{i}$ and $\zeta_{i}$ in the mixed basis,
respectively. For simplicity, we have defined the matrices $M_{RL}$ and $M_{LR}$ as combination of Left- and Right-components :
\begin{eqnarray}
M_{ij}^{RL}&:=&\frac{Y_{ij} \, v \, R + Z_{ij} \, u \, L}{\sqrt{2}}  \; ,
\nonumber \\
M_{ij}^{LR}&:=&\frac{Y_{ij} \, v \, L + Z_{ij} \, u \, R}{\sqrt{2}}  \; .
\end{eqnarray}

The hermitian conjugate of (\ref{eqnuzeta}) is written below:
\begin{eqnarray}\label{eqnuzetaconj}
\bar{u}_{\nu_{i}}(p)\left( \, \slash{\!\!\!p} - m_{\nu_{i}} \, \right) -\bar{u}_{\zeta_{j}}(p) \, M_{ij}^{LR \, \dagger} &=& 0
\nonumber \\
\bar{u}_{\zeta_{i}}(p)\left( \, \slash{\!\!\!p} - m_{\zeta_{i}} \, \right)- \bar{u}_{\nu_{j}}(p) \, M_{ij}^{RL \, \dagger} &=& 0
\; .
%\hspace{0.7cm}
\end{eqnarray}
If we substitute $p^{\mu}$ by $p^{\prime\mu}$ in (\ref{eqnuzetaconj}), we can combine the hermitian conjugate equations
with the equations (\ref{eqnuzeta}) to obtain the following tree-level Gordon decompositions:
\begin{eqnarray}\label{IDGordonnu}
&&
\bar{u}_{\nu_{i}}(p') \, \gamma^{\mu} \, u_{\nu_{i}}(p) = \bar{u}_{\nu_{i}}(p') \!\left( \frac{\ell^{\mu}}{2m_{\nu_{i}}}+i\, \frac{\sigma^{\mu\nu}q_{\nu}}{2m_{\nu_{i}}}\right)\!u_{\nu_{i}}(p)
\nonumber \\
&&
%-\frac{|t_{ij}|u}{4\sqrt{2} \, m_{\nu}m_{\zeta}}
- \bar{u}_{\nu_{i}}(p') \! \left( \hspace{-0.25cm} \phantom{\frac{1}{2}}
\, \mu_{ij} \, \frac{\ell^{\mu}}{2m_{\nu_{i}}}+ \mu_{ij} \, i \, \frac{\sigma^{\mu\nu}q_{\nu}}{2m_{\nu_{i}}}
%\right.
%\nonumber \\
%&&
%\left.
%+
+ \eta_{ij} \, \frac{q^{\mu}\gamma_{5}}{2m_{\nu_{i}}}
+ \eta_{ij} \, i \, \frac{\sigma^{\mu\nu}\ell_{\nu}}{2m_{\nu_{i}}} \, \gamma^{5} \hspace{-0.25cm} \phantom{\frac{1}{2}} \, \right) u_{\zeta_{j}}(p)
+\mbox{h. c.}
%\nonumber \\
%&&
%-\frac{|t_{ij}|u}{4\sqrt{2} \, m_{\nu}m_{\zeta}}
%-\mu_{ij}^{(\nu)} \, \bar{u}_{\zeta_{i}}(p')\!
%\left( \hspace{-0.25cm} \phantom{\frac{1}{2}} \, \ell^{\mu}
%+i\,\sigma^{\mu\nu}q_{\nu}
%+
%\right.
%\nonumber \\
%&&
%\left.
%+q^{\mu}\gamma_{5}+i\,\sigma^{\mu\nu}\ell_{\nu} \, \gamma^{5} \hspace{-0.25cm} \phantom{\frac{1}{2}} \, \right) u_{\nu_{j}}(p)
%\nonumber \\
%&&
%+\frac{|z_{ik}||t_{kj}|uv}{2m_{\nu}m_{\zeta}} \! \left[ \hspace{-0.25cm} \phantom{\frac{1}{2}} \, \bar{u}_{\nu_{i}}(p') \, \gamma^{\mu}  u_{\nu_{j}}(p)
%+ \, \bar{u}_{\zeta_{i}}(p')  \gamma^{\mu} \, u_{\zeta_{j}}(p) \hspace{-0.25cm} \phantom{\frac{1}{2}} \right]
\, ,
%\hspace{-0.4cm}
%\nonumber \\
\end{eqnarray}
and
\begin{eqnarray}\label{IDGordonzeta}
&&
\bar{u}_{\zeta_{i}}(p') \, \gamma^{\mu} \, u_{\zeta_{i}}(p) = \bar{u}_{\zeta_{i}}(p')\left( \frac{\ell^{\mu}}{2m_{\zeta_{i}}}+i\, \frac{\sigma^{\mu\nu}q_{\nu}}{2m_{\zeta_{i}}}\right) u_{\zeta_{i}}(p)
\nonumber \\
&&
%-\frac{|t_{ij}|u}{4\sqrt{2} \, m_{\zeta}^{2}}
- \bar{u}_{\zeta_{i}}(p') \!
\left( \hspace{-0.25cm} \phantom{\frac{1}{2}} \, \mu_{ij} \, \frac{\ell^{\mu}}{2m_{\zeta_{i}}}
+ \mu_{ij} \, i \, \frac{\sigma^{\mu\nu}q_{\nu}}{2m_{\zeta_{i}}}
%\right.
%\nonumber \\
%&&
%\left.
+\eta_{ij} \, \frac{q^{\mu}\gamma_{5}}{2m_{\zeta_{i}}}
+ \eta_{ij} \, i \, \frac{\sigma^{\mu\nu}\ell_{\nu}}{2m_{\zeta_{i}}} \, \gamma_{5} \phantom{\frac{1}{2}} \hspace{-0.25cm} \right) u_{\nu_{j}}(p)
+\mbox{h. c.} \; ,
%\nonumber \\
%&&
%-\mu_{ij}^{(\zeta)} \, \bar{u}_{\nu_{i}}(p') \!
%\left( \hspace{-0.25cm} \phantom{\frac{1}{2}} \, \ell^{\mu}+i \, \sigma^{\mu\nu}q_{\nu}
%\right.
%\nonumber \\
%&&
%\left.
%+q^{\mu}\gamma_{5}
%+ i \, \sigma^{\mu\nu}\ell_{\nu} \gamma_{5} \phantom{\frac{1}{2}} \hspace{-0.25cm} \right) u_{\zeta_{j}}(p) \, ,
%\nonumber \\
%&&
%+\frac{|z_{ik}||t_{kj}|uv}{2\,m_{\zeta}^{2}} \! \left[ \hspace{-0.25cm} \phantom{\frac{1}{2}} \bar{u}_{\nu_{i}}(p') \, \gamma^{\mu}  u_{\nu_{j}}(p)
%+ \, \bar{u}_{\zeta_{i}}(p')  \gamma^{\mu} \, u_{\zeta_{j}}(p) \right]
%\hspace{-0.4cm}
%\nonumber \\
\end{eqnarray}
where $q^{\mu}=p^{\mu}-p^{\prime \mu}$ is the photon's transfer momentum, and $\ell^{\mu}:=p^{\mu}+p^{\prime\mu}$
is the total $4$-momentum. These expressions yield the currents of neutrinos with the $\zeta_{i}$-fermions of the model, written in
momentum space, that we refer to as transition terms.
The coefficients $\mu_{ij}$ and $\eta_{ij}$ are matrix elements that depend on the Yukawa complex constant coupling and the
$u$-VEV scale :
\begin{eqnarray}
\mu_{ij}&=& \frac{1}{2\sqrt{2}} \, \frac{Z_{ij} \, u+Y_{ij} \, v}{m_{\zeta_{i}}+m_{\nu_{i}}} \; ,
\nonumber \\
\eta_{ij}&=&\frac{1}{2\sqrt{2}} \, \frac{Z_{ij} \, u-Y_{ij} \, v}{m_{\zeta_{i}}+m_{\nu_{i}}} \; .
\end{eqnarray}

Here, if we use that $m_{\zeta_{i}} \gg m_{\nu_{i}}$,
and $u \gg v$, so we can approximate $Y_{ij} \, v+Z_{ij} \, u \approx Z_{ij} \, u$, and the coefficients $\mu_{ij}$ and $\eta_{ij}$
are approximately equals, $\mu_{ij}\simeq\eta_{ij}$.
We also neglect terms of order $Z_{ij}^{2} \approx 0$ in relation to the linear terms of $Z_{ij}$ in (\ref{IDGordonnu}) and (\ref{IDGordonzeta}).
%, we can neglect the effect of two
%last terms in (\ref{IDGordonnu}) and (\ref{IDGordonzeta}).
We observe the emergence of TMDM for neutrinos and $\zeta_{i}$-fermions in both expressions (\ref{IDGordonnu}) and (\ref{IDGordonzeta}).
If we multiply these currents by $e \, A_{\mu}$, and using the representation $q_{\nu} \rightarrow i \, \partial_{\nu}$
for the photon momentum, the terms $\sigma^{\mu\nu}q_{\nu}$ have the following TMDM for the $\nu_{i}$-neutrino :
\begin{eqnarray}\label{muneutrino}
\mu_{ij}^{\, (\nu_{i})} &=& \frac{e \, \mu_{ij}}{2 \, m_{\nu_{i}}}=\frac{e \, Z_{ij} \, u}{4\sqrt{2} \, m_{\nu_{i}} \, m_{\zeta_{i}}}
%\simeq 4.8 \times 10^{-7} \, \mbox{eV}^{-1}
\simeq \frac{Z_{ij}}{4\sqrt{2}} \, \left(\frac{2.8 \, \mbox{TeV}}{m_{\zeta_{i}}}\right) \,  \left( \frac{1 \, \mbox{MeV}}{m_{\nu_{i}}} \right) \, \mu_{B} \; .
\end{eqnarray}
We have a result at tree-level for Dirac neutrinos that depends on the $\nu_{i}$-neutrino mass and the $\zeta_{i}$-mass. This result allows
us to fix an small estimate  for the $Z_{ij}$ coupling constant using the known result $|\mu^{(\nu)}| \, \lesssim \, 8 \, \times \, 10^{-15} \, \mu_{B}$ in the literature \cite{BellPRL2005}. We consider the mass spectrum of $m_{\nu_{i}} \, \sim \, 1 \, \mbox{eV}$ for neutrinos, the mass of $m_{\zeta_{1}}=0.5 \, \mbox{TeV}$ for $\zeta_{1}$-fermion, and $u=2.8 \, \mbox{TeV}$. In so doing, the expression (\ref{muneutrino}) yields the $Z_{ij}$ coupling constant below :
\begin{eqnarray}
|Z_{ij}| \, \lesssim \, 7.5 \, \times \, 10^{-21} \; .
\end{eqnarray}
The TMDM for $\zeta_{i}$-hidden fermion is given by expression
\begin{eqnarray}
\mu_{ij}^{\, (\zeta_{i})}=\frac{e \, \mu_{ij}}{2 \, m_{\zeta_{i}}}
= \frac{Z_{ij}}{4\sqrt{2}} \, \left(\frac{2.8 \, \mbox{TeV}}{m_{\zeta_{i}}}\right) \,  \left( \frac{1 \, \mbox{MeV}}{m_{\zeta_{i}}} \right) \, \mu_{B}  \; .
\end{eqnarray}
Therefore, we can estimate the TMDM for the $\zeta_{1}$-hidden fermion :
\begin{eqnarray}
|\mu_{ij}^{\, (\zeta_{1})}| \, \lesssim \, 1.6 \, \times \, 10^{-26} \, \mu_{B} \; ,
\end{eqnarray}
where $\mu_{B}=3 \times 10^{-7} \, \mbox{eV}^{-1}$ is the Bohr magneton, in natural units $c=\hbar=1$.
%For this estimative, we use $u=2.8 \, \mbox{TeV}$, $Z_{ij} \simeq 8.3 \, \times \, 10^{-8}$ and
%$m_{\zeta_{1}}=0.5 \, \mbox{TeV}$.

%%%%%%%%%%%%%%%%%%%

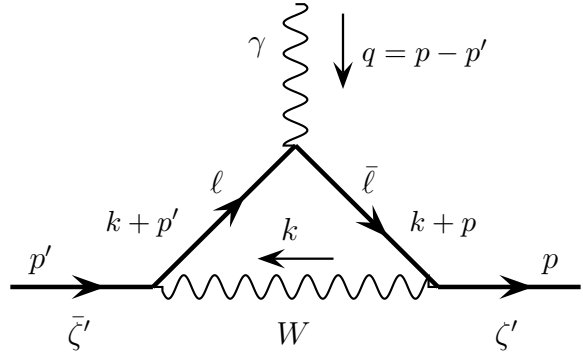
\begin{figure}[!h]%\label{figvm}
\begin{center}
\newpsobject{showgrid}{psgrid}{subgriddiv=1,griddots=10,gridlabels=6pt}
\begin{pspicture}(0,0.5)(8.8,5)
%\showgrid
\psset{arrowsize=0.2 2}
\psset{unit=1.25}
%
%\put(-4.5,0){$g_{1}f^{abc}\partial_{\mu}\bar{\eta}^{a}G^{\mu\;c}\eta^{b}\;:$}
%
%%%%%%%%%%%%%%%%%%%%%%%%%%%%%%%%%%%%%%%%%%%%%%%%%%%%%%%%%%%%%
%
%\pscoil[coilarm=0,coilaspect=0,coilwidth=0.5,coilheight=1.0,linecolor=black](1,2)(5,2)
%
\psline[linecolor=black,linewidth=0.6mm,ArrowInside=->,ArrowInsidePos=0.6](0.5,1)(2,1)
%
%\psline[linecolor=black,linewidth=0.6mm](1.2,1.5)(2.4,1.5)
%
%\psline[linestyle=dashed,linecolor=black,linewidth=0.6mm](2.2,1.5)(4.7,1.5)
%
\psline[linecolor=black,linewidth=0.6mm,ArrowInside=->,ArrowInsidePos=0.6](5,1)(6.5,1)
%
%\psline[linecolor=black,linewidth=0.6mm](5,1.5)(6,1.5)
%
%%%%%%%%%%%%%%%%%%%%%%%%%%%%%%%%%%%%%%%%%%%%%%%%%%%%%%
%
%\psline[linecolor=black,linewidth=0.6mm]{->}(2.4,1.5)(2.4,2.6)
%
%\psline[linecolor=black,linewidth=0.6mm](2.4,2)(2.4,3)
%
%\psline[linecolor=black,linewidth=0.6mm](4.5,1.5)(4.5,2.2)
%
%\psline[linecolor=black,linewidth=0.6mm]{<-}(4.5,2.1)(4.5,3)
%
\pscoil[coilaspect=0,coilarm=0.1,coilwidth=0.25,coilheight=1.3,linecolor=black](2,1)(5,1)
\psline[linecolor=black,linewidth=0.6mm,ArrowInside=->,ArrowInsidePos=0.6](2,1)(3.5,2.5)
\psline[linecolor=black,linewidth=0.6mm,ArrowInside=->,ArrowInsidePos=0.6](3.5,2.5)(5,1)
%
%\psarc[linecolor=black,linewidth=0.6mm](3.45,1.5){1.1}{0}{45}
%
%\psarc[linecolor=black,linewidth=0.6mm](3.45,1.5){1.1}{42}{135}
%
%\psarc[linecolor=black,linewidth=0.6mm](3.45,1.5){1.1}{132}{180}
%
\pscoil[coilaspect=0,coilarm=0,coilwidth=0.25,coilheight=1.3,linecolor=black](3.5,2.5)(3.5,4)
%
%%%%%%%%%%%%%%%%%%%%%%%%%%%%%%%%%%%%%%%%%%%%%%%%%%%%
%
%\put(-1,0){\Large$+\;\;\; {\cal O}\,(e^{4})$}
%
%Momenta
%
%\psline[linecolor=black,linewidth=0.3mm]{->}(2.2,2.7)(3.8,2.7)
%\put(1,3.1){\large$q=p^{\prime}-p$}
%
%\put(2.8,2.7){\large$\gamma$}
%
%\psline[linecolor=black,linewidth=0.3mm]{->}(7.8,1.2)(7.8,2.8)
%
%\put(3.3,2.4){\large$W$}
%
\put(3.3,0.4){\large$W$}
\put(1.1,0.4){\large$\bar{\zeta}'$}
\put(5.6,0.4){\large$\zeta'$}
\put(2.6,2){\large$\ell$}
\put(4.2,2){\large$\bar{\ell}$}
\put(3,3.5){\large$\gamma$}
\put(4.2,3.4){$q=p-p^{\prime}$}
\psline[linecolor=black,linewidth=0.3mm]{->}(4,3.9)(4,3.1)
\put(0.7,1.2){\large$p^{\prime}$}
\put(4.7,1.6){$k+p$}
\put(1.5,1.6){$k+p^{\prime}$}
\put(6.1,1.2){\large$p$}
\psline[linecolor=black,linewidth=0.3mm]{<-}(3.1,1.3)(3.9,1.3)
\put(3.35,1.5){\large$k$}
\end{pspicture}
%
%\vspace{0.2cm}
%
\caption{\scshape{The first contribution at one-loop for the MDM of $\zeta'_{i}$-fermion combining the
$W^{\pm}-\zeta'-\ell'$ vertex with the external photon.}}\label{Vertex1}
\end{center}
\end{figure}

The important contribution for the MDM
%and EDM
of $\zeta_{i}$ combines two external lines of $\zeta'_{i}$ with one external line of photon
in a one-loop diagram. This possible loop diagram emerges when we work in the mass basis $\zeta'_{i}$ due to the vertex
$W^{\pm}-\zeta'-\ell'$. This vertex depends on the $\theta$-mixing angle, and we belief that it must
be a small effect. Obviously, the vertex goes to zero, when $\theta \, \rightarrow \, 0$.
%It yields the dipole magnetic momentum of $\zeta'$-fermion.
The vertex at one loop is illustrated in the figure (\ref{Vertex1}).
%
%%%%%%%%%%%%%%%%%%%%%%%%%%%%%%%%%%%%%%%%%%%%%%
%
%
%%%%%%%%%%%%%%%%%%%%%%%%%%%%%%%%%%%%%%%%%%%%%%%%%%%%%%%%%%%%%%%%%%%%%%%%%%%%%%%%%%%%%%%%%%
%
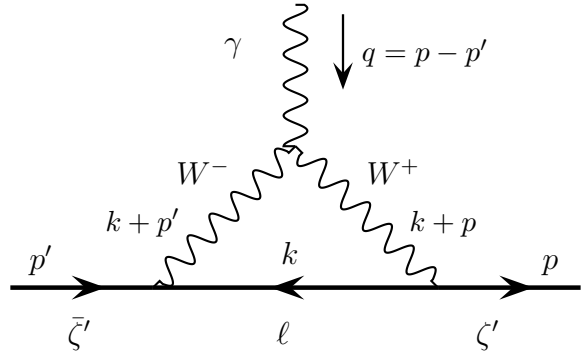
\begin{figure}[!h]%\label{figvm}
\begin{center}
\newpsobject{showgrid}{psgrid}{subgriddiv=1,griddots=10,gridlabels=6pt}
\begin{pspicture}(0,0.5)(8.8,5)
%\showgrid
\psset{arrowsize=0.2 2}
\psset{unit=1.25}
%
%\put(-4.5,0){$g_{1}f^{abc}\partial_{\mu}\bar{\eta}^{a}G^{\mu\;c}\eta^{b}\;:$}
%
%%%%%%%%%%%%%%%%%%%%%%%%%%%%%%%%%%%%%%%%%%%%%%%%%%%%%%%%%%%%%
%
%\pscoil[coilarm=0,coilaspect=0,coilwidth=0.5,coilheight=1.0,linecolor=black](1,2)(5,2)
%
\psline[linecolor=black,linewidth=0.6mm,ArrowInside=->,ArrowInsidePos=0.6](0.5,1)(2,1)
%
%\psline[linecolor=black,linewidth=0.6mm](1.2,1.5)(2.4,1.5)
%
%\psline[linestyle=dashed,linecolor=black,linewidth=0.6mm](2.2,1.5)(4.7,1.5)
%
\psline[linecolor=black,linewidth=0.6mm,ArrowInside=->,ArrowInsidePos=0.6](5,1)(6.5,1)
%
%\psline[linecolor=black,linewidth=0.6mm](5,1.5)(6,1.5)
%
%%%%%%%%%%%%%%%%%%%%%%%%%%%%%%%%%%%%%%%%%%%%%%%%%%%%%%
%
%\psline[linecolor=black,linewidth=0.6mm]{->}(2.4,1.5)(2.4,2.6)
%
%\psline[linecolor=black,linewidth=0.6mm](2.4,2)(2.4,3)
%
%\psline[linecolor=black,linewidth=0.6mm](4.5,1.5)(4.5,2.2)
%
%\psline[linecolor=black,linewidth=0.6mm]{<-}(4.5,2.1)(4.5,3)
%
%\pscoil[coilaspect=0,coilarm=0.1,coilwidth=0.25,coilheight=1.3,linecolor=black](2,1)(5,1)
%
\pscoil[coilaspect=0,coilarm=0.1,coilwidth=0.25,coilheight=1.3,linecolor=black](2,1)(3.5,2.5)
\pscoil[coilaspect=0,coilarm=0.1,coilwidth=0.25,coilheight=1.3,linecolor=black](3.5,2.5)(5,1)
%
%\psline[linecolor=black,linewidth=0.6mm,ArrowInside=->,ArrowInsidePos=0.6](2,1)(3.5,2.5)
%
%\psline[linecolor=black,linewidth=0.6mm,ArrowInside=->,ArrowInsidePos=0.6](3.5,2.5)(5,1)
%
%\psarc[linecolor=black,linewidth=0.6mm](3.45,1.5){1.1}{0}{45}
%
%\psarc[linecolor=black,linewidth=0.6mm](3.45,1.5){1.1}{42}{135}
%
%\psarc[linecolor=black,linewidth=0.6mm](3.45,1.5){1.1}{132}{180}
%
\psline[linecolor=black,linewidth=0.6mm,ArrowInside=->,ArrowInsidePos=0.55](5,1)(2,1)
\pscoil[coilaspect=0,coilarm=0,coilwidth=0.25,coilheight=1.3,linecolor=black](3.5,2.5)(3.5,4)
%
%%%%%%%%%%%%%%%%%%%%%%%%%%%%%%%%%%%%%%%%%%%%%%%%%%%%
%
%\put(-1,0){\Large$+\;\;\; {\cal O}\,(e^{4})$}
%
%Momenta
%
%\psline[linecolor=black,linewidth=0.3mm]{->}(2.2,2.7)(3.8,2.7)
%\put(1,3.1){\large$q=p^{\prime}-p$}
%
%\put(2.8,2.7){\large$\gamma$}
%
%\psline[linecolor=black,linewidth=0.3mm]{->}(7.8,1.2)(7.8,2.8)
%
%\put(3.3,2.4){\large$W$}
%
\put(3.3,0.4){\large$\ell$}
\put(1.1,0.4){\large$\bar{\zeta}'$}
\put(5.4,0.4){\large$\zeta'$}
\put(2.25,2.1){\large$W^{-}$}
\put(4.25,2.1){\large$W^{+}$}
\put(2.75,3.5){\large$\gamma$}
\put(4.2,3.4){$q=p-p^{\prime}$}
\psline[linecolor=black,linewidth=0.3mm]{->}(4,3.9)(4,3.1)
\put(0.7,1.2){\large$p^{\prime}$}
\put(4.7,1.6){$k+p$}
\put(1.5,1.6){$k+p^{\prime}$}
\put(6.1,1.2){\large$p$}
%
%\psline[linecolor=black,linewidth=0.3mm]{->}(3.1,1.3)(3.9,1.3)
%
\put(3.35,1.25){\large$k$}
%
%\put(2.4,3.8){\large$\bar{\ell}$}
%
%\put(4.3,3.8){\large$\ell$}
%
%
\end{pspicture}
%
%\vspace{0.2cm}
%
\caption{\scshape{The second contribution at one-loop for the MDM of the$\zeta'_{i}$-fermion with the $W^{+}W^{-}$-photon vertex of the GSW model.}}\label{Vertex2}
\end{center}
\end{figure}

\noindent
Following the previous rules, the one-loop vertex (\ref{Vertex1}) is represented by the integral
\begin{eqnarray}\label{IntLambda}
\Lambda_{(1)ij}^{\mu}(p,p') &=& -\frac{g^2\theta^{2}}{8} \, V_{ik}V_{kj} \int \frac{d^4k}{(2\pi)^4} \, \gamma^{\alpha}\left(1- \gamma_{5}\right)
%\nonumber \\
%&&
%\times \,
\frac{i\left(\slash{\!\!\!k}+\slash{\!\!\!p}'+m\right)}{(k+p')^{2}-m^{2}} \, \times
\nonumber \\
&&
\times \, \left(ie\gamma^{\mu}\right) \, \frac{i\left(\slash{\!\!\!k}+\slash{\!\!\!p}+m\right)}{(k+p)^{2}-m^{2}}
%\nonumber \\
%&&
%\times
\, \gamma_{\alpha}\left(1- \gamma_{5}\right) \frac{-i}{k^{2}-m_{W}^{2}} \; .
\end{eqnarray}
The second contribution comes from the combination of $W^{\pm}-\zeta'-\ell'$ interaction with the vertex $W^{\pm}$-photon
of the GSW model. It is illustrated in the figure (\ref{Vertex2}). The Feynman rules yield the momentum space loop integral below
\begin{eqnarray}\label{IntGamma}
\Lambda_{(2)ij}^{\mu}(p,p')= - \frac{g^2\theta^{2}}{8} \, V_{ik}V_{kj} \int \frac{d^4k}{(2\pi)^4} \, \gamma_{\alpha} \left(1-\gamma_{5}\right)
%\nonumber \\
%&&
%\hspace{-0.5cm}
%\times
\,
\frac{-i}{\left(k+p'\right)^{2}-m_{W}^{2}}
\, \times
\nonumber \\
\times \, V^{\mu\alpha\beta}\left(q,k+p',-k-p\right)
\frac{-i}{\left(k+p\right)^{2}-m_{W}^{2}} \, \frac{i\left(\slash{\!\!\!k}+m\right)}{k^{2}-m^{2}} \, \gamma_{\beta}\left(1-\gamma_{5}\right) \; .
\end{eqnarray}
We have used the $W^{\pm}$-propagator in the Feynman gauge in the expressions (\ref{IntLambda}) and $(\ref{IntGamma})$, $V^{\mu\alpha\beta}$
sets the $W^{\pm}-\gamma$ vertex following the GSW model rules.
%
%Since the vertex $\gamma-W^{\pm}$ does not depend on $\gamma$'s Dirac matrices,
%the trivial identity $\left(1-\gamma_{5}\right)\gamma_{\beta}\left(1-\gamma_{5}\right)=0$ insurances that $\Gamma^{\mu}(p,p')=0$.
Well-known techniques to deal with Feynman integrals are introduced to calculate the finite part of these integrals and, then, the
contributions for the MDM of $\zeta_{i}$-fermions. The sum of these two contributions is denoted by $\Gamma^{\mu}=\Lambda_{(1)}^{\mu}+\Lambda_{(2)}^{\mu}$, so the finite part of $\Gamma^{\mu}$ at one-loop is written into the form
\begin{equation}
J_{\zeta'_{i}(em)}^{\; \mu}=\bar{u}_{\zeta'_{i}} \, \Gamma_{ij}^{\, \mu} \, u_{\zeta'_{j}}=\bar{u}_{\zeta'_{i}}(p') \left[ \phantom{\frac{1}{2}} \hspace{-0.25cm} f_{1}(q^2) \, \gamma^{\mu}+f_{2}(q^2) \gamma_{5} \, \gamma^{\mu}
+f_{3}(q^2) \, i \, \sigma^{\mu\nu}q_{\nu}+f_{4}(q^2) \, q^{\mu} \gamma_{5} \, \right]_{ij} \!\! u_{\zeta'_{j}}(p) \; ,
\end{equation}
where $f_{1}$, $f_{2}$, $f_{3}$ and $f_{4}$ are the form factors of the previous diagrams in this order. The current conservation implies that
$q_{\mu}J_{\zeta'_{i}(em)}^{\mu}=0$, then under this condition we obtain the relation $f_{2}(q^2)=-q^2 \, f_{4}(q^2)/2M_{\zeta'_{i}}$.
Thereby, the EM-current of $\zeta'_{i}$ is reduced to expression
\begin{equation}\label{currentEMzeta}
J_{\zeta'_{i}(em)}^{\; \mu}=\bar{u}_{\zeta'_{i}} \, \Gamma_{ij}^{\, \mu} \, u_{\zeta'_{j}}=\bar{u}_{\zeta'_{i}}(p') \left[ \phantom{\frac{1}{2}} \hspace{-0.25cm} f_{1}(q^2) \, \gamma^{\mu}+f_{A}(q^2) \gamma_{5} \left(q^{\mu} \, \slash{\!\!\!q}-q^2 \, \gamma^{\mu} \right)
+f_{3}(q^2) \, i \, \sigma^{\mu\nu}q_{\nu} \, \right]_{ij} \!\! u_{\zeta'_{j}}(p) \; .
\end{equation}
%

%
%\begin{eqnarray}
%\bar{u}_{\zeta'_{i}}(p') \, \Gamma_{ij}^{\, \mu}(q) \, u_{\zeta'_{j}}(p)&=&\bar{u}_{\zeta'_{i}}(p') \left[ \phantom{\frac{1}{2}} \hspace{-0.25cm} %\gamma^{\mu}f_{Q}(q^2)_{ij}
%+
%\right.
%\nonumber \\
%&&
%\left.
%+f_{M}(q^2)_{ij} \, i \, \sigma^{\mu\nu}q_{\nu}
%\right.
%\nonumber \\
%&&
%\left.
%+f_{E}(q^2)_{ij} \, i \, \sigma^{\mu\nu}\ell_{\nu}\gamma_{5}
%\right.
%\nonumber \\
%&&
%\left.
%+f_{A}(q^2)_{ij}\left(q^{2}\gamma^{\mu}-q^{\mu} \slash{\!\!\!q} \right) \hspace{-0.3cm} \phantom{\frac{1}{2}} \right] u_{\zeta'_{j}}(p) \; ,
%\end{eqnarray}
%
%where $f_{Q}$, $f_{M}$, $f_{E}$ and $f_{A}$ are charge, dipole magnetic, dipole electric and anapole form factors for the
%$\zeta$-fermion one-loop vertex with the EM-photon , respectively.
We have also used here the mass on-shell conditions for $\zeta'_{i}$-fermions : $\slash{\!\!\!p} \, u_{\zeta'_{i}}(p)=M_{\zeta'_{i}} \, u_{\zeta'_{i}}(p)$, $\bar{u}_{\zeta'_{i}}(p') \, \slash{\!\!\!p}'=\bar{u}_{\zeta'_{i}}(p') \, M_{\zeta'_{i}}$ and $p^{2}=p^{\prime2}=M_{\zeta'_{i}}^{2}$. The amplitudes $\bar{u}_{\zeta'_{i}}(p')$ and $u_{\zeta'_{i}}(p)$ stand for the plane wave solutions in the diagonal basis of $\zeta'_{i}$-fermions. The $f_{1}$-form factor is the contribution to electric charge given by
\begin{eqnarray}\label{fQ}
f_{1}(q^2)_{ij}=\frac{eg^{2}\theta^{2}}{32\pi^{2}} \, V_{ik}V_{kj} \int_{0}^{1} dx \, dy \, dz \, \delta(x+y+z-1) \, \times
\nonumber \\
\times \, \left[ \, \frac{M_{\zeta_{i}^{\prime}}^{2} \, z(4-z)-m^{2}}{M_{\zeta'_{i}}^{2} \, z(1-z)-m_{W}^{2}z-m^{2} \, z(1-z)+q^{2} xy}
\right.
\nonumber \\
\left.
-\frac{M_{\zeta_{i}^{\prime}}^{2} \, (1-z)(2+3z)}{M_{\zeta'_{i}}^{2} \, z(1-z)-m_{W}^{2}(1-z)-m^{2}z+q^{2} xy} \, \right] \, .
\end{eqnarray}
The second term in (\ref{currentEMzeta}) is known as the anapole (or toroidal momentum) term with the $f_{A}$-form factor that follows: 
\begin{eqnarray}
f_{A}(q^2)_{ij}=\frac{f_{4}(q^2)_{ij}}{2M_{\zeta'_{i}}}= \frac{eg^{2}\theta^{2}}{32\pi^{2}} \, V_{ik}V_{kj} \int_{0}^{1} dx \, dy \, dz \, \delta(x+y+z-1) \, \times
\nonumber \\
\times \, \left[ \, \frac{2+2z-2(x-y)^2}{M_{\zeta'_{i}}^{2} \, z(1-z)-m_{W}^{2}z-m^{2} \, z(1-z)+q^{2} xy}
\right.
\nonumber \\
\left.
-\frac{3(1-z)-2(x-y)^{2}}{M_{\zeta'_{i}}^{2} \, z(1-z)-m_{W}^{2}(1-z)-m^{2}z+q^{2} xy} \, \right] \, .
\end{eqnarray}
The $f_{3}$-form factor is the contribution to the $\zeta'_{i}$-fermion MDM :
%given by integrals in terms of Feynman's parameters
%
\begin{eqnarray}\label{fM}
f_{3}(q^2)_{ij}= -\frac{e g^{2} \theta^{2}}{ 32 \pi^{2}} \, M_{\zeta'_{i}} \, V_{ik}V_{kj} \int_{0}^{1} dx \, dy \, dz \, \delta(x+y+z-1) \times
\nonumber \\
\times
\left[ \, \frac{z(1-z)}{M_{\zeta'_{i}}^{2} \, z(1-z)-m_{W}^{2}z-m^{2}z(1-z)+q^{2} xy}
\right.
\nonumber \\
\left.
-\frac{(1-z)(1/2-z)}{M_{\zeta'_{i}}^{2} \, z(1-z)-m_{W}^{2}(1-z)-m^{2}z+q^{2} xy} \, \right]  \, .
\hspace{0.5cm}
\end{eqnarray}
%
%The contribution for the $\zeta$-fermion MDE is
%
%\begin{eqnarray}\label{fE}
%f_{E}(q^2)_{ij}= \frac{e g^{2} \theta^{2}}{ 64 \pi^{2}} \, V_{ik}V_{kj} \int_{0}^{1} dx \, dy \, dz \, \delta(x+y+z-1) \times
%\nonumber \\
%\times
%\left[ \, \frac{M_{\zeta'}z(2+z)-m^{2}/M_{\zeta'}-q^2/M_{\zeta'}(1-x)(1-y)}{M_{\zeta'}^{2}z(1-z)-m_{W}^{2}z-m^{2}z(1-z)+q^{2} xy}
%\right.
%\nonumber \\
%\left.
%-\frac{M_{\zeta'}(1-z)(3/2-z)-q^2/M_{\zeta'}(xy-x-y)}{M_{\zeta'}^{2}z(1-z)-m_{W}^{2}(1-z)-m^{2}z+q^{2} xy} \, \right]  \, .
%\hspace{0.5cm}
%\end{eqnarray}
%
The on-shell condition for the external photon imposes $q^{2}=0$. Therefore, all the form factors depend on the masses of $\zeta'_{i}$-fermion,
$W^{\pm}$ and the lepton mass, where we take $M_{\zeta'_{i}} \, , \, m_{W} \gg m$ and also consider $M_{\zeta'_{i}} > m_{W}$ .
Under these conditions, the elements of the $f_{1}$-factor are
\begin{eqnarray}
f_{1}(0)_{ij}=-\frac{e g^2 \theta^{2}}{16\pi^{2}} \, V_{ik}V_{kj} \, \left(1+\frac{m_{W}^2}{M_{\zeta'_{i}}^{2}} \right)
%\, \times
%\nonumber \\
%\times
\left[2-\left(1-\frac{m_{W}^{2}}{M_{\zeta'_{i}}^2}\right)\ln\left(\frac{M_{\zeta'_{i}}^2}{m_{W}^{2}}-1\right) \right] \; .
\end{eqnarray}
The interaction of $A^{\mu}$-field with
the $\zeta'_{i}$-current yields the form factor in the limit $q^{2} \, \rightarrow \, 0$
\begin{equation}
\bar{u}_{\zeta'_{i}}(p') \, \Gamma_{ij}^{\mu}(q) \, u_{\zeta'_{j}}(p) \, A_{\mu} \!\stackrel{q^{2} \rightarrow 0}{=}
%\bar{u}_{\zeta'}(p') \, i \, f_{m}(0) \, \sigma^{\mu\nu}q_{\nu}A_{\mu} \, u_{\zeta'}(p)
%\nonumber \\
%=
%\frac{1}{2} \, f_{m}(0) \, \bar{u}_{\zeta'}(p') \, \sigma^{\mu\nu} F_{\mu\nu} \, u_{\zeta'}(p)
%\nonumber \\
-f_{3}(0)_{ij} \, \bar{u}_{\zeta'_{i}}(p') \, \vec{\sigma} \cdot \vec{{\bf B}} \, u_{\zeta'_{j}}(p) \; ,
\end{equation}
where the transfer momentum is represented in an operator form, $q_{\nu} \rightarrow i \, \partial_{\nu}$. We have considered the
EM tensor as $F^{\mu\nu}=\left(0, \epsilon^{ijk}B^{k} \right)$ with an external magnetic field, and $\sigma^{ij}=\varepsilon^{ijk}\sigma^{k}$.
We identify the elements of $\zeta'_{i}$-MDM as $\mu_{M \, ij}^{\, (\zeta'_{i})}=f_{3}(0)_{ij}$, thus, the form factor $f_{3}(0)_{ij}$ can be written
in terms of electron's mass and the Bohr magneton :
\begin{eqnarray}
%&&
\mu_{M \, ij}^{\, (\zeta'_{i})}=
%f_{M}(0)=
-\frac{3 G_{F} m_{e} M_{\zeta'_{i}}}{4\sqrt{2} \pi^{2}} \, \theta^{2} \, V_{ik}V_{kj} \, \frac{m_{W}^{2}}{M_{\zeta'_{i}}^{2}}
%\, \times
%\nonumber \\
%&&
%\times
\left[1-\frac{1}{3}\ln\left(\frac{M_{\zeta'_{i}}^{2}}{m_{W}^{2}}-1 \right)
+\frac{m_{W}^{2}}{M_{\zeta'_{i}}^{2}} \right] \, \mu_{B} \, .
%\hspace{0.7cm}
\end{eqnarray}
This depends on the $M_{\zeta'_{i}}$-mass and on the ratio $m_{W}/M_{\zeta'_{i}}$, if we use $m_{W}=80 \, \mbox{GeV}$
and $M_{\zeta'_{1}}=0.5 \, \mbox{TeV}$, we have $m_{W}/M_{\zeta'_{1}}\simeq 0.14$, and $\theta\simeq -9 \, \times \, 10^{-8}$.
With these values, the MDM for the $\zeta'_{1}$-hidden fermion gets the one-loop contribution 
\begin{eqnarray}
\mu_{M \, ij}^{\, (\zeta'_{1})}\simeq 1.2 \, \times \, 10^{-21} \; V_{ik} V_{kj} \, \mu_{B} \; .
\end{eqnarray}
%
%The contribution for the $\zeta$-fermion MDE of $f_{E}(0)$ has the result
%
%\begin{equation}
%\mu_{E \, ij}^{\, (\zeta')}=\frac{e \, G_{F} \theta^{2}}{16\pi^{2}} \, M_{\zeta'}
%\, V_{ik} V_{kj} \, \frac{m_{W}^{2}}{M_{\zeta'}^2}
%\times
%\nonumber \\
%\times
%\left[ \, 9 -4 \, \frac{m_{W}^{2}}{M_{\zeta'}^2}
%\right.
%\nonumber \\
%\left.
%+\left(3
%-11\frac{m_{W}^{2}}{M_{\zeta'}^2}+4\frac{m_{W}^{4}}{M_{\zeta'}^4}\right) \ln\left(\frac{M_{\zeta'}^2}{m_{W}^{2}}-1 \right) \right] \; ,
%\end{equation}
%
%and using the numerical values, we obtain the estimative
%
%\begin{eqnarray}
%\mu_{E}^{\, (\zeta')}\simeq 6 \, \times \, 10^{-25} \; V_{ik} V_{kj} \, \left(\frac{M_{\zeta'}}{\mbox{GeV}}\right) \, \mu_{B} \; .
%\end{eqnarray}
%
Therefore, the elements $ij$ of $\zeta'_{1}$-MDM depend on the $V$-matrix elements as given in (\ref{PMNSMatrix}). For example,
for the diagonal elements, $\mu_{M \, ii}^{\, (\zeta'_{1})}\simeq 1.2 \, \times \, 10^{-21} \; V_{ik} V_{ki} \, \mu_{B} $
with the implicit sum runs on the $k$-index. For $i=1$, the element $\mu_{M \, 11}^{\, (\zeta'_{1})}$ depends on the cosines of mixing angles
$\theta_{12}$ and $\theta_{13}$, so we obtain the upper bound
\begin{eqnarray}
\mu_{M \, 11}^{\, (\zeta'_{1})}\simeq 1.2 \, \times \, 10^{-21} \, c_{12}^{\, 2} \, c_{13}^{\, 2} \; \mu_{B} \lesssim 1.2 \, \times \, 10^{-21} \, \mu_{B} \; .
\end{eqnarray}
\section{The X-boson scenario at the MeV-scale and below}
\renewcommand{\theequation}{6.\arabic{equation}}
\setcounter{equation}{0}

Contrary to Section II, the $X$-boson scenario is introduced through the SSB pattern-II mechanism
in which the new VEV-scale is at a lower scale with respect to the EW model, $u \ll v=246 \, \mbox{GeV}$.
This SSB pattern-II can be followed in more details in \cite{MJNeves2017Annalen}; the original
$SU_{L}(2) \times U_{R}(1)_{J} \times U(1)_{K}$-symmetry is broken as below :
%
%This correspond to
%
%with the new Higgs and its associated SSB, we here analyze the case of a VEV-scale such that $u \ll v$.
%This corresponds to a SSB , $u$, with respect to  $v = 246 \, \mbox{GeV}$ of the EW model. To this end, we re-start from the
%original , but we first evaluate the SSB in the Higgs of EW model as done in (\ref{LHiggs}),
%and consequently, we break the remaining symmetry to get the $X$-boson mass. After the sequence of the SSBs, we obtain the final symmetry
%
\begin{equation}
SU_{L}(2) \times U_{R}(1)_{J} \times U(1)_{K}
\stackrel{\langle \Phi \rangle_{0}}{\longmapsto}
%\nonumber \\
U(1)_{G} \times U(1)_{K} \stackrel{\langle \Xi \rangle_{0}}{\longmapsto} U(1)_{em} \; ,
\end{equation}
where we have the VEV scale-$u$ that defines a lightest massive boson, that can describe the physics of the dark-photon with mass bound of $m_{A'} < 8 \, \mbox{GeV}$ \cite{BaBarPRL2017}, the $X$-boson of mass $17 \, \mbox{MeV}$, or also can describe the para-photon physics at the $\mbox{Sub-eV}$-scale.
We call the $U(1)_{G}$-group as the result from the mixing of $SU_{L}(2) \times U_{R}(1)_{J}$. The generator $G$ of $U(1)_{G}$ is given by $G=I^{3}+J$.
In both cases, the extra-sector $U(1)_{K}$ of $C^{\mu}$ couples kinetically with $B^{\mu}$ of $U_{R}(1)_{J}$ by means
of mixing parameter $\chi$ given by
\begin{equation}\label{Lgauge}
{\cal L}_{gauge}=-\frac{1}{2} \, \mbox{tr}\left(F_{\mu\nu}^{\; 2}\right)
-\frac{1}{4} \, B_{\mu\nu}^{\; 2}
-\frac{1}{4} \, C_{\mu\nu}^{\; 2}
+\frac{\chi}{2} \, B_{\mu\nu} \, C^{\mu\nu} \; .
\end{equation}
%
%where $\chi$ is a real parameter that mixes the Abelian gauge fields of the $U_{R}(1)_{J} \times U(1)_{K}$-subgroup.
Its currently estimated value is $10^{-6} < \chi < 10^{-3}$ for models that discuss hidden photons as dark matter \cite{Arias2010}.

The sector of fermions is given by the Lagrangian (\ref{Lleptons}), but the covariant derivatives must exhibit the millicharged coupling
of extra boson $C^{\mu}$ of $U(1)_{K}$ with the fermions of the Standard Model. Thus, we modify these couplings introducing the parameters
$\varepsilon_{\Psi}$ and $\varepsilon_{\zeta}$
\begin{eqnarray}\label{DmuPsi}
\!\!\!\!\!\!D_{\mu}\Psi_{L} &=& \left( \partial_{\mu} + i \, g \, A_{\mu}^{\, a} \frac{\sigma^{a}}{2} + i \, J_{L} \, g^{\prime} \, B_{\mu}
%\right.
%\nonumber \\
%&&
%\left.
+ i \, K_{L} \, \varepsilon_{\Psi} \, g^{\prime \prime} \, C_{\mu} \phantom{\frac{1}{2}} \!\!\!\!\! \right)\Psi_{L} \, ,
\nonumber \\
D_{\mu} \Psi_{R} &=& \left( \phantom{\frac{1}{2}} \!\!\!\! \partial_{\mu} + i \, J_{R} \, g^{\prime} \, B_{\mu}
+ i \, K_{R} \, \varepsilon_{\Psi} \, g^{\prime \prime} \, C_{\mu} \right) \Psi_{R} \, ,
\nonumber \\
D_{\mu} \nu_{iR} &=& \left( \phantom{\frac{1}{2}} \!\!\!\! \partial_{\mu} + i \, J_{R} \, g^{\prime} \, B_{\mu}
+ i \, K_{R} \, \varepsilon_{\nu} \, g^{\prime \prime} \, C_{\mu} \right) \nu_{iR} \, ,
\nonumber \\
D_{\mu}\zeta_{i} &=& \left( \phantom{\frac{1}{2}} \!\!\!\! \partial_{\mu}+ i \, J_{\zeta} \, g^{\prime} \, B_{\mu}
+ i \, K_{\zeta} \, \varepsilon_{\zeta} \, g^{\prime \prime} \, C_{\mu}  \right) \zeta_{i} \, ,
%\hspace{-0.5cm}
\end{eqnarray}
that represent the magnitude of the weaker interaction with the $X$-boson mentioned in (\ref{Jproto}).
Furthermore, the $\Xi$-Higgs sector has the modified covariant derivative
\begin{eqnarray}\label{DmuXiProto}
D_{\mu} \Xi(x) =\left( \phantom{\frac{1}{2}} \!\!\!\! \partial_{\mu} + i \, J_{\Xi} \, g^{\prime} \, B_{\mu}+i \, K_{\Xi}
\, \varepsilon_{\Xi} \, g^{\prime\prime} \, C_{\mu} \right) \Xi(x) \, .
%\hspace{0.5cm}
%\nonumber \\
\end{eqnarray}
%
%Therefore, the interactions are modified by
%
%\begin{eqnarray}\label{LintLeptonsAZX}
%{\cal L}^{\, int} = -\bar{\Psi}\left( g \, \slash{\!\!\!\!A}^{3} I^{3} + J \, g^{\prime} \, \slash{\!\!\!\!B} \!+\! K \, \varepsilon_{\Psi} \, %g^{\prime\prime} \, \slash{\!\!\!\!C}\right)\Psi
%-\chi \, \bar{\Psi}\left( J \, g^{\prime} \, \slash{\!\!\!\!B} \,\right) \Psi
%\; .
%\hspace{0.5cm}
%\end{eqnarray}
%
%
%The first Higgs sector couples the gauge fields of the symmetry $SU_{L}(2)\times U_{R}(1)_{J}$
%
The fermion notation is kept like in the previous $Z'$-model, but the particle content is modified to insure an anomaly-free model:
\begin{eqnarray}
%L_{i}
%\!=\!
%\left(
%\begin{array}{c}
%\nu_{i} \\
%\ell_{i} \\
%\end{array}
%\right)_{L}
%\hspace{0.2cm}
L_{i}&=&
\left(
\begin{array}{c}
\nu_{e} \\
\ell_{e} \\
\end{array}
\right)_{L} ,
%\hspace{-0.1cm}
%L_{2}
%\!=\!
\left(
\begin{array}{c}
\nu_{\mu} \\
\ell_{\mu} \\
\end{array}
\right)_{L},
%\hspace{-0.1cm}
%L_{3}
%\!=\!
\left(
\begin{array}{c}
\nu_{\tau} \\
\ell_{\tau} \\
\end{array}
\right)_{L}
\hspace{-0.1cm}
: \left(\underline{{\bf 2}}, -\frac{1}{2}, -\frac{1}{2} \right) \, ,
\nonumber \\
Q_{iL}&=&
\left(
\begin{array}{c}
u \\
d \\
\end{array}
\right)_{L} ,
%\hspace{-0.1cm}
%Q_{2L}\!=\!
\left(
\begin{array}{c}
c \\
s \\
\end{array}
\right)_{L} ,
%\hspace{-0.1cm}
%Q_{3L}\!=\!
\left(
\begin{array}{c}
t \\
b \\
\end{array}
\right)_{L}
\hspace{-0.1cm}
: \left(\underline{{\bf 2}}, +\frac{1}{6}, +\frac{1}{6} \right) \, ,
%\hspace{0.45cm}
\nonumber
\end{eqnarray}
\vspace{-0.5cm}
\begin{eqnarray}
\ell_{iR} &=& \left\{ \, e_{R} \, , \, \mu_{R} \, , \, \tau_{R} \, \right\} : \left(\underline{{\bf 1}}, -1, -1 \right) \, ,
\nonumber \\
Q_{iR} &=& \left\{ \, u_{R} \, , \, c_{R} \, , \, t_{R} \, \right\} : \left(\underline{{\bf 1}}, +\frac{2}{3}, +\frac{2}{3} \right) \, ,
\nonumber \\
q_{iR} &=& \left\{ \, d_{R} \, , \, s_{R} \, , \, b_{R} \, \right\} : \left(\underline{{\bf 1}}, -\frac{1}{3}, -\frac{1}{3} \right) \, ,
\nonumber \\
\nu_{iR} &=& \left\{ \, \nu_{eR} \, , \, \nu_{\mu R} \, , \, \nu_{\tau R} \, \right\} : \left(\underline{{\bf 1}}, 0, 0 \right) \, ,
\nonumber \\
\zeta_{iL} &=& \left\{ \, \zeta_{1L} \, , \, \zeta_{2L} \, , \, \zeta_{3L} \, \right\} : \left(\underline{{\bf 1}}, 0, +\frac{3}{2} \right) \, ,
\nonumber \\
\zeta_{iR} &=& \left\{ \, \zeta_{1R} \, , \, \zeta_{2R} \, , \, \zeta_{3R}  \, \right\} : \left(\underline{{\bf 1}}, 0, -\frac{3}{2} \right) \, .
\end{eqnarray}
After the SSBs, the free Lagrangian for the neutral gauge fields reads as follows :
\begin{eqnarray}\label{LgaugemassZAC}
{\cal L}_{gauge}=
-\frac{1}{4} \, F_{\mu\nu}^{\,2}
-\frac{1}{4} \, \tilde{\tilde{Z}}_{\mu\nu}^{\, 2}
-\frac{1}{4} \, \tilde{\tilde{X}}_{\mu\nu}^{\,2}
%\nonumber \\
%&&
%\hspace{-0.5cm}
-\frac{\chi_{W}}{2} \, \tilde{\tilde{Z}}_{\mu\nu} \tilde{\tilde{X}}^{\mu\nu}
+\frac{1}{2} \, m_{\tilde{\tilde{Z}}}^{\, 2} \, \tilde{\tilde{Z}}_{\mu}^{\, 2}
%+\frac{\chi_{W}}{2} \, \tilde{\tilde{Z}}_{\mu\nu} \, \tilde{A}^{\mu\nu}
+\frac{1}{2} \, m_{\tilde{\tilde{X}}}^{\, 2} \, \tilde{\tilde{X}}_{\mu}^{\, 2} \; ,
%+ x \; m_{Z} \, \tilde{\tilde{Z}}_{\mu} \, \right)^{2} \; ,
\hspace{0.4cm}
\end{eqnarray}
where we have defined
\begin{eqnarray}
\chi_{W} := \frac{\chi\, \sin\theta_{W}}{\sqrt{1-\chi^{2}\cos^{2}\theta_{W}}} \; ,
%\nonumber \\
%x \!&=&\! 2 \; \frac{u}{v} \; \sin^{2}\theta_{W} \; ,
\end{eqnarray}
and the masses of $\tilde{\tilde{Z}}^{\mu}$ and $\tilde{\tilde{X}}^{\mu}$ are given by
\begin{eqnarray}\label{mX}
m_{\tilde{\tilde{Z}}}=\frac{e \, v}{\sin2\theta_{W}}
\hspace{0.4cm} \mbox{and} \hspace{0.4cm}
m_{\tilde{\tilde{X}}}= 3 \, |\varepsilon_{\Xi}| \, e \, u \; .
\end{eqnarray}
For more details on the diagonalization procedure, see \cite{MJNeves2017Annalen}. The sector of neutral gauge bosons
is composed by two Proca fields with the mixing parameter $\chi_{W}$ in the kinetic term.

%
%\begin{eqnarray}\label{LgaugeU1XU1}
%{\cal L}_{gauge} \!\!&=&\!\! - \frac{1}{4} \, \left( A_{\mu\nu}^{3} \right)^{2}
%%-\frac{1}{4} \, B_{\mu\nu}^{\,2}-\frac{1}{4} \, C_{\mu\nu}^{\,2}
%\nonumber \\
%&&
%\hspace{-0.3cm}
%+\frac{\chi}{2} \, B_{\mu\nu} \, C^{\mu\nu}
%+\frac{1}{2} \, \frac{v^{2}}{4} \left( \phantom{\frac{1}{2}} \!\!\!\! g^{\prime} \, B_{\mu}-g \, A_{\mu}^{3} \right)^{2}
%\nonumber \\
%&&
%\hspace{-0.3cm}
%+ \frac{u^2}{2} \left( \phantom{\frac{1}{2}} \!\!\!\! J_{\Xi} \, g^{\prime} \, B_{\mu}
%+ K_{\Xi} \, \varepsilon_{\Xi} \, g^{\prime \prime} \, C_{\mu} \right)^2 \; ,
%\end{eqnarray}
%
%where $A_{\mu\nu}^{3}:=\partial_{\mu}A_{\nu}^{\, 3}-\partial_{\nu}A_{\mu}^{\, 3}$ . To diagonalize the mixed terms in (\ref{LgaugeU1XU1}),
%we introduce the following transformations
%
%\begin{eqnarray}\label{transfABZG}
%A_{\mu}^{\, 3} \!\!&=&\!\! \cos\theta_{W} \, \tilde{\tilde{Z}}_{\mu} + \sin\theta_{W} \, G_{\mu}
%\nonumber \\
%B_{\mu} \!\!&=&\!\! -\sin\theta_{W} \, \tilde{\tilde{Z}}_{\mu} + \cos\theta_{W} \, G_{\mu} \; ,
%\nonumber \\
%G_{\mu} \!\!&=&\!\!  A_{\mu}+\frac{\chi \cos\theta_{W} \tilde{\tilde{X}}_{\mu} }{\sqrt{1-\chi^{2} \cos^{2}\theta_{W}}}
%\nonumber \\
%C_{\mu} \!\!&=&\!\! \frac{\tilde{\tilde{X}}_{\mu}}{\sqrt{1-\chi^{2} \cos^{2}\theta_{W}}} \, ,
%\end{eqnarray}
%
%where the
Here, the Weinberg mixing angle $\theta_{W}$ satisfies the parametrization
\begin{eqnarray}\label{eggg}
e=g^{\prime \prime}= g \, \sin\theta_{W}=g^{\prime} \, \cos\theta_{W} \; .
\end{eqnarray}
%
%Here, $G^{\mu}$ is the vector field associated with the $U(1)_{G}$ subgroup.
%After this first SSB, the $G$-field remains massless, while
%
In the gauge sector, $\tilde{\tilde{Z}}^{\mu}$ acquires a mass due to VEV-scale of $v=246 \, \mbox{GeV}$,
and $\tilde{\tilde{X}}^{\mu}$ mass is due to lower VEV-scale $u$.
For convenience, we use here the notation $\tilde{\tilde{Z}}^{\mu}$ and $\tilde{\tilde{X}}^{\mu}$
because they are not the physical $Z$- and $X$-bosons yet. The real $Z$-boson will be mixed with the $X$-boson,
in which a full diagonalization of (\ref{LgaugemassZAC}) implies to shift fields
\begin{eqnarray}\label{transfZX}
\tilde{\tilde{X}}^{\mu} &=& X^{\mu}+\frac{\chi_{W} \, Z^{\mu}}{\sqrt{1-\chi_{W}^{\, 2}}} \; ,
\nonumber \\
\tilde{\tilde{Z}}^{\mu} &=& \, \frac{Z^{\mu}}{\sqrt{1-\chi_{W}^{\, 2}}} \; ,
\end{eqnarray}
and it gives the masses of the physical $Z$- and $X$-bosons with the correction induced by the $\chi_{W}$-parameter :
%
%in which the $X$-boson mass is $M_{X}\simeq 3 \, |\varepsilon| \, e \, u$ at the leading $\chi^2$-order,
%and the $Z$-mass has the
%
\begin{eqnarray}
M_{Z} &\simeq& \frac{m_{\tilde{\tilde{Z}}}}{\sqrt{1-\chi_{W}^{\, 2}}} \left( 1 + \frac{1}{2} \, \frac{m_{X}^{\, 2}}{m_{Z}^{\, 2}} \, \chi_{W}^{\, 2} \right) \; ,
\hspace{0.5cm}
\nonumber \\
M_{X} &\simeq& \frac{m_{\tilde{\tilde{X}}}}{\sqrt{1-\chi_{W}^{\, 2}}} \, \left( 1 - \frac{\chi_{W}^{2}}{2} \right) \; .
\end{eqnarray}
The parametrization (\ref{eggg}) suggests that the electric charge generator can be defined by $Q_{em}=G=I^{3}+J$.
For convenience, the new $\Xi$-Higgs carries the charges $J_{\Xi}=0$ and $K_{\Xi}=+3$ to generate the lightest new fermion
and also to avoid stable charged matter, see \cite{Duerr2014,Duerr2015}.
Here, the hypercharge $Y$ is not given by the sum of the charges $J$ and $K$.
It is so defined by the proper $J$-generator, {\it i. e.}, $J=Y$, and the $K$-generator has independent
values of the $Y$-charges. The simplest charge values are displayed in the \ref{Table2},
and the model in the X-boson scenario is also anomaly-free. All matter content of the SM has $K=0$.
The $\zeta_{i}$-fermions do not carry electric charge, such that it just sets the $K$-charge
as an example of a hidden charge. Thus, this fact supports the viewpoint of the $\zeta_{i}$-fermions as dark matter candidates.
Other important point is that the gauge symmetry forbids the mixed Yukawa interactions in (\ref{LHiggs}),
so we must take the coupling constants, $Y_{ij}=Z_{ij} \rightarrow 0$. Therefore,
the gauge symmetry and the anomaly cancellation condition lead us to the values for $K_{\zeta}$-charges :
$K_{\zeta_{L}}=-K_{\zeta_{R}}=+3/2$.
%
%In this stage, the model for the $X$-boson description proposed here reduces to the analogous case
%of the $U_{B}(1)$-symmetry also studied in \cite{JFengPRD2017}. All terms of neutral bosons are gathered
%together in the Lagrangian
%
%
\begin{table}
\centering
\begin{tabular}{|l|l|l|l|l|l|l|}
\hline
%after \\: \hline or \cline{col1-col2} \cline{col3-col4} ...
\mbox{Fields} & $Q_{em}$ & $I^{3}$ & $Y=J$ & $K$ \\
\hline
\mbox{leptons-left} & $-1$ & $-1/2$ & $-1/2$ & $-1/2$  \\
\hline
\mbox{leptons-right} & $-1$ & $0$ & $-1$ & $-1$ \\
\hline
\mbox{neutrinos-left} & $0$ & $+1/2$ & $-1/2$ & $-1/2$ \\
\hline
\mbox{neutrinos-right} & $0$ & $0$ & $0$ & $0$ \\
\hline
$\zeta_{i}$-\mbox{fermions left} & $0$ & $0$ & $0$ & $+3/2$ \\
\hline
$\zeta_{i}$-\mbox{fermions right} & $0$ & $0$ & $0$ & $-3/2$ \\
\hline
\mbox{u-quark-left} & $+2/3$ & $+1/2$ & $+1/6$ & $+1/6$  \\
\hline
\mbox{d-quark-left} & $-1/3$ & $-1/2$ & $+1/6$ & $+1/6$ \\
\hline
\mbox{s-quark-left} & $-1/3$ & $-1/2$ & $+1/6$ & $+1/6$  \\
\hline
\mbox{u-quark-right} & $+2/3$ & $0$ & $+2/3$ & $+2/3$ \\
\hline
\mbox{d-quark-right} & $-1/3$ & $0$ & $-1/3$ & $-1/3$ \\
\hline
\mbox{s-quark-right} & $-1/3$ & $0$ & $-1/3$ & $-1/3$  \\
\hline
$W^{\pm}$-\mbox{bosons} & $\pm \, 1$ & $\pm \, 1$ & $0$ & $0$ \\
\hline
\mbox{neutral bosons} & $0$ & $0$ & $0$ & $0$  \\
\hline
$\Xi$-\mbox{Higgs} & $0$ & $0$ & $0$ & $+3$ \\
\hline
$\Phi$-\mbox{Higgs} & $0$ & $-1/2$ & $+1/2$ & $0$ \\
\hline
\end{tabular}
\caption{The simplest anomaly-free particle content for the $U(1)_{K}$-dark photon model.}\label{Table2}
\end{table}
%
%Thereby, the full diagonal Lagrangian is more explicitly written as
%
%\begin{equation}\label{LgaugeZG}
%{\cal L}_{gauge} \!=\!
%- \frac{1}{4} F_{\mu\nu}^{\,2}
%- \frac{1}{4} Z_{\mu\nu}^{\, 2}
%+ \frac{1}{2} M_{Z}^{\, 2} Z_{\mu}^{\, 2}
%\nonumber \\
%&&
%- \frac{1}{4} X_{\mu\nu}^{\,2} + \frac{1}{2} M_{X}^{\, 2} X_{\mu}^{\, 2} \; ,
%\end{equation}
%
The interactions between $Z-$ and $X$-bosons and any quiral fermion $\Psi$ of the model are cast below :
%
%, we substitute all the previous transformations from (\ref{transfABZG}) and (\ref{transfZX}) into the fermion sector, and using the parametrization (\ref{eggg}); this yields :
%
\begin{equation}
{\cal L}^{\, int} \!=\! - \, e\, Q_{em} \, \bar{\Psi} \, \, \slash{\!\!\!\!A} \, \Psi
- e \, Q_{Z} \, \bar{\Psi} \, \, \slash{\!\!\!\!Z} \, \Psi
%\nonumber \\
%&&
%\hspace{-0.5cm}
- e \, Q_{X} \, \bar{\Psi} \,\, \slash{\!\!\!\!X} \, \Psi \; .
\end{equation}
The charge generator $Q_{X}$ is defined as follows
\begin{eqnarray}
%Q_{Z}&=& \frac{\sqrt{2}}{\sin\theta_{W} \, \cos\theta_{W}} \, \left( \, I^{3}-\sin^{2}\theta_{W} \, Q_{em}+\sin^{2}\theta_{W} \, K \, \right) \; ,
%\nonumber \\
Q_{X}:=
%\tilde{\chi} \, \sin\theta_{W} \, Q_{Z} -
+\tilde{\chi} \, \cos\theta_{W} \, Q_{em} +K_{\Psi} \, \tilde{\varepsilon}_{\Psi}  \; ,
%\hspace{-0.5cm}
%\nonumber \\
\end{eqnarray}
where $\tilde{\chi}:=\chi/\sqrt{1-\chi^{2}}$, $\tilde{\varepsilon}_{\Psi}:=\varepsilon_{\Psi}/\sqrt{1-\chi^{2}}$.
%
%and $\chi_{\Psi}$ is given by
%
%\begin{eqnarray}
%\chi_{\Psi}=- \, Q_{em} \, \frac{\chi \, \cos\theta_{W}}{\sqrt{1-\chi^{2} \, \cos^2\theta_{W}}}
%+K_{\Psi} \, \frac{\varepsilon_{\Psi}}{\sqrt{1-\chi^{2} \, \cos^2\theta_{W}}} \; .
%\end{eqnarray}
%
We observe a resulting millicharged current in the interaction of any fermion with the $X$-boson,
whose coupling constant is given by $e \, Q_{X}$. Thereby, the interaction of $X$-boson
with the non-chiral components of $f$-fermions can be written as
\begin{equation}\label{LintX}
{\cal L}^{int}_{X}=-e \, \bar{f} \, \, \slash{\!\!\!\!X} \, \left(c^{f}_{V}-c^{f}_{A} \, \gamma_{5}\right) f \; ,
\end{equation}
where the coefficients $c^{f}_{V}$ and $c^{f}_{A}$ are defined by
\begin{eqnarray}
c^{f}_{V}=Q_{X}^{\,f_{L}}+Q_{X}^{\,f_{R}}
\hspace{0.5cm} \mbox{and} \hspace{0.5cm}
c^{f}_{A}=Q_{X}^{\,f_{L}}-Q_{X}^{\,f_{R}} \; .
\end{eqnarray}
The $X$-$f$ interaction is not CP-invariant. This shall have some consequence when we are going to calculate the MDM of the charged leptons.
%
%The $X$-vertex with the $f$-fermions of model is setting by
%
\begin{figure}[!h]%\label{figvm}
\begin{center}
\newpsobject{showgrid}{psgrid}{subgriddiv=1,griddots=10,gridlabels=6pt}
\begin{pspicture}(5,1)(9.5,2.4)
%\showgrid
\psset{arrowsize=0.2 2}
\psset{unit=0.8}
%
%%%%%%%%%%%%%%%%%%%% Vertice Boson X - Leptons %%%%%%%%%%%%%%%%%%%%%%%%%%%%%%%%
%
\pscoil[coilarm=0,coilwidth=0.3,coilheight=0.7,linecolor=black](6.5,1.05)(6.5,3)
\psline[linecolor=black,linewidth=0.5mm]{-}(5,1)(8,1)
\psline[linecolor=black,linewidth=0.5mm]{->}(5,1)(6,1)
\psline[linecolor=black,linewidth=0.5mm]{->}(7,1)(7.55,1)
\put(6.8,2.8){\large$X$}
\put(4.95,1.2){\large$\bar{f}$}
\put(7.8,1.2){\large$f$}
\put(8.5,1){\large$\Gamma_{X}^{\, \mu}=- \, i \, e \, \gamma^{\mu}\left( c_{V}^{f}-c_{A}^{f} \, \gamma_{5} \right) .$}
\end{pspicture}
%
%\vspace{0.2cm}
%
%\caption{\scshape{The one loop correction of the vertex diagram.}}\label{figvm}
%
\end{center}
\end{figure}
%
%
%The $f$-sum in (\ref{LintX}), as previously, runs to all fermions of the model.
Thereby, we list all expressions of $c^{f}_{V}$ and $c^{f}_{A}$ in terms of $\varepsilon$-parameters,
following the charges in the table (\ref{Table2}) :
\begin{eqnarray}
c^{\ell}_{V} &=& - \, \chi \, \cos\theta_{W}-\frac{1}{2}\left(\frac{1}{2} \, \varepsilon_{\ell_{L}}+ \varepsilon_{\ell_{R}} \right) \; ,
\nonumber \\
c^{\ell}_{A} &=& +\frac{1}{2}\left( -\frac{1}{2} \, \varepsilon_{\ell_{L}}+ \varepsilon_{\ell_{R}} \right) \; ,
\nonumber \\
c^{\nu}_{V} &=& c^{\nu}_{A}= -\frac{1}{4} \, \varepsilon_{\nu_{L}}
\hspace{0.2cm} , \hspace{0.2cm}
%\nonumber \\
\nonumber \\
c^{u}_{V} &=& +\frac{2}{3} \, \chi \, \cos\theta_{W}
+\frac{1}{2}\left( \frac{1}{6} \, \varepsilon_{u_{L}}+\frac{2}{3} \, \varepsilon_{u_{R}} \right)
\hspace{0.2cm} , \hspace{0.2cm}
\nonumber \\
c^{u}_{A} &=& +\frac{1}{2}\left( \frac{1}{6} \, \varepsilon_{u_{L}}-\frac{2}{3} \, \varepsilon_{u_{R}} \right)
\nonumber \\
c^{d}_{V} &=& -\frac{1}{3} \, \chi \, \cos\theta_{W}+\frac{1}{2}\left( \frac{1}{6} \, \varepsilon_{d_{L}}-\frac{1}{3} \, \varepsilon_{d_{R}} \right)
\hspace{0.2cm} , \hspace{0.2cm}
\nonumber \\
c^{d}_{A} &=& +\frac{1}{2}\left( \frac{1}{6} \, \varepsilon_{d_{L}}+\frac{1}{3} \, \varepsilon_{d_{R}} \right)
\nonumber \\
c^{\zeta}_{V} &=& +\frac{3}{2} \, \left(\varepsilon_{\zeta_{L}}-\varepsilon_{\zeta_{R}}\right)
\hspace{0.1cm} , \hspace{0.1cm}
%\nonumber \\
c^{\zeta}_{A}= +\frac{3}{2} \, \left(\varepsilon_{\zeta_{L}}+\varepsilon_{\zeta_{R}}\right) \, .
\end{eqnarray}
Here, we have 11 $\varepsilon$-parameters so that we can choose the convenient way to describe the phenomenology of the $X$-boson
interacting with the SM matter. We consider the simplest case for both vector and axial currents type of coupling with leptons
in which the parameters are fixed as : $\varepsilon_{\ell_{L}}=-2 \, \varepsilon_{\ell_{R}}$, and consequently,
$c_{V}^{\ell}=-\chi \, \cos\theta_{W}$ and $c_{A}^{\ell}=-\varepsilon_{\ell_{L}}/2$. In the quark sector, the $X$-boson has only
axial coupling to avoid an extra flavour mixing of $u$ and $s$, so it is convenient to fix $c_{V}^{Q}=0$ that implies in the conditions :
\begin{eqnarray}\label{condaxial}
\varepsilon_{u_{L}}+2 \, \varepsilon_{d_{L}} &=& 4 \left( \varepsilon_{d_{R}}- \varepsilon_{u_{R}} \right)
\nonumber \\
\varepsilon_{u_{L}}+2 \, \varepsilon_{s_{L}} &=& 4 \left( \varepsilon_{s_{R}}- \varepsilon_{u_{R}} \right)
\nonumber \\
\varepsilon_{d_{L}}- \varepsilon_{s_{L}} &=& 2 \left( \varepsilon_{d_{R}}- \varepsilon_{s_{R}} \right) \; .
\end{eqnarray}
The allowed region to explain the $8 \, \mbox{Be}$-anomaly requires for quarks couplings the conditions $c_{A}^{u}<0$, $c_{A}^{d}>0$
and $c_{A}^{d}=c_{A}^{s}$ that fix the bounds \cite{KozaczukPRD2017}
\begin{eqnarray}
&&
10^{-4} \lesssim |\varepsilon_{d_{L}}+2 \, \varepsilon_{d_{R}}| \lesssim 10^{-3}
\nonumber \\
&&
10^{-4} < |\varepsilon_{u_{L}}-4 \, \varepsilon_{u_{R}}| \lesssim 10^{-3} \; .
\end{eqnarray}
So, by adopting the range $10^{-4} < |\varepsilon_{\Xi}| < 10^{-3}$,
we fix the value $M_{X}=17 \, \mbox{MeV}$, and the VEV-scale $u$ is estimated in the range
$u \sim 1.9-19 \, \mbox{GeV}$.
%for mass of $X$-boson into the expression (\ref{mX})
%
%\begin{eqnarray}
%u \simeq \frac{19}{|\varepsilon_{\Xi}|} \, \mbox{MeV} \simeq 6.3 \, \mbox{GeV} \; .
%\end{eqnarray}
%
%In so doing, we obtain
%the VEV $u= 25 \, \mbox{MeV}$,
%
In the Higgs sector, whenever $u \ll v$, the masses of $H$- and $F$-scalars fields are
\begin{eqnarray}
M_{H} &\simeq& \sqrt{ 2 \, \lambda_{\Phi} \, v^{2}}
\, \left( 1+\frac{\lambda^2}{8 \, \lambda_{\Phi}^2} \, \frac{u^2}{v^2} \right)=125 \, \mbox{GeV} \; ,
\nonumber \\
M_{F} &\simeq& \sqrt{ 2 \, \lambda_{\Xi} \, u^{2}}
\, \left( 1-\frac{\lambda^2}{8 \, \lambda_{\Phi} \, \lambda_{\Xi}} \right) \; ,
\end{eqnarray}
%
%and mass of the corresponding  $F$-Higgs is comes out in the range $0.35 \, \mbox{GeV} \lesssim M_{F} \lesssim 1 \, \mbox{GeV}$.
and the $F$-mass is estimated in the range
%
%\begin{eqnarray}
$1.9 \, \mbox{GeV} \lesssim M_{F} \lesssim 19 \, \mbox{GeV}$.
%\end{eqnarray}
%
Under these conditions, the neutrino and $\zeta_{i}$-fermions mass matrices are given by the
\begin{eqnarray}
M_{ij}^{(\nu)} = \frac{v \, X_{ij}}{\sqrt{2}}
\hspace{0.3cm} \mbox{and} \hspace{0.3cm}
M_{ij}^{(\zeta)} = \frac{u \, W_{ij}}{\sqrt{2}} \, ,
%\hspace{0.8cm}
\end{eqnarray}
in which $M_{ij}^{(\zeta)}$ must have lighter elements in the $X$-Boson scenario if compared with the corresponding case of the $Z'$ scenario.
For example, the estimate here satisfies the Treimaine-Gunn lower bound of $> 1.5 \, \mbox{KeV}$ for the mass
of a dark matter candidate, see \cite{DasPRD2012}. ASt this stage, the diagonalization procedure is like in the early case of the
$Z'$-scenario. The diagonalization of mass matrix $M^{(\zeta)}$ is carried out by a unitary transformation such that the elements
of the diagonal matrix are the masses $M_{\zeta'_{i}} \, (i=1,2,3)$. If we impose the previous coupling constants
$W_{\zeta'_{1}}=0.28$, $W_{\zeta'_{2}}=0.4$ and $W_{\zeta'_{3}}=0.5$, we obtain the following masses for the light-fermions
$M_{\zeta'_{1}} \simeq 0.3-3.8 \, \mbox{GeV}$, $M_{\zeta'_{2}}\simeq 0.5-5.3 \, \mbox{GeV}$ and $M_{\zeta'_{3}}=0.6-6.6 \, \mbox{GeV}$.
\section{The $X$-boson phenomenology}
\renewcommand{\theequation}{7.\arabic{equation}}
\setcounter{equation}{0}
\subsection{The X-boson decays processes}

The $X$-boson phenomenology points out to the decay $X \, \rightarrow \, e^{+} \, e^{-}$ as one of main processes.
Furthermore, we will obtain here the others possible processes of $X$-decay into the SM fermions. Thereby,
using the QFT rules for $X$-interaction in (\ref{LintX}), the decay width for the process $X \rightarrow \bar{f} \, f$
is given by
\begin{eqnarray}\label{XffDecay}
\Gamma(X \rightarrow \bar{f} \, f )=\frac{\alpha \, M_{X} }{6}
\left( |c^{f}_{V}|^2+|c^{f}_{A}|^2 \right)
%\times
%\nonumber \\
%&&
%\hspace{-0.5cm}
%\times \,
\sqrt{1-\frac{4M_{f}^{\, 2}}{M_{X}^{\, 2}}} \left( 1
- \frac{3}{4}\frac{M_{f}^{\, 2}}{M_{X}^{\, 2}} \right) \; ,
%\hspace{0.3cm}
\end{eqnarray}
with the condition $M_{X} > 2 \, M_{f}$. It is clear that, for $X$-boson of $M_{X}=17 \, \mbox{MeV}$ some decays processes
are forbidden according to the result of (\ref{XffDecay}), as for example, the $X$-decays into the muon and tau-particles in the leptons sector,
and $X$-decays into the heavier quarks.

The decay processes into the $X \, \rightarrow \, e^{+} \, e^{-}$ and $X \, \rightarrow \, \bar{\nu}_{e} \, \nu_{e}$ yield the $X$-decay widths
\begin{eqnarray}\label{DecayWX}
\Gamma(X \rightarrow e^{+} \, e^{-} ) &=& \frac{\alpha \, M_{X} }{6}\left( \chi^2 \, \cos^2\theta_{W}+ \frac{\varepsilon_{e_{L}}^{2}}{4} \right) \; ,
\nonumber \\
\Gamma(X \rightarrow \bar{\nu}_{e} \, \nu_{e} ) &=& \frac{\alpha \, M_{X} }{48} \, \varepsilon_{\nu_{L}}^{\, 2} \; ,
\end{eqnarray}
in which we have used $M_{X} \gg \left\{ \, M_{e} \, , \, M_{\nu} \, \right\}$. The KLOE-2 collaboration sets the limit
$|\varepsilon_{e_{L}}| \lesssim 4 \times 10^{-3}$ \cite{KLOE2}, and neutrino constraint is in the scattering
$\bar{\nu}_{e} \, e \, \rightarrow \, \bar{\nu}_{e} \, e$, that fixes the bound $|\varepsilon_{\nu_{L}}| < 2.2 \times 10^{-5}$ \cite{BilmisPRD2015}.
The decays widths (\ref{DecayWX}) are estimated by the results below
\begin{eqnarray}
\Gamma(X \rightarrow e^{+} \, e^{-} ) &=& 8.2 \times 10^{-8} \, \mbox{MeV} \; ,
\nonumber \\
\Gamma(X \rightarrow \bar{\nu}_{e} \, \nu_{e} ) &=& 10^{-12} \, \mbox{MeV} \; .
\end{eqnarray}
The allowed decay process in the quarks sector has the decay width
\begin{eqnarray}
\Gamma(X \rightarrow \bar{u} \, u ) = \frac{\alpha \, M_{X} }{864} \, \left| \varepsilon_{u_{L}}-4 \, \varepsilon_{u_{R}} \right|^{2}
%\nonumber \\
%&&
%\hspace{-0.5cm}
\simeq 10^{-12}-10^{-10} \, \mbox{MeV} \; .
%\nonumber \\
%\Gamma(X \rightarrow \bar{d} \, d ) \!\!&=&\!\! \frac{\alpha \, M_{X} }{864} \, \left| \varepsilon_{d_{L}}+2 \, \varepsilon_{d_{R}} \right|^{2}
%\nonumber \\
%&&
%\hspace{-0.5cm}
%\simeq 10^{-12}-10^{-10} \, \mbox{MeV} \; .
\end{eqnarray}
%

%
%and we obtain the sum of two decays as
%%
%\begin{eqnarray}
%\Gamma(X \rightarrow e^{+} \, e^{-} )+\Gamma(X \rightarrow \bar{\nu} \, \nu )=\frac{\alpha}{2}\frac{M_{X} \, \tilde{\chi}^2}{\cos^2\theta_{W}} \; .
%\end{eqnarray}
%
%Thus, the $X$-boson decay time with these two contributions is estimated by
%%
%\begin{eqnarray}
%\tau= ??? \mbox{s} \; .
%\end{eqnarray}
%
%In the case $X \rightarrow \bar{\zeta}_{i} \, \zeta_{i}$, it decay width depends on the
%$\varepsilon_{\zeta}$-parameters :
%
%\begin{eqnarray}
%\Gamma(X \, \rightarrow \, \bar{\zeta} \, \zeta )=\frac{3\alpha}{4} \, M_{X} \, \left( \varepsilon_{\zeta_{L}}^2+\varepsilon_{\zeta_{R}}^2 \right)
%\nonumber \\
% =0.09 \, \left( \varepsilon_{\zeta_{L}}^2+\varepsilon_{\zeta_{R}}^2 \right) \, \mbox{MeV} \; .
%\end{eqnarray}
%

%
\subsection{The $X$-boson scattering}

\begin{figure}[!h]%\label{figvm}
\begin{center}
\newpsobject{showgrid}{psgrid}{subgriddiv=1,griddots=10,gridlabels=6pt}
\begin{pspicture}(0,-0.6)(2.5,3.3)
%\showgrid
\psset{arrowsize=0.2 2}
\psset{unit=1.2}
%
%%%%%%%%%%%%%%%%%%%%%%%%%%% Bosons Z' Z W %%%%%%%%%%%%%%%%%%%%%%%%%%%%%%%%%%
%
\pscoil[coilaspect=0,coilarm=0,coilwidth=0.25,coilheight=1.3,linecolor=black](0,0.3)(2.99,0)
\pscoil[coilarm=0,coilwidth=0.25,coilheight=1.3,linecolor=black](0,1.7)(2.99,2)
\put(1.5,2.1){$X$}
\put(1.5,-0.3){$\gamma$}
%
%\psline[linecolor=black,linewidth=0.3mm]{->}(5,2)(6.5,3.5)
%
\psline[linecolor=black,linewidth=0.3mm]{->}(0,0.3)(0,1.2)
\psline[linecolor=black,linewidth=0.3mm]{-}(0,0.9)(0,1.7)
\put(-0.45,1){$e^{-}$}
%
%%%%%%%%%%%%%%%%%%%%%%%%%%%%%%%%%%%%%%
%
\psline[linecolor=black,linewidth=0.3mm]{->}(-2,-0.5)(-0.9,-0.05)
\psline[linecolor=black,linewidth=0.3mm]{-}(-1,-0.1)(0,0.3)
\put(-1,2.3){$e^{-}$}
\psline[linecolor=black,linewidth=0.3mm]{->}(0,1.7)(-1.1,2.105)
\psline[linecolor=black,linewidth=0.3mm]{-}(-1,2.08)(-2,2.5)
\put(-1,-0.4){$e^{+}$}
%
%%%%%%%%%%%%%%%%%%%%%%%%%%%%%%%%%%%%
%
\psline[linecolor=black,linewidth=0.3mm](3,2)(3.75,2.37)
\psline[linecolor=black,linewidth=0.3mm]{<-}(3.4,2.2)(4,2.5)
\put(3.5,2.6){$\bar{\zeta}'_{1}$}
\psline[linecolor=black,linewidth=0.3mm]{->}(3,2)(3.75,1.63)
\psline[linecolor=black,linewidth=0.3mm](3.5,1.75)(4,1.5)
\put(3.5,1.3){$\zeta'_{1}$}
%
%%%%%%%%%%%%%%%%%%%%%%%%%%%%%%%%%%%%%%%%%%%%%%%%%%%%
%
\psline[linecolor=black,linewidth=0.3mm]{->}(3,0)(3.75,0.37)
\psline[linecolor=black,linewidth=0.3mm](3.4,0.2)(4,0.5)
\put(3.5,0.5){$e^{-}$}
\psline[linecolor=black,linewidth=0.3mm](3,0)(3.75,-0.37)
\psline[linecolor=black,linewidth=0.3mm]{<-}(3.4,-0.2)(4,-0.5)
\put(3.5,-0.7){$e^{+}$}
%
%\psline[linecolor=black,linewidth=0.3mm]{<-}(0,1)(-1.5,-0.5)
%\psline[linecolor=black,linewidth=0.3mm]{->}(1,2)(-0.4,3.4)
%
%
%Momenta
%
%\psline[linecolor=black,linewidth=0.3mm]{->}(2.2,2.7)(3.8,2.7)
%\put(3,2.6){\Large$X^{\mu}$}
%
%\psline[linecolor=black,linewidth=0.3mm]{->}(7.8,1.2)(7.8,2.8)
%\put(3,1){\Large$k$}
%
%\put(5.5,3.6){\Large$e^{-}$}
%\put(5.4,-0.1){\Large$e^{+}$}
%\put(6.7,3.1){\Large$q$}
%\put(6.7,0.7){\Large$q^{\prime}$}
%
%\put(-0.2,3.6){\Large$e^{-}$}
%\put(-0.4,-0.1){\Large$e^{+}$}
%\put(-1.1,3.1){\Large$q$}
%\put(-1.1,0.7){\Large$q^{\prime}$}
%
%\put(0.6,-0.3){$p$}
%
%\put(1.3,0){$\Gamma_{\mu}^{(3)\;abc}(p)=g_{1}f^{abc}p_{\mu}$}
%
%
\end{pspicture}
%
%\vspace{0.2cm}
%
\caption{\scshape{The $X$-decay scattering into a possible dark matter scenario with the $\zeta'_{1}$-fermion of mass
$M_{\zeta'_{1}}=1 \, \mbox{GeV}$.}} \label{XScattering}
\end{center}
\end{figure}
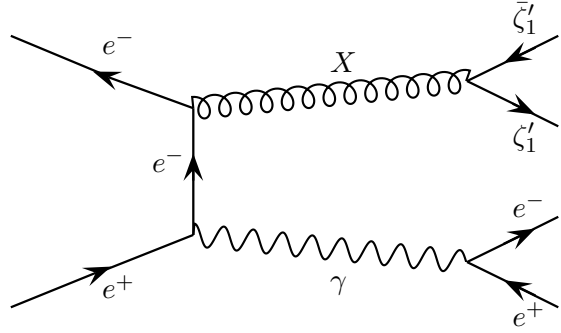

A possible scenario for the $X$-boson scattering lies in its connection with the $\zeta_{i}$-fermions. The recent bound for dark photons $A'$ of mass $m_{A'}< 8 \, \mbox{GeV}$ in the process $e^{+} \, e^{-} \, \rightarrow \, \gamma \, A'$ could establish a connection with a dark matter scenario through the interaction of $A'$ with a dark sector. Therefore, we propose the study of the scattering $e^{+} \, e^{-} \, \rightarrow \, X \, \rightarrow \, \bar{\zeta}'_{1} \, \zeta'_{1}$ in which the $X$-boson acts like a dark photon, and we choose the mass of $\zeta'_{1}$ as $M_{\zeta'_{1}}=1 \, \mbox{GeV}$. The process is illustrated in the figure (\ref{XScattering}).
%
%%%%%%%%%%%%%%%%% The X-scattering into zeta %%%%%%%%%%%%%%%%%%%%%%%%%
%

\noindent
The amplitude for this scattering is depicted in the figure (\ref{ScatteringX}).
%

%%%%%%%%%%%%%%%%% The Z'-scatterring into DM %%%%%%%%%%%%%%%%%%%%%%%%%
%
\begin{figure}[!h]%\label{figvm}
\begin{center}
\newpsobject{showgrid}{psgrid}{subgriddiv=1,griddots=10,gridlabels=6pt}
\begin{pspicture}(0,0.5)(2.5,4)
%\showgrid
\psset{arrowsize=0.2 2}
\psset{unit=1.3}
%
%%%%%%%%%%%%%%%%%%%%%%%%%%% Bosons Z' Z W %%%%%%%%%%%%%%%%%%%%%%%%%%%%%%%%%%
%
%\pscoil[coilaspect=0,coilarm=0,coilwidth=0.25,coilheight=1.3,linecolor=black](0,0.3)(2.99,0)
%
\pscoil[coilarm=0,coilwidth=0.25,coilheight=1.3,linecolor=black](0,1.7)(2,1.7)
\put(0.9,2.1){$X$}
%
%\put(1.5,-0.5){$W/Z$}
%
%\psline[linecolor=black,linewidth=0.3mm]{->}(5,2)(6.5,3.5)
%
%\psline[linecolor=black,linewidth=0.3mm]{->}(0,0.3)(0,1.2)
%\psline[linecolor=black,linewidth=0.3mm]{-}(0,0.9)(0,1.7)
%
%\put(-0.45,1){$q$}
%
%%%%%%%%%%%%%%%%%%%%%%%%%%%%%%%%%%%%%%
%
\psline[linecolor=black,linewidth=0.5mm,ArrowInside=->,ArrowInsidePos=0.5](0,1.7)(-1.5,0.3)
%\psline[linecolor=black,linewidth=0.3mm]{-}(-1,-0.1)(0,0.3)
%
\put(-1.1,2.8){\large$e^{-}$}
\psline[linecolor=black,linewidth=0.5mm,ArrowInside=->,ArrowInsidePos=0.65](-1.5,3)(0,1.7)
%\psline[linecolor=black,linewidth=0.3mm]{-}(-1,2.08)(-2,2.5)
%
\put(-1.1,0.4){\large$e^{+}$}
%
%%%%%%%%%%%%%%%%%%%%%%%%%%%%%%%%%%%%
%
\psline[linecolor=black,linewidth=0.5mm,ArrowInside=->,ArrowInsidePos=0.65](2,1.7)(3,3)
%\psline[linecolor=black,linewidth=0.3mm]{<-}(3.4,2.2)(4,2.5)
%
\put(2.4,2.8){\large$\zeta'_{1}$}
\psline[linecolor=black,linewidth=0.5mm,ArrowInside=->,ArrowInsidePos=0.65](3,0.5)(2,1.7)
%\psline[linecolor=black,linewidth=0.3mm](3.5,1.75)(4,1.5)
%
\put(2.4,0.5){\large$\bar{\zeta}'_{1}$}
%
%%%%%%%%%%%%%%%%%%%%%%%%%%%%%%%%%%%%%%%%%%%%%%%%%%%%
%
%\Psline[linecolor=black,linewidth=0.3mm]{->}(3,0)(3.75,0.37)
%\psline[linecolor=black,linewidth=0.3mm](3.4,0.2)(4,0.5)
%
%\put(3.5,0.55){$q$}
%
%\psline[linecolor=black,linewidth=0.3mm](3,0)(3.75,-0.37)
%\psline[linecolor=black,linewidth=0.3mm]{<-}(3.4,-0.2)(4,-0.5)
%
%\put(3.5,-0.8){$\bar{q}$}
%
%\psline[linecolor=black,linewidth=0.3mm]{<-}(0,1)(-1.5,-0.5)
%\psline[linecolor=black,linewidth=0.3mm]{->}(1,2)(-0.4,3.4)
%
%
%Momenta
%
%\psline[linecolor=black,linewidth=0.3mm]{->}(2.2,2.7)(3.8,2.7)
%\put(3,2.6){\Large$X^{\mu}$}
%
%\psline[linecolor=black,linewidth=0.3mm]{->}(7.8,1.2)(7.8,2.8)
%\put(3,1){\Large$k$}
%
\put(-1.5,1.1){$(k,t)$}
\put(-1.5,2.2){$(p,s)$}
\put(2.7,1.1){$(k',t')$}
\put(2.7,2.2){$(p',s')$}
%
%\put(-0.2,3.6){\Large$e^{-}$}
%\put(-0.4,-0.1){\Large$e^{+}$}
%\put(-1.1,3.1){\Large$q$}
%\put(-1.1,0.7){\Large$q^{\prime}$}
%
%\put(0.6,-0.3){$p$}
%
%\put(1.3,0){$\Gamma_{\mu}^{(3)\;abc}(p)=g_{1}f^{abc}p_{\mu}$}
%
%
\end{pspicture}
%
%\vspace{0.2cm}
%
\caption{\scshape{The electron-positron scattering $e^{+} \, e^{-} \, \rightarrow \, X \, \rightarrow \, \bar{\zeta}'_{1} \, \zeta'_{1}$ that can connect
the sector of leptons with the dark matter content via hidden photon or $X$-boson.}} \label{ScatteringX}
\end{center}
\end{figure}
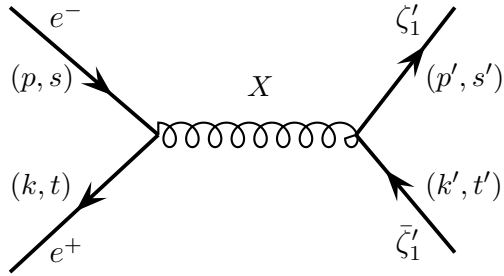
\begin{figure}[h]
\centering
\includegraphics[scale=0.6]{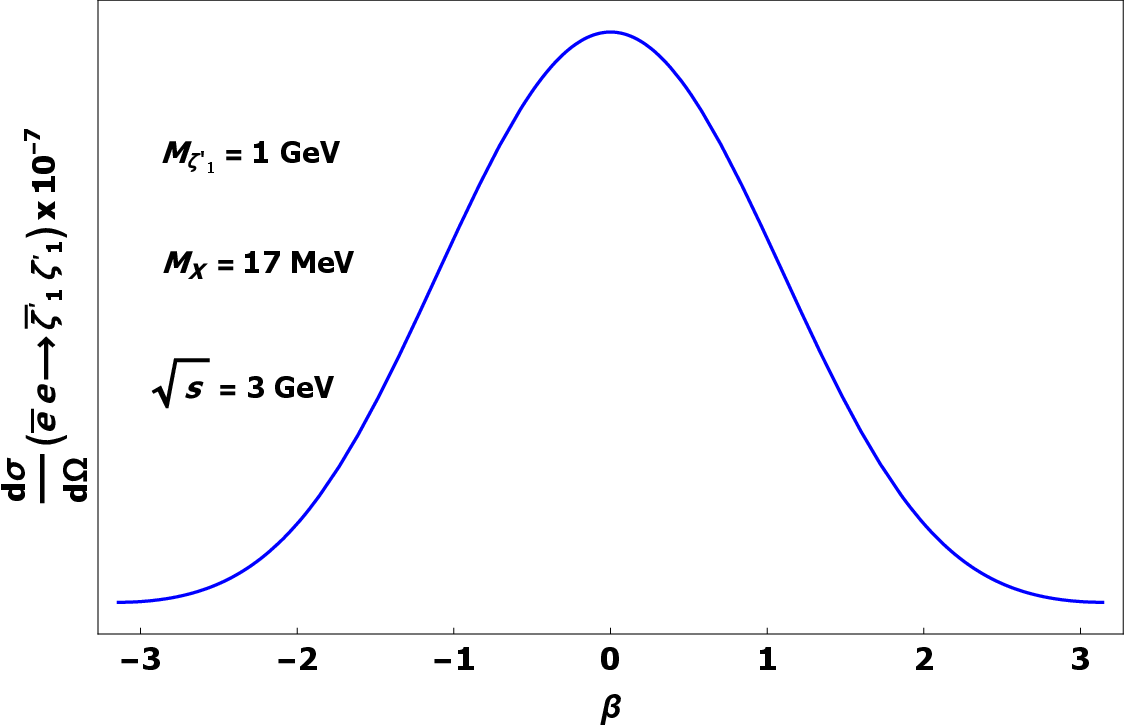}
\caption{The differential cross section of $e^{+} \, e^{-} \, \rightarrow \, \bar{\zeta}'_{1} \, \zeta'_{1} $ as function of the $\beta$-angle.
We use the masses $M_{\zeta'_{1}}=1 \, \mbox{GeV}$
and $M_{X}=17 \, \mbox{MeV}$, the CM-energy as $\sqrt{s}=3 \, \mbox{GeV}$.}
\label{SecdiffbetaX}
\end{figure}

\begin{eqnarray}
i{\cal M}_{X}&=&\bar{v}(k,t) \, i \, e \, \gamma^{\mu} \left( \, c_{V}^{e}-c_{A}^{e} \, \gamma_{5} \, \right) \, u(p,s)
%\nonumber \\
%&&
%\times \,
\, \frac{-i \, \eta_{\mu\nu}}{(k+p)^2-M_{X}^{2}} \, \times
\nonumber \\
&&
\times \, \bar{u}(p',s') \, i \, e \, \gamma^{\nu} \left( \, c_{V}^{\zeta}-c_{A}^{\zeta} \, \gamma_{5} \, \right) \, v(k',t') \, .
\end{eqnarray}
The differential cross section for the CM collision is
\begin{eqnarray}\label{diffsecX}
&&
\frac{d\sigma}{d\Omega}(e^{+} \, e^{-} \rightarrow \bar{\zeta}'_{1} \, \zeta'_{1})=\frac{9\alpha^{2}}{2s}\left(1-\frac{M_{X}^{2}}{s}\right)^{-2}
\times
\nonumber \\
&&
\hspace{-0.5cm}
\times \, \left\{ \left[ \varepsilon_{\zeta_{L}}\left(\chi \cos\theta_{W}+\frac{\varepsilon_{e_{L}}}{2}\right)
-\varepsilon_{\zeta_{R}}\left(\chi \cos\theta_{W}-\frac{\varepsilon_{e_{L}}}{2}\right) \right]^2 \, \times
\right.
\nonumber \\
&&
\hspace{-0.5cm}
\left.
\times \, \left[3+2 \, \sqrt{1-\frac{4M_{\zeta'_{1}}^2}{s}}\cos\beta+\left(1-\frac{4M_{\zeta'_{1}}^2}{s}\right)\cos^{2}\beta \right]+
\right.
\nonumber \\
&&
%\hspace{-0.5cm}
\left.
%+\left[ \phantom{\frac{1}{2}} \hspace{-0.25cm}
+ \, \chi^{2}\cos^{2}\theta_{W} \left( \varepsilon_{\zeta_{L}}- \varepsilon_{\zeta_{R}}\right)^{2}
- \frac{\varepsilon_{e_{L}}^{\,2}}{4} \, \left( \varepsilon_{\zeta_{L}}+\varepsilon_{\zeta_{R}} \right)^{2}
%\right]
\right\} \, ,
\hspace{0.5cm}
\end{eqnarray}
where $s>4 \, M_{\zeta'_{1}}^{2}$. If we use the $\zeta'_{1}$-mass as $M_{\zeta'_{1}}=1 \, \mbox{GeV}$,
the condition on the CM-energy is $\sqrt{s} \, > \, 2 \, \mbox{GeV}$. The simplest case of (\ref{diffsecX})
assumes that the $\varepsilon$-parameters of the left-components are of the same order as the right-components, {\it i. e.},
$\varepsilon_{\zeta_{L}} \simeq \varepsilon_{\zeta_{R}}$; so, the previous expression is reduced to
\begin{eqnarray}\label{diffsecX}
&&
\frac{d\sigma}{d\Omega}(e^{+} \, e^{-} \, \rightarrow \, \bar{\zeta}'_{1} \, \zeta'_{1})=\frac{9\alpha^{2}}{s}\left(1-\frac{M_{X}^{2}}{s}\right)^{-2}
\varepsilon_{e_{L}}^{2} \, \varepsilon_{\zeta_{L}}^2
\times
\nonumber \\
&&
%\hspace{-0.5cm}
\times \, \left[1+\sqrt{1-\frac{4M_{\zeta'_{1}}^2}{s}}\cos\beta+\frac{1}{2}\left(1-\frac{4M_{\zeta'_{1}}^2}{s}\right)\cos^{2}\beta \right] \, ,
\hspace{0.5cm}
\end{eqnarray}
The differential cross section (\ref{diffsecX}) is illustrated in figure (\ref{SecdiffbetaX}).
The corresponding total cross section is obtained upon integration over the solid angle:
\begin{eqnarray}
&&
\sigma(e^{+} \, e^{-} \, \rightarrow \, \bar{\zeta}'_{1} \, \zeta'_{1})=\frac{42\pi \, \alpha^{2}}{s}\frac{\varepsilon_{e_{L}}^{2} \, \varepsilon_{\zeta_{L}}^{2}}{\left(1-M_{X}^{2}/s \right)^{2}} \left(1-\frac{4}{7}\frac{M_{\zeta'_{1}}^2}{s}\right) \; .
\end{eqnarray}
Using the previous masses, the C.M. energy of $\sqrt{s}=3 \, \mbox{GeV}$ and $|\varepsilon_{e_{L}}| \lesssim 4 \times 10^{-3}$,
the $\sigma$-function has the following estimate
\begin{eqnarray}
\sigma(e^{+} \, e^{-} \, \rightarrow \, \bar{\zeta}'_{1} \, \zeta'_{1})= 1.2 \, \times \, 10^{-11} \, \varepsilon_{\zeta_{L}}^{2} \, \mbox{GeV}^{-2} \; .
\end{eqnarray}
%

%
%is plotted as function of $CM$-energy scale $\sqrt{s}$ in the figure
%(\ref{SecTotalMX}).
%%
%\begin{figure}[h]
%\centering
%\includegraphics[scale=0.55]{SecTotalMX.eps}
%\caption{The total cross section of $e^{+} \, e^{-} \, \rightarrow \, \bar{\zeta}'_{1} \, \zeta'_{1} $
%as function of the CM-energy $\sqrt{s}$.}
%Here, we adopt the CM-energy values by $\sqrt{s}=13 \, \mbox{TeV}$ and $\sqrt{s}=8 \, \mbox{TeV}$. }
%\label{SecTotalMX}
%\end{figure}
%

%
\subsection{The electron and muon $g-2$ factors}

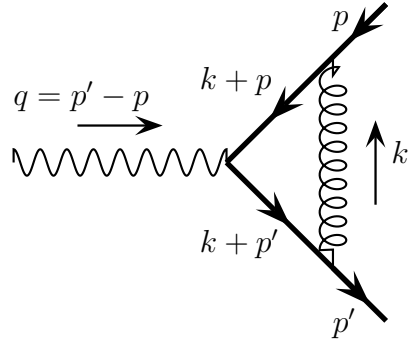
\begin{figure}[!h]%\label{figvm}
\begin{center}
\newpsobject{showgrid}{psgrid}{subgriddiv=1,griddots=10,gridlabels=6pt}
\begin{pspicture}(0,-0.7)(7,3.5)
%\showgrid
\psset{arrowsize=0.2 2}
\psset{unit=0.7}
%
%\put(-4.5,0){$g_{1}f^{abc}\partial_{\mu}\bar{\eta}^{a}G^{\mu\;c}\eta^{b}\;:$}
%
%%%%%%%%%%%%%%%%%%%%%%%%%%%%%%%%%%%%%%%%%%%%%%%%%%%%%%%%%%%%%
%
\pscoil[coilarm=0,coilaspect=0,coilwidth=0.5,coilheight=1.0,linecolor=black](1,2)(5,2)
\psline[linecolor=black,linewidth=0.7mm]{->}(8,5)(7.3,4.3)
\psline[linecolor=black,linewidth=0.7mm]{->}(8,5)(5.8,2.8)
\psline[linecolor=black,linewidth=0.7mm](6,3)(5,2)
\psline[linecolor=black,linewidth=0.7mm]{->}(5,2)(6.2,0.8)
\psline[linecolor=black,linewidth=0.7mm](6,1)(8,-1)
\psline[linecolor=black,linewidth=0.7mm]{->}(7,0)(7.7,-0.7)
\pscoil[coilarm=0.35,coilwidth=0.5,coilheight=1,linecolor=black](7,0)(7,4)
%
%%%%%%%%%%%%%%%%%%%%%%%%%%%%%%%%%%%%%%%%%%%%%%%%%%%%
%
%\put(-1,0){\Large$+\;\;\; {\cal O}\,(e^{4})$}
%
%Momenta
%
\psline[linecolor=black,linewidth=0.3mm]{->}(2.2,2.7)(3.8,2.7)
\put(1,3.1){\large$q=p^{\prime}-p$}
\psline[linecolor=black,linewidth=0.3mm]{->}(7.8,1.2)(7.8,2.8)
\put(8.1,2){\large$k$}
\put(7,4.6){\large$p$}
\put(7,-1.2){\large$p^{\prime}$}
\put(4.5,3.4){\large$k+p$}
\put(4.5,0.3){\large$k+p^{\prime}$}
%\put(0.6,-0.3){$p$}
%
%\put(1.3,0){$\Gamma_{\mu}^{(3)\;abc}(p)=g_{1}f^{abc}p_{\mu}$}
%
%
\end{pspicture}
%
%\vspace{0.2cm}
%
\caption{\scshape{The one-loop correction of the $X$-boson to the QED vertex diagram.}}\label{VertexX}
\end{center}
\end{figure}

The $g-2$ factor is an important experimental physical quantity to estimate the
$\varepsilon$-parameters of the leptonic sector in interaction with the $X$-boson.
With those, we can obtain the form factors associated to the vertex
$\ell^{\pm}$-photon including the correction of the $X$-boson propagator.
This correction is illustrated in the figure (\ref{VertexX}).
%
%%%%%%%%%%%%%%%%%%%% Vertex with the X-Boson correction %%%%%%%%%%%%%%%%%%%%%%%%
%
%
With that, we write the QED vertex of lepton-photon corrected by the diagram (\ref{VertexX}) as the sum : $\Gamma^{\mu}(p,p^{\prime})=
\Gamma_{QED}^{\mu}(p,p^{\prime})+\Lambda^{\mu}(p,p^{\prime})$. The diagram is represented by the loop $4$-momentum integral:
\begin{eqnarray}\label{IntLambdaX}
\Lambda^{\mu}(p,p')&=& \int \frac{d^4k}{(2\pi)^4} \, i \, e \, \gamma_{\alpha}\left( c_{V}^{\ell}-c_{A}^{\ell} \, \gamma_{5} \right)
%\nonumber \\
%&&
%\times \,
\frac{i\left(\slash{\!\!\!k}+\slash{\!\!\!p}'+m\right)}{(k+p')^{2}-m^{2}}
\left(\, i \, e \, \gamma^{\mu} \, \right) \, \times
\nonumber \\
&&
\times \,
\frac{i\left(\slash{\!\!\!k}+\slash{\!\!\!p}+m\right)}{(k+p)^{2}-m^{2}}
\, i \, e \, \gamma^{\alpha}\left( c_{V}^{\ell}-c_{A}^{\ell} \, \gamma_{5} \right) \frac{-i}{k^{2}-M_{X}^{2}} \; ,
\end{eqnarray}
where $m$ stands for the electron or muon mass. Obviously, the integral is divergent in the
ultraviolet regime, but we are interested in its finite part. To simplify the previous integral,
we use here that the vector coefficient is of the same order of the axial coefficient for Leptons,
{\it i. e.}, $c_{V}^{\ell} \simeq c_{A}^{\ell}=-\chi \, \cos\theta_{W}=-\varepsilon_{\ell_{L}}/2=+\varepsilon_{\ell_{R}}$.
The integral can be calculated by using the known methods of Feynman integrals; then, the finite part of the vertex
can be extracted and gives the leptonic current $\bar{u}\Gamma^{\mu}(q)u$, that can be written as
\begin{equation}\label{Jmuleptons}
J^{\mu}_{em}=\bar{u}(p^{\prime})\Gamma^{\mu}(q)u(p)=\bar{u}(p^{\prime})
\left[ F_{1}\left(q^{2}\right) \gamma^{\mu}
+F_{2}(q^2) \, \gamma^{5} \gamma^{\mu}
+F_{3}(q^{2}) \, i \, \frac{\sigma^{\mu\nu}q_{\nu}}{2m}
%\right.
%\nonumber \\
%\left.
%\hspace{-0.5cm}
+  F_{4}(q^2) \, \frac{q^{\mu}\gamma^{5}}{2m}
\right] u(p) \; ,
\end{equation}
%
%
%\begin{equation}
%\bar{u}(p^{\prime})\Gamma^{\mu}(q)u(p)=\bar{u}(p^{\prime})
%\left[ \gamma^{\mu} F_{1}\left(q^{2}\right)
%+i \, \frac{\sigma^{\mu\nu}q_{\nu}}{2m} \, F_{2}(q^{2})
%\right.
%\nonumber \\
%\left.
%\hspace{-0.5cm}
%-i \, \frac{\sigma^{\mu\nu}\ell_{\nu}}{2m} \, \gamma^{5} \, F_{E}(q^2)
%+\frac{q^{\mu}\gamma^{5}}{2m} \, F_{3}(q^2) \right] u(p) \; ,
%\end{equation}
%
where $q^{\mu}$ stands for the momentum transfer and $\ell^{\mu}=p^{\mu}+p^{\prime \, \mu}$, as before.
%, and a term in $\gamma_{5}$ is dropped out if we adopt the Lorenz gauge.
Here, we obtain the QED form factors with the correction of $\varepsilon$-parameters
\begin{eqnarray}
F_{1}(q^2)&=&F_{1}^{QED}(q^{2})
+\frac{\alpha \, \varepsilon_{\ell_{L}}^{2}}{4\pi}
%\times
%\nonumber \\
%\times \,
\int_{0}^{1}dxdydz \, \delta(x+y+z-1)
\, \times
\nonumber \\
&&
\times \,
\frac{4z-1-z^2-(1-x)(1-y)q^{2}/m^{2}}{(1-z)^{2}+zM_{X}^{2}/m^{2}-xyq^{2}/m^{2}} \; ,
\end{eqnarray}
and
\begin{eqnarray}
F_{3}(q^{2})&=&\frac{\alpha}{2\pi}\int_{0}^{1}dxdydz \, \delta(x+y+z-1) \, \frac{2z(1-z)}{(1-z)^{2}-xy \, q^{2}/m^{2}}
\nonumber \\
&&
\hspace{-1.2cm}
-\frac{\alpha \, \varepsilon_{\ell_{L}}^{2}}{16\pi^2}
%\times
%\nonumber \\
%\times \,
\int_{0}^{1}dxdydz \, \delta(x+y+z-1) \,
%\times
%\nonumber \\
%\times \,
\frac{2z(1-z)}{(1-z)^{2}+zM_{X}^{2}/m^{2}-xy \, q^{2}/m^{2}}
\; .
\end{eqnarray}
%
%The others form factors emerge due to $X$-boson axial current :
%
%\begin{eqnarray}
%F_{E}(q^{2})&=&\frac{\alpha \, \varepsilon_{\ell_{L}}^{2}}{16\pi^{2}}
%\int_{0}^{1}dxdydz \, \delta(x+y+z-1) \,
%\times
%\nonumber \\
%\times \,
%\frac{1-2z-z^2+(1-x)(1-y) \, q^{2}/m^{2}}{(1-z)^{2}+zM_{X}^{2}/m^{2}-xy \, q^{2}/m^{2}}
%\nonumber \\
%F_{3}(q^{2})&=& \frac{\alpha \, \varepsilon_{\ell_{L}}^{2}}{16\pi^{2}}
%\int_{0}^{1}dxdydz \, \delta(x+y+z-1) \,
%\times
%\nonumber \\
%&&
%\times \,
%\frac{1+4z+z^2-2(x-y)^{2}-(1-x)(1-y) \, q^{2}/m^{2}}{(1-z)^{2}+zM_{X}^{2}/m^{2}-xy \, q^{2}/m^{2}}
%\; .
%\hspace{0.5cm}
%\end{eqnarray}
%
The $F_{2}$ form factor is associated with the CP-violation of the $X$-fermion coupling.
The conservation current yields us the condition : $F_{2}(q^2)=-q^{2} \, F_{4}(q^2)/4m^{2}$;
in which the current (\ref{Jmuleptons}) can be rewritten as
\begin{equation}
J^{\mu}_{em}=\bar{u}(p^{\prime})\Gamma^{\mu}(q)u(p)=\bar{u}(p^{\prime})
\left[ F_{1}\left(q^{2}\right) \gamma^{\mu}
+F_{4}(q^2) \, \gamma^{5} \left(q^{\mu} \, \slash{\!\!\!q}-q^{2}\gamma^{\mu}\right)
+F_{3}(q^{2}) \, i \, \frac{\sigma^{\mu\nu}q_{\nu}}{2m}
%\right.
%\nonumber \\
%\left.
%\hspace{-0.5cm}
\right] u(p) \; ,
\end{equation}
Explicitly, we have the anapole term with the $F_{A}$-form factor given by
\begin{equation}
F_{A}(q^2)=\frac{F_{4}(q^2)}{2m}= \frac{\alpha \, \varepsilon_{\ell_{L}}^{2}}{8 \pi}
\int_{0}^{1}dxdydz \, \delta(x+y+z-1) \,
%\times
%\nonumber \\
%\times \,
\frac{1+z-(x-y)^2}{(1-z)^{2}+zM_{X}^{2}/m^{2}-xy \, q^{2}/m^{2}} \; .
\end{equation}
The interesting analysis now is the situation where $q^{2}=0$ for the external photon
in the form factor $F_{3}$. The first term yields the
$(g-2)_{\ell}$ factor for the lepton with the correction of the $c_{V}^{\ell}$
coefficient, which in turn, depends on $\varepsilon_{\ell}$-parameters.
%The $F_{E}$-factor yields the lepton's Electric Dipole Momentum (EDM) and
%it is proportional to the $\varepsilon_{\ell_{L}}$ coefficient. 
For the electron case, the $F_{3}(0)$ is given by
\begin{eqnarray}
F_{3}^{e}(0)\simeq\frac{\alpha}{2\pi}\left(1-\frac{\varepsilon_{e_{L}}^{2}}{12\pi} \, \frac{m_{e}^{2}}{M_{X}^{2}} \right) \; ,
\end{eqnarray}
where we have used that $M_{X} \gg m_{e}$. Using the $(g-2)_{e}$ uncertainty in the literature,
we obtain the upper bound on $\varepsilon_{e_{L}}$-parameter
\begin{eqnarray}
|\varepsilon_{e_{L}}| \, \lesssim \, 5.3 \, \times \, 10^{-5} \; ,
\end{eqnarray}
and, thus, we estimate the other parameters $|\varepsilon_{e_{R}}| \, \lesssim \, 2.6 \, \times \, 10^{-5}$
and $|\chi| \, \lesssim \, 3 \, \times \, 10^{-5}$. For the muon, we obtain
\begin{eqnarray}
F_{3}^{\mu}(0)\simeq\frac{\alpha}{2\pi}\left(1-\frac{\varepsilon_{\mu_{L}}^{2}}{8\pi} \right) \; ,
\end{eqnarray}
and using the corresponding $(g-2)_{\mu}$ uncertainty, we obtain the following upper bound
\begin{eqnarray}
|\varepsilon_{\mu_{L}}| \, \lesssim \, 8.7 \, \times \, 10^{-5} \; .
\end{eqnarray}
%
%Now we use these estimative to obtain the $F_{E}$ factor for electron and muon,
%respectively,
%
%\begin{eqnarray}
%F_{E}^{e}(0) &=& \frac{\alpha \, \varepsilon_{e_{L}}^{2}}{16\pi^{2}} \frac{m_{e}^{2}}{M_{X}^{2}} \,
%\int_{0}^{1}dz \,
%\times
%\nonumber \\
%\times \,
%\frac{(1-z)(1-2z-z^2)}{z+(1-z)^2 \, m_{e}^{2}/M_{X}^{2}} = \frac{\alpha \, \varepsilon_{e_{L}}^{2}}{16\pi^{2}} \, \times \, 4.3 \, \times \, 10^{-3}
%\lesssim 5.5 \, \times \, 10^{-16} \; ,
%\nonumber \\
%F_{E}^{\mu}(0) &=& \frac{\alpha \, \varepsilon_{\mu_{L}}^{2} }{16\pi^{2}} \,
%\int_{0}^{1}dz \,
%\times
%\nonumber \\
%\times \,
%\frac{(1-z)(1-2z-z^2)}{(1-z)^{2}+z \, M_{X}^{2}/m_{\mu}^{2}}=-5.41 \, \times \, \frac{\alpha \, \varepsilon_{\mu_{L}}^{2} }{16\pi^{2}} \lesssim
%- 1.8 \, \times \, 10^{-12} \; .
%\hspace{0.5cm}
%\end{eqnarray}
%
%The expression for the MDE is $d^{(\ell)}=e \, F_{E}^{(\ell)}(0)/2m_{\ell}$, so we obtain the upper bounds for MDE of electron and muon, respectively
%
%\begin{eqnarray}
%d^{(e)}&=&\frac{e}{2 \, m_{e}} \, F_{E}^{(e)}(0) \lesssim 1.08 \, \times \, 10^{-26} \, \mbox{e} \cdot \mbox{cm} \; ,
%\nonumber \\
%d^{(\mu)}&=&\frac{e}{2 \, m_{\mu}} \, F_{E}^{(\mu)}(0) \lesssim -1.6 \, \times \, 10^{-25} \, \mbox{e} \cdot \mbox{cm}  \; .
%\end{eqnarray}
%

%
\section{Concluding Comments}

We have made efforts to discuss a model with an extra $U(1)_{K}$-factor that may describe a possible scenario of a Particle Physics beyond the Standard Model (SM).
The proposal is based on the gauge group $SU_{L}(2) \times U_{R}(1)_{J}\times U(1)_{K}$, where the extra $U(1)_{K}$-factor introduces
a (new) massive and neutral vector boson. The mass of the latter is generated upon a spontaneous symmetry breaking mechanism that defines an extra VEV
scale, beyond that VEV of $246 \, \mbox{GeV}$ associated with the usual Higgs field of SM. In our model, the Higgs sector displays two
scalar fields and the gauge symmetry allows interactions between them. This mechanism can be introduced in two ways :
in the first case, the SSB pattern-I takes place through a VEV scale-$u$, where $u > 246 \, \mbox{GeV}$, and, as consequence, we fix a mass
for the hypothetical $Z'$-boson around 2 TeV. Next, the SM Higgs acquires its VEV of $246 \, \mbox{GeV}$, breaks the
electroweak symmetry to give the masses for $W^{\pm}$ and $Z$. The sequence of SSB mechanism is as follows:
\begin{eqnarray*}
SU_{L} \! \times \! U_{R}(1)_{J} \! \times \! U(1)_{K}
%\nonumber \\
%&&
\stackrel{\langle \Xi \rangle_{0}}{\longmapsto}
SU_{L}(2) \!\times\! U_{Y}(1) \stackrel{\langle \Phi \rangle_{0}}{\longmapsto} U_{em}(1) \; .
\end{eqnarray*}
The result for the mass of $Z'$ is
%in agreement with the mass values for the boson $Z'$
estimated by the ATLAS and CMS Collaborations as a possible particle at the $\mbox{TeV}$-scale.
The maximum VEV scale for $M_{Z'}=2 \, \mbox{TeV}$ is $u=2.8 \, \mbox{TeV}$, and it allows an estimation of the
mass of the new Higgs within the range $1.2 \, \mbox{TeV} < M_{F} < 3.7 \, \mbox{TeV}$.
%that is next the diphoton resonance at $750 \, \mbox{GeV}$.
Furthermore, the subgroup $U(1)_{K}$ also introduces a new family of fermions in the TEV-scale,
which we call $\zeta_{i}$-fermions $(i=1,2,3)$, that could be candidates to the dark matter particles.
It is a set of neutral heavy fermions associated with the VEV scale of $2.8 \, \mbox{TeV}$ and they
guarantee that model be free from the chiral anomaly. We use the masses for $\zeta_{i}$ in the range of
$0.5-1 \, \mbox{TeV}$, motivated by recent simulations of dark matter fermions scenario in the
CMS-Collaboration. Furthermore, these new fermions also mix with a Dirac's right-neutrino component.
This new mixing motivated us to investigate the MDM of the $\zeta_{i}$-fermions,
and the transition MDMs of the $\zeta_{i}$ mixed with neutrinos. Since $\zeta_{i}$
are neutral fermions, their MDMs depend on their masses and also on the mixing with the right-neutrinos.
At one-loop, the diagonal $\zeta_{1}$-MDM for mass of $0.5 \, \mbox{TeV}$ is estimated by $\mu_{M}^{(\zeta'_{1})}=1.2 \, \times \, 10^{-21} \, \mu_{B}$.
%and $\mu_{E}^{(\zeta')}=7.3 \, \times \, 10^{-11} \, \theta^{2} \, \mu_{B}$, respectively.
%

%
In the second scenario, the SSB mechanism pattern-II is introduced to break the composite subgroup $SU_{L}(2) \times U_{R}(1)_{J}$
in the VEV scale of the SM, $246 \, \mbox{GeV}$, to yield the usual $Z$-mass, and so, a further SBB defines the VEV scale-$u$, $u < 246 \, \mbox{GeV}$.
In this second situation, we have the scenario of a lighter extra gauge boson; it may describe a phenomenology of elementary
particle physics at a lower energy scale. The scheme for the pattern-II SSB is displayed below:
\begin{eqnarray*}
SU_{L} \times U_{R}(1)_{J} \times U(1)_{K}
%\nonumber \\
%&&
\stackrel{\langle \Phi \rangle_{0}}{\longmapsto}
U(1)_{G} \times U(1)_{K} \stackrel{\langle \Xi \rangle_{0}}{\longmapsto} U(1)_{em} \; .
\end{eqnarray*}
This proposal may be discussed in connection with the light X-Boson, dark photon or para-photon phenomenology, at $\mbox{MeV}$-scale or lower,
which can be related to dark matter particles. This context includes the recent description of a new boson needed to explain the excited
$8$-Beryllium nuclear decay. In this case, the mass of the $X$-Boson is fixed at the mass scale of $m_{X}=17 \, \mbox{MeV}$.
Therefore, the results of SSB lead us to the VEV scale of $u\simeq 1.9-19 \, \mbox{GeV}$
that must be origin of a hidden Higgs within the range mass $1.9 \, \mbox{GeV} \lesssim M_{F} \lesssim 19 \, \mbox{GeV}$. In this context,
the $\zeta_{i}$-fermions are interpreted as light particles with masses in the range of $0.3-6.6 \, \mbox{GeV}$. It can be a candidate
to dark matter constituent. After the SSB pattern-II, we study the phenomenology of the $X$-Boson through its
decay into electron, neutrino and light $u$-quark, and the scattering $e^{+} \, e^{-} \rightarrow X \rightarrow \bar{\zeta}'_{1} \, \zeta'_{1}$,
with $M_{\zeta'_{1}}=1 \, \mbox{GeV}$; this can connect the SM with the dark sector with a hidden photon mediating the interaction.
To conclude, we calculate the correction to the Quantum Electrodynamics (QED) vertex due to axial interaction of $X$-Boson with the leptons of the SM; 
in particular, for the electron and muon cases.
The results impose the bounds $|\varepsilon_{e_{L}}| \, \lesssim \, 5.3 \, \times \, 10^{-5}$ and
$|\varepsilon_{\mu_{L}}| \, \lesssim \, 8.7 \, \times \, 10^{-5}$ for the electron and muon, respectively, which measures the
magnitude of the weak interaction of the electron/muon with the $X$-Boson. The Electric Dipole Momentum of the charged leptons
must emerge at higher loop orders, in both cases of the $Z'$ and $X$-boson scenarios. This is an issue we are now investigating and we shall report on it 
in a further paper.

%The decay of new Higgs and the influence on the $X$-boson can be the motivation to a forthcoming paper.

A unification scheme including $W'$- and $Z'$-bosons is in the framework of an left-right symmetric model 
$SU_{L}(2)\times SU_{R}(2) \times U(1)_{J} \times U(1)_{K}$ may also be the subject of a further investigation to be pursued 
in the presence of dark matter fermion content.
%

%\lllll

%\newpage

%

%
\end{document}